\documentclass[a4paper,fleqn,usenatbib]{mnras}
\usepackage[T1]{fontenc}
\usepackage{ae,aecompl}

\usepackage{graphicx}	
\usepackage{amsmath}	
\usepackage{amssymb}
\usepackage{nicefrac}
\usepackage[none]{hyphenat}

\def\Msun{M$_{\odot}$}

\def\Rearth{R$_{\oplus}$}
\def\Mearth{M$_{\oplus}$}

\title[Impact of Binaries on Planet Statistics - I.]{Impact of Binary Stars on Planet Statistics - ~~~~~~~~~~~~~~~~~ I. Planet Occurrence Rates, Trends with Stellar Mass, and Wide Companions to Hot Jupiter Hosts}

\author[Moe \& Kratter]{Maxwell Moe$^{1}$\thanks{E-mail: moem@arizona.edu} and
Kaitlin M. Kratter$^{1}$ \\
$^{1}$Steward Observatory, University of Arizona, 933~N.~Cherry~Ave., Tucson, AZ 85721, USA
}

\date{Accepted XXX. Received YYY; in original form ZZZ}
\pubyear{2019}

\begin{document}
\label{firstpage}
\pagerange{\pageref{firstpage}--\pageref{lastpage}}
\maketitle

\begin{abstract}
 Close binaries suppress the formation of circumstellar (S-type) planets and therefore significantly bias the inferred planet occurrence rates and statistical trends. After compiling various radial velocity and high-resolution imaging surveys, we determine that binaries with $a$~$<$~1~au fully suppress S-type planets, binaries with $a$~=~10\,au host close planets at 15$_{-12}^{+17}$\% the occurrence rate of single stars, and wide binaries with $a$~$>$~200\,au have a negligible effect on close planet formation. We show that $F$~$=$~43\%\,$\pm$\,7\% of solar-type primaries do not host close planets due to suppression by close stellar companions. By removing spectroscopic binaries from their samples, radial velocity surveys for giant planets boost their detection rates by a factor of 1/(1-$F$)~$=$~1.8\,$\pm$\,0.2 compared to transiting surveys. This selection bias fully accounts for the discrepancy in hot Jupiter occurrence rates inferred from these two detection methods. Correcting for both planet suppression by close binaries and transit dilution by wide binaries, the occurrence rate of small planets orbiting single G-dwarfs is 2.1\,$\pm$\,0.3 times larger than the rate inferred from all G-dwarfs in the {\it Kepler} survey. Additionally, about half (but not all) of the observed increase in small, short-period planets toward low-mass hosts can be explained by the corresponding decrease in the binary fraction. Finally, we demonstrate that the apparent enhancement of wide stellar companions to hot Jupiter hosts is due to multiple selection effects. Very close binaries, brown dwarf companions, and massive planets with $M_2$~$>$~7\,M$_{\rm J}$ within $a$~$<$~0.2~au preferentially have metal-poor hosts and wide tertiary companions, but genuine hot Jupiters with $M_{\rm p}$~=~0.2\,-\,4\,M$_{\rm J}$ that formed via core accretion instead favor metal-rich hosts and do not exhibit a statistically significant excess of wide stellar companions.
\end{abstract}


\section{Introduction}
\label{Introduction}

About half of solar-type main-sequence (MS) primaries have stellar companions, and the binary fraction of young pre-MS stars is even larger \citep{Duquennoy1991,Raghavan2010,Duchene2013,Tokovinin2014,Moe2017}. A complete picture of planet statistics must therefore be examined in the context of how binaries influence planet formation and thus occurrence rates.  Very close pre-MS binaries ($a$~$<$~0.5~au; $\approx$\,5\% of solar-type systems) accrete from circumbinary disks, which can form circumbinary P-type planets such as the 10 systems discovered by {\it Kepler} \citep{Thebault2015,Bromley2015,Kratter2017,Martin2018,Czekala2019}. Meanwhile, components in wide binaries ( $a$~$>$~200~au; $\approx$\,15\% of field systems) accrete nearly independently from their own relatively unperturbed circumprimary and circumsecondary disks \citep{White2001,Tobin2016,Lee2019}, which may produce planets in circumstellar S-type configurations. Binaries with intermediate separations ($a$~=~0.5\,-\,200~au; $\approx$\,30\% of solar-type systems) open a large inner gap in their circumbinary disks, which feed material onto smaller circumprimary and circumsecondary disks \citep{Artymowicz1994,Bate1997}. 

Theoretical models have shown that such binaries with intermediate separations sculpt and suppress planet formation, either by increasing turbulence in the disks, truncating the mass and radius of the circumprimary disk, and/or by accreting from or clearing out disk material on timescales faster than the planets can form \citep{Artymowicz1994,Haghighipour2007,Thebault2008,Xie2010,Silsbee2015,Rafikov2015a,Rafikov2015b}. For example, within the same star-forming environment, the observed disk fraction of resolved close binaries ($a$~=~1\,-\,50~au) is substantially lower than the disk fractions measured for wide binaries and single stars \citep{Kraus2012,Harris2012,Cheetham2015}. Similarly, the bias-corrected frequency of double-lined spectroscopic binaries (SB2s; $a$~$<$~10~au) is lower among Class~II T~Tauri stars with disks compared to Class~III T~Tauri stars without disks, suggesting more massive companions with $q$~=~$M_2$/$M_1$~$>$~0.6 quickly consume or disrupt their disks, thereby accelerating the Class II phase \citep{Kounkel2019}. 

Radial velocity (RV) and high-resolution imaging surveys of various types of planet hosts reveal a close binary fraction that is smaller than that observed in the field \citep{Knutson2014,Wang2014b, Wang2015c, Wang2015d, Ngo2016, Kraus2016, Ziegler2019}, qualitatively consistent with the theoretical expectations of planet suppression. Based on adaptive optics (AO) observations of {\it Kepler} objects of interest (KOIs), \citet{Kraus2016} measured a factor of 2.9 deficit of stellar companions within $a$~$<$~47~au compared to the field population, and they concluded that 19\% of solar-type primaries do not host planets due to suppression by close binaries. As emphasized in these studies, the deficit of close stellar companions to planet hosts is not simply due to dynamical stability alone.  Hosts of hot Jupiters with $P_{\rm p}$~$<$~10~days can have stellar companions across $a$~=~1\,-\,10~au that are completely dynamical stable, but their occurrence rate is substantially reduced compared to close binaries in the field \citep{Knutson2014,Wang2015c,Ngo2016}. Instead, during the planet formation process, close binaries inhibit the growth of solid material into large planets via one of the mechanisms outlined above. 

The RV and imaging surveys cited above examined different types of planet hosts with varying degrees of sensitivity, and therefore reported different results. \citet{Matson2018} in particular did not find a statistically significant difference between the close binary fractions of K2 planet hosts and field stars in their speckle imaging survey. They therefore concluded that close binaries do not substantially suppress planet formation as previously claimed.  The major goal of the first part of this paper (Section~\ref{part1}) is to compile and homogeneously analyze the different RV and imaging samples, and to discuss the implications for planet suppression by close binaries. In Sections~\ref{closebinfrac} and Appendix~\ref{Appendix}, we measure the bias-corrected binary fraction and period distribution as a function of stellar mass\footnote{ The analysis of how binary properties vary with metallicity and the consequences for planet statistics is the subject of Paper~II.}. After compiling the various RV and imaging surveys of planet hosts (Section~\ref{Suppression}), we find that planet suppression is a continuous function of binary separation, not a step function as previously modeled. In Sections \ref{CloseBinary} and \ref{Implications}, we demonstrate that 33\%\,$\pm$\,5\% and 43\%\,$\pm$\,7\% of solar-type primaries in volume-limited and magnitude-limited samples, respectively, do not host close planets due to suppression by close binaries. These values are considerably larger than the estimate of 19\% reported in \citet{Kraus2016}. 

 We subsequently examine whether binaries can account for the previously reported discrepancies and trends in planet occurrence rates (Section~\ref{bias}). For example, the hot Jupiter occurrence rate inferred from RV surveys is twice the rate determined by the transit method, inconsistent at the 3$\sigma$ level \citep[][references therein]{Winn2015}. Although previous studies have speculated that the binary fractions are different between the RV and transit samples \citep{Wright2012,Wang2015b}, none have hitherto demonstrated quantitatively how binaries can account for the observed factor of two disparity in the hot Jupiter occurrence rates. In Section~\ref{HJRate}, we show that RV surveys for giant planets boost their detection rates by a factor of 1.8\,$\pm$\,0.2 by systematically removing spectroscopic binaries from their samples.  Indeed, we prove that the binary star selection bias fully resolves the discrepancy in the hot Jupiter occurrence rates inferred from RV versus transit methods. 

We then discuss the impact of an increasing close binary fraction with respect to stellar mass on the occurrence rates of giant and small planets in Sections~\ref{GiantPlanet} and \ref{SmallPlanet}, respectively. In particular, we show that the occurrence rate of small planets orbiting single solar-type stars is 2.1\,$\pm$\,0.3 times larger than the overall rate inferred in previous studies. Close binaries bias the measured frequency $\eta_{\oplus}$ of Earth-sized planets in the habitable zone and the resulting trends with host mass. For example, {\it Kepler} revealed that M-dwarfs host 3.0\,-\,3.5 times more close, small planets than F-dwarfs  \citep{Howard2012,Dressing2015,Mulders2015a,Mulders2015b}. Although binaries alone are unlikely to explain the full factor of 3.0\,-\,3.5 variation, we conclude that the decrease in the binary fraction toward later spectral types accounts for roughly half of the observed increase in close, small planets.

We dedicate the final part of this paper to a detailed analysis of wide companions to hot Jupiter hosts. \citet{Ngo2016} measured a deficit of close stellar companions to hosts of hot Jupiters ($M_{\rm p}$~=~0.2\,-\,4\,M$_{\rm J}$), consistent with the other studies, but also discovered an excess of wide stellar companions compared to the field. After correcting for incompleteness in their AO survey, they reported that 47\%\,$\pm$\,7\% of hot Jupiter hosts have stellar companions across $a$~=~50\,-\,2,000~au, which is 2.9 times larger than their adopted solar-type field binary fraction of 16\% across the same separation interval at the 4.4$\sigma$ confidence level. \citet{Ngo2016} and \citet{Evans2018} also speculated that the wide binary fraction of hot Jupiter hosts might be even larger after accounting for the potential transit dilution selection bias against discovering hot Jupiters in binaries with bright companions. Imaging surveys of giant planet KOIs \citep{Law2014,Wang2015c,Ziegler2018}, {\it TESS} giant planet candidates \citep{Ziegler2019}, hosts of very massive planets and brown dwarfs with $M_2$~=~7\,-\,60\,M$_{\rm J}$ \citep{Fontanive2019}, and very close binaries in general \citep{Tokovinin2006} also reveal an excess of wide stellar companions relative to the field.  \citet{Ngo2016} and \citet{Fontanive2019} concluded that most of the stellar companions to planet hosts are too wide to induce Kozai-Lidov cycles, but instead argued that the additional mass necessary to make wide stellar companions also facilitated in the formation and migration of close giant planets. Wide stellar companions can also induce spiral density waves in the protoplanetary disks, which may trap dust particles and seed the growth of planetesimals \citep{Rice2006,Dong2015,Carrera2015}. 

Given our detailed analysis of multiplicity statistics and planet suppression by close binaries, we are excellently poised to determine the intrinsic effect of wide binaries on giant planet formation  (Section~\ref{HJwide}).   In Section~\ref{HJq}, we first compare the mass-ratio distributions of wide stellar companions to hot Jupiter hosts to wide field binaries. We determine that the transit dilution selection bias against discovering hot Jupiters hosts with wide bright companions is negligible. In Section~\ref{HJcomprate}, we account for various selection effects in the \citet{Ngo2016} sample. We show that the wide binary fraction of hot Jupiter hosts is fully consistent with expectations, i.e., wide binaries do not enhance the formation of hot Jupiters at a statistically significant level. Meanwhile, we find in Section~\ref{HJvBin} that very close binaries and hosts of close brown dwarf companions exhibit a real excess of tertiary companions, consistent with the conclusions in \citet{Tokovinin2006} and \citet{Fontanive2019}, respectively. Samples of {\it Kepler} and {\it TESS} giant planet candidates that do not have dynamical masses confirmed by spectroscopic RV monitoring (e.g., \citealt{Law2014,Wang2015c,Ziegler2018,Ziegler2019}) are substantially contaminated by eclipsing binary (EB) false positives, thereby leading to the spurious enhancement of wide stellar companions. We conclude in Section~\ref{Fragmentation} that very close binaries and sub-stellar companions with masses $M_2$~$>$~7\,M$_{\rm J}$ formed via fragmentation of gravitationally unstable disks, are metal-poor, and exhibit a large 5$\sigma$ excess of tertiary companions, whereas hosts of hot Jupiters with $M_{\rm p}$~$=$~0.2\,-\,4\,M$_{\rm J}$ that formed via core accretion are metal-rich and do not exhibit a statistically significant excess of wide stellar companions. We summarize our main results in Section~\ref{Summary}.

\section{Impact of Close Binaries on Planet Occurrence Rates}
\label{part1}

\subsection{Binary Fractions within 10 and 100 AU}
\label{closebinfrac}

To deduce the impact of close binaries on planet occurrence rates, we begin by measuring the close binary fraction of main-sequence (MS) stars. It is well established that the close binary fraction increases with primary mass $M_1$ \citep{Abt1990,Raghavan2010,Sana2012,Duchene2013,Moe2017}.  In Appendix~\ref{Appendix}, we compile literature results to quantify more precisely how the fractions $F_{\rm a<10au}$ and $F_{\rm a<100au}$ of primaries with stellar companions within $a$~$<$~10\,au and $a$~$<$~100\,au, respectively, vary according to spectral type. 

We correct for incompleteness down to the hydrogen-burning MS limit of  $M_2$~=~0.08\,\Msun. M-type \citep{Dieterich2012,Duchene2013}, solar-type  \citep{Grether2006,Santerne2016}, and A-type \citep{Murphy2018} primaries all exhibit an intrinsic dearth of close brown dwarf companions ( $M_2$~$<$~0.08\,\Msun) commonly known as the brown dwarf desert.  The brown dwarf desert is observed within  $a$~$<$~1~au \citep{Grether2006,Csizmadia2015,Santerne2016,Murphy2018,Shahaf2019}, and only 2\%\,$\pm$\,1\% of stars have brown dwarf companions across intermediate separations of $a$~=~10\,-\,100~au \citep{Kraus2008,Kraus2011,Dieterich2012,Wagner2019,Nielsen2019}. Even if brown dwarf companions are capable of suppressing planet formation, their contribution is statistically insignificant compared to close stellar companions. 

\begin{figure}
\centerline{
\includegraphics[trim=0.6cm 0.1cm 0.3cm 0.1cm, clip=true, width=3.3in]{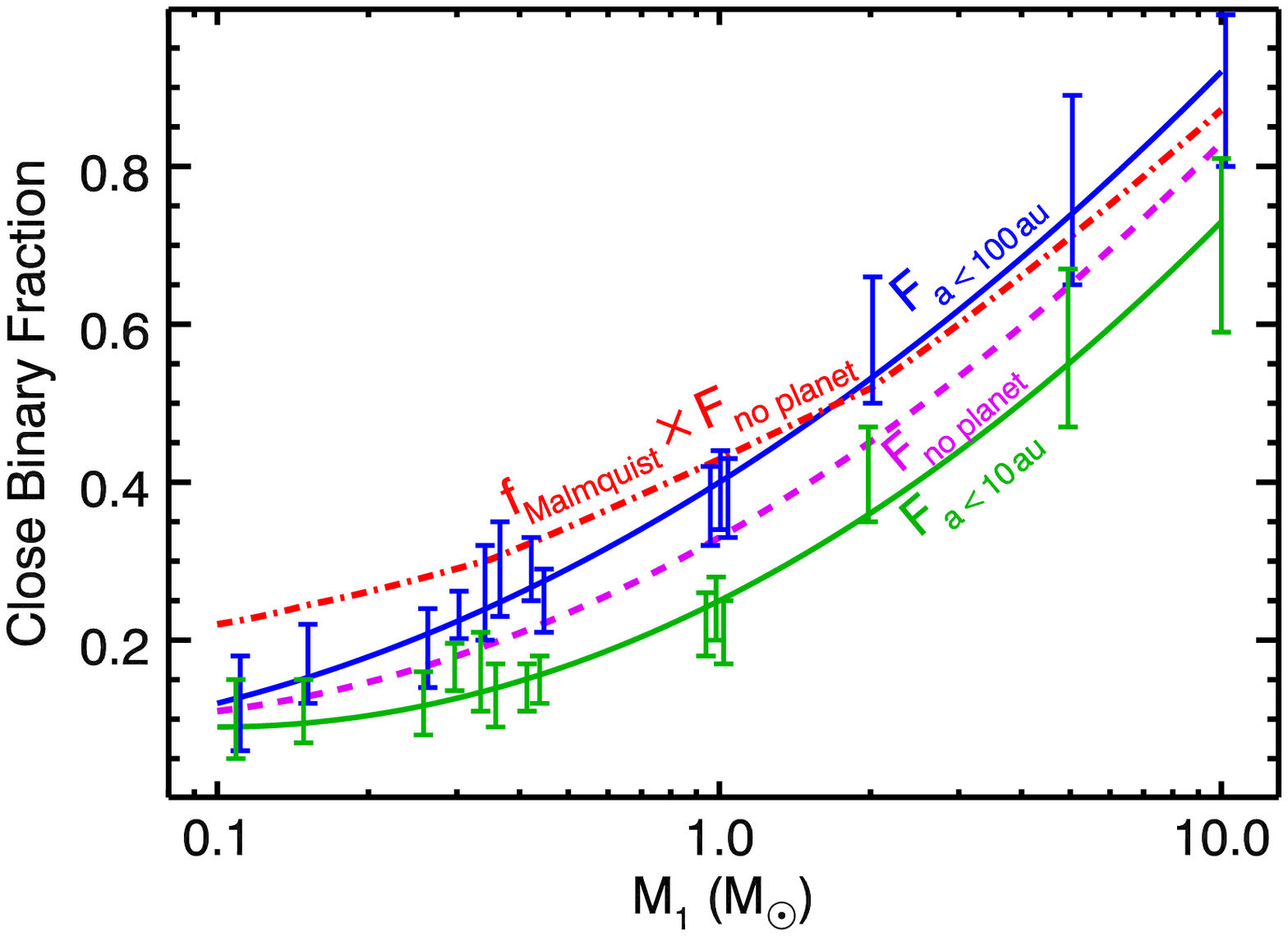}}
\caption{For the mean field metallicity of $\langle$[Fe/H]$\rangle$~=~$-$0.1, the bias-corrected, volume-limited stellar binary fractions inside orbital separations of $a$~$<$~10~au (green) and $a$~$<$~100~au (blue) as a function of primary mass $M_1$.  We correct for incompleteness down to $M_2$~=~0.08\,\Msun, and we include WD companions to solar-type and early-type primaries. For volume-limited samples, the dashed magenta line shows that $F_{\rm no\,planet}$~=~18\%, 33\%, and 45\% of M, G, and A primaries, respectively, do not host close planets due to suppression by close stellar companions. The dash-dotted red line shows that Malmquist bias in magnitude-limited samples boosts the observed fractions of stars without close planets to $f_{\rm Malmquist}$\,$\times$\,$F_{\rm no\,planet}$~=~28\%, 43\%, and 52\% for M, G, and A primaries, respectively.} 
\label{closebin}
\end{figure}

We incorporate close, unresolved  white dwarf (WD) companions in a manner that mimics the expected contamination in planet surveys.  For example, WD companions are significantly fainter than, and have temperatures comparable to, early-type primaries.  Only the youngest and hottest $\approx$\,10\% of close WD companions to solar-type primaries produce a detectable UV excess \citep{Moe2017}. The majority of close WD companions to solar-type and early-type primaries are therefore unnoticeable prior to multi-epoch RV monitoring or high-resolution imaging \citep{Holberg2016,Moe2017,Toonen2017}. Thus planet transiting surveys like {\it Kepler} and {\it TESS} are not biased against AFGK primaries with WD companions, and so we include such Sirius-like binaries in our analysis. Meanwhile, WD companions are substantially hotter than M-dwarf primaries and generally dominate at bluer optical wavelengths.  Unresolved WD companions to M-dwarfs are easily discernible according to their composite spectra or B-band excess. Planet surveys therefore do not typically target M-dwarfs with close WD companions, and so we do not include such systems in our binary statistics. We emphasize that our completeness-corrected multiplicity statistics, which include WD companions to solar-type and early-type primaries, specifically cater to planet-finding surveys in the field and should not be used in studies of star formation.

The inclusion of WD companions significantly increases the binary fraction of solar-type systems \citep{Moe2017}. For their sample of 164 F7-G9\,IV-VI primaries ($\langle M_1 \rangle$~=~1.02\,\Msun), \citet{Duquennoy1991} reported a bias-corrected total binary fraction of $F_{\rm bin}$~=~57\% after adding $\Delta F_{\rm bin}$~=~6\% to account for faint WD companions that hid below their detection threshold. Alternatively, \citet{Raghavan2010} did not correct for incompleteness of WD companions, and therefore reported a lower binary fraction of $F_{\rm bin}$~=~46\%\,$\pm$\,2\% for their sample of 454 F6-K3\,IV/V primaries ($\langle M_1 \rangle$~=~0.95\,\Msun). The remaining difference of $\Delta F_{\rm bin}$~=~5\% between the \citet{Duquennoy1991} and 
\citet{Raghavan2010} samples is mostly due to the difference $\Delta \langle M_1 \rangle$~=~0.07\,\Msun\ between their mean primary masses. The separation distribution of B/A-type binaries peaks near a~=~10~au \citep{Abt1990,Rizzuto2013,Moe2017}, and so roughly half of the WD companions to solar-type stars are located within $a$~$<$~10~au. For FG primaries, 20\%\,$\pm$\,6\% of the stellar companions within $a$~$<$~10~au are WDs \citep{Moe2017,Murphy2018}.

The majority of the studies investigated in Appendix~\ref{Appendix} survey the solar neighborhood, and all have mean metallicities consistent with the field population.  The mean field metallicity varies slightly with spectral type and the details of the target selection criteria. Stars observed near the mid-plane of the galactic disk tend to be more metal-rich. For example, solar-type stars monitored by the main {\it Kepler} mission span small galactic latitudes  ($b$~=~6$^{\circ}$\,-\,21$^{\circ}$) and are therefore slightly metal rich ($\langle$[Fe/H]$\rangle$~=~$-$0.05; \citealt{Dong2014,Zong2018}) compared to a volume-limited sample of solar-type stars ( $\langle$[Fe/H]$\rangle$~=~$-$0.15; \citealt{Nordstrom2004,Raghavan2010}).  Nevertheless, the mean field metallicity in all cases is found within the narrow interval $-$0.2~$<$~$\langle$[Fe/H]$\rangle$~$<$~0.0.  Across this small metallicity interval, the close binary fraction varies by less than $\Delta F_{\rm a<10au}$/$F_{\rm a<10au}$~$<$~10\% \citep{Moe2019}, which is well within the measurement uncertainties (see also Paper~II). In Fig.~\ref{closebin}, we compare our bias-corrected results for $F_{\rm a<10au}$ and $F_{\rm a<100au}$ as a function of $M_1$ at the mean field metallicity of $\langle$[Fe/H]$\rangle$~=~$-$0.1. 

\begin{figure}
\centerline{
\includegraphics[trim=0.7cm 0.1cm 0.7cm 0.3cm, clip=true, width=3.3in]{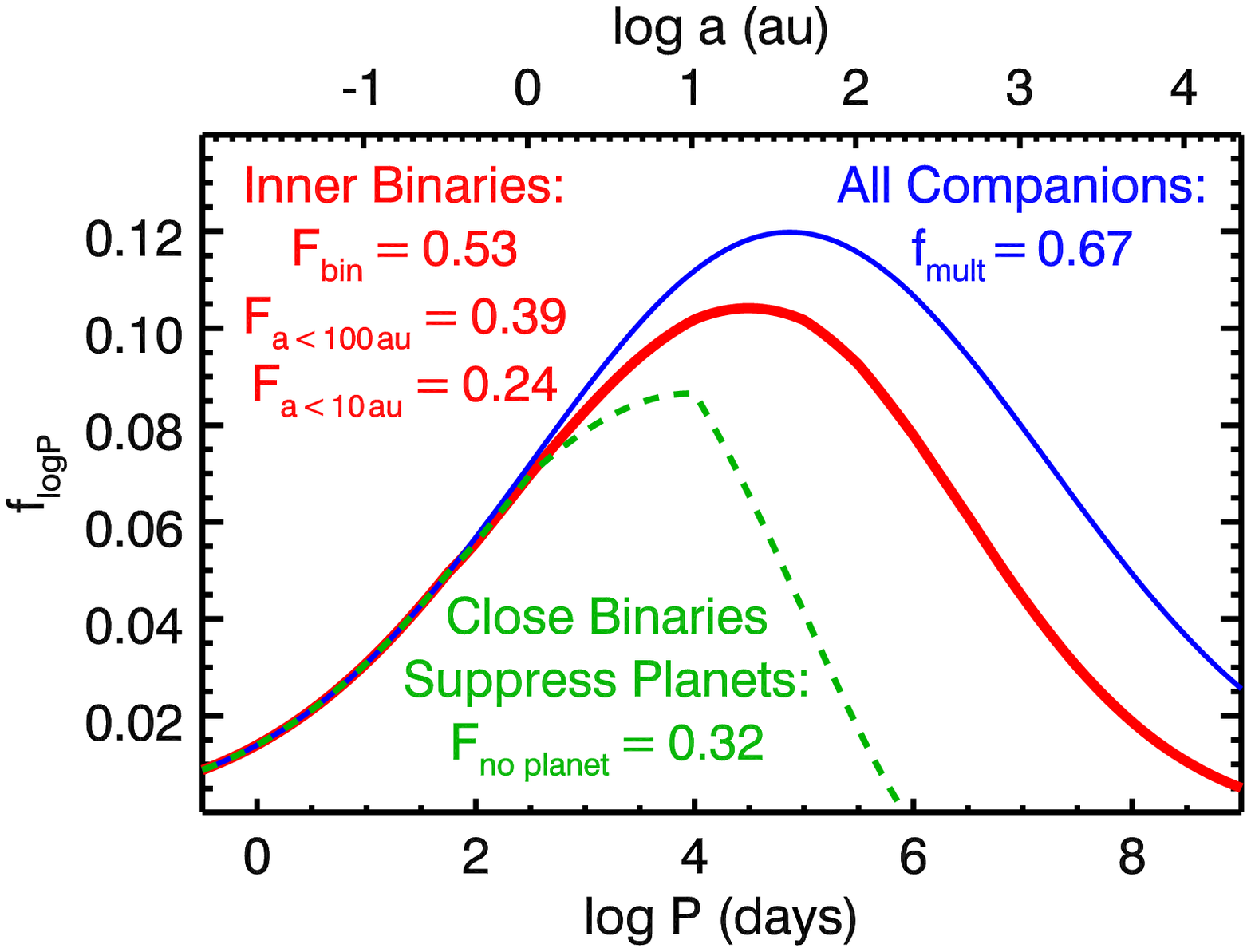}}
\caption{The frequency of all stellar-mass companions (including WDs) per decade of orbital period for a volume-limited sample of field solar-type primaries ( $M_1$~=~1.0\,\Msun; $\langle$[Fe/H]$\rangle$~=~$-$0.1). We display the canonical log-normal period distribution ($\mu_{\rm logP}$~=~4.9, $\sigma_{\rm logP}$~=~2.3) scaled to a multiplicity frequency of $f_{\rm mult}$~=~0.67 companions per primary across $-$0.5~$<$~log\,$P$\,(days)~$<$~9.0 (blue). Nearly all very close companions with $a$~$<$~1~au are inner binaries while half of wide companions with $a$~$>$~1,000~au are outer tertiaries in hierarchical triples. The resulting inner binary period distribution is skewed toward shorter separations (thick red), providing in an overall binary fraction of $F_{\rm bin}$~=~53\% and close binary fractions of $F_{\rm a<100au}$~=~39\% and $F_{\rm a<10au}$~=~24\%. In our model (dashed green), very close binaries with $a$~$<$~1~au completely suppress S-type planets while wide binaries with $a$~$>$~200~au have no effect on close planet formation.  In volume-limited samples, $F_{\rm no\,planet}$~=~32\% of field solar-type primaries do not host close planets due to suppression by close stellar companions.}
\label{solarperiod}
\end{figure}

The binary fraction within $a$~$<$~100~au closely resembles the overall binary fraction.  The separation distribution of late-M binaries ( $M_1$~$<$~0.3\,\Msun) narrowly peaks near $a$~=~10~au \citep{Basri2006,Law2008,Janson2012,Dieterich2012,Duchene2013,Winters2019}, and so the late-M overall binary fraction of $F_{\rm bin}$~=~20\% is only marginally larger than the fraction $F_{\rm a<100au}$~=~15\% measured within $a$~$<$~100~au. Nearly all wide companions ($a$~$>$ 100~au) to B-type primaries are outer tertiaries in hierarchical triples \citep{Abt1990,Rizzuto2013,Moe2017}. The measured fractions $F_{\rm bin}$~$\approx$~$F_{\rm a<100au}$~=~70\%\,-\,90\% for B-type stars are therefore similar to each other. Companions to solar-type primaries peak near $a$~=~40\,au \citep{Duquennoy1991,Raghavan2010}, but nearly half of the wide companions are outer tertiaries \citep[][see Fig.~\ref{solarperiod}]{Tokovinin2014,Chini2014,Moe2017}.  The overall solar-type binary fraction of $F_{\rm bin}$~=~53\%\,$\pm$\,5\% (including WD companions) is slightly larger than the fraction $F_{\rm a<100au}$~=~39\%\,$\pm$\,4\% inside of $a$~$<$~100~au. Combining these measurements together, we fit an empirical quadratic relation as a function of primary mass:

\begin{equation}
 F_{\rm a<100au} = 0.40 + 0.40\,{\rm log}(M_1/{\rm M}_{\odot})+0.12[{\rm log}(M_1/{\rm M}_{\odot})]^2,
\end{equation}

\noindent which we display as the solid blue line in Fig.~\ref{closebin}. For the various samples of mid-M dwarfs through A-dwarfs ($M_1$ = 0.3\,-\,2.4\,\Msun) investigated in Appendix~\ref{Appendix}, the relative measurement uncertainties span $\Delta F_{\rm a<100au}$/$F_{\rm a<100au}$ = 12\%\,-\,21\% (see individual blue error bars in Fig.~\ref{closebin}). These samples are neither fully independent nor completely overlapping, and so the uncertainty in $F_{\rm a<100au}$ at a given $M_1$ is slightly smaller than the individual measurement uncertainties. We adopt $\delta F_{\rm a<100au}$/$F_{\rm a<100au}$ = 0.10,  and so our quadratic model fit yields $F_{\rm a<100au}$ = 40\%\,\,$\pm$\,4\% for $M_1$~=~1\,\Msun\ primaries. This model fit result is marginally larger than the value of  $F_{\rm a<100au}$~=~39\%\,$\pm$\,4\% measured for solar-type primaries (see Appendix~\ref{Appendix} and Fig.~\ref{solarperiod}), but the difference (1\%) is negligible compared to the uncertainty (4\%).

The binary fraction inside of $a$~$<$~10~au is noticeably flatter across $M_1$~=~0.1\,-\,1.0\,\Msun.  The solar-type measurement of $F_{\rm a<10au}$~=~24\%\,$\pm$\,4\% is slightly larger than the early-M value of $F_{\rm a<10au}$~=~14\%\,$\pm$\,3\% mainly because the former includes WD companions. Toward larger masses, $M_1$~$>$~1\,\Msun, the binary fraction $F_{\rm a<10au}$ increases in parallel with $F_{\rm a<100au}$, nearly tripling by $M_1$~=~10\,\Msun.  We display our fit:

\begin{equation}
 F_{\rm a<10au} = 0.25 + 0.32\,{\rm log}(M_1/{\rm M}_{\odot})+0.16[{\rm log}(M_1/{\rm M}_{\odot})]^2
\end{equation}

\noindent as the solid green line in Fig.~\ref{closebin}.  For $M_1$ = 0.3\,-\,2.4\,\Msun\ primaries, the measurement uncertainties span $\Delta F_{\rm a<10au}$/$F_{\rm a<10au}$ = 16\%\,-\,27\% for the individual samples (see Appendix~\ref{Appendix} and green error bars in Fig.~\ref{closebin}). We adopt an overall error of $\delta F_{\rm a<10au}$/$F_{\rm a<10au}$ = 0.15 for our model fit, resulting in $F_{\rm a<10au}$ = 25\%\,$\pm$\,4\% for $M_1$~=~1\,\Msun\ primaries.

\subsection{Model for Planet Suppression}
\label{Suppression}

\begin{figure}
\centerline{
\includegraphics[trim=0.5cm 0.5cm 0.3cm 0.2cm, clip=true, width=3.3in]{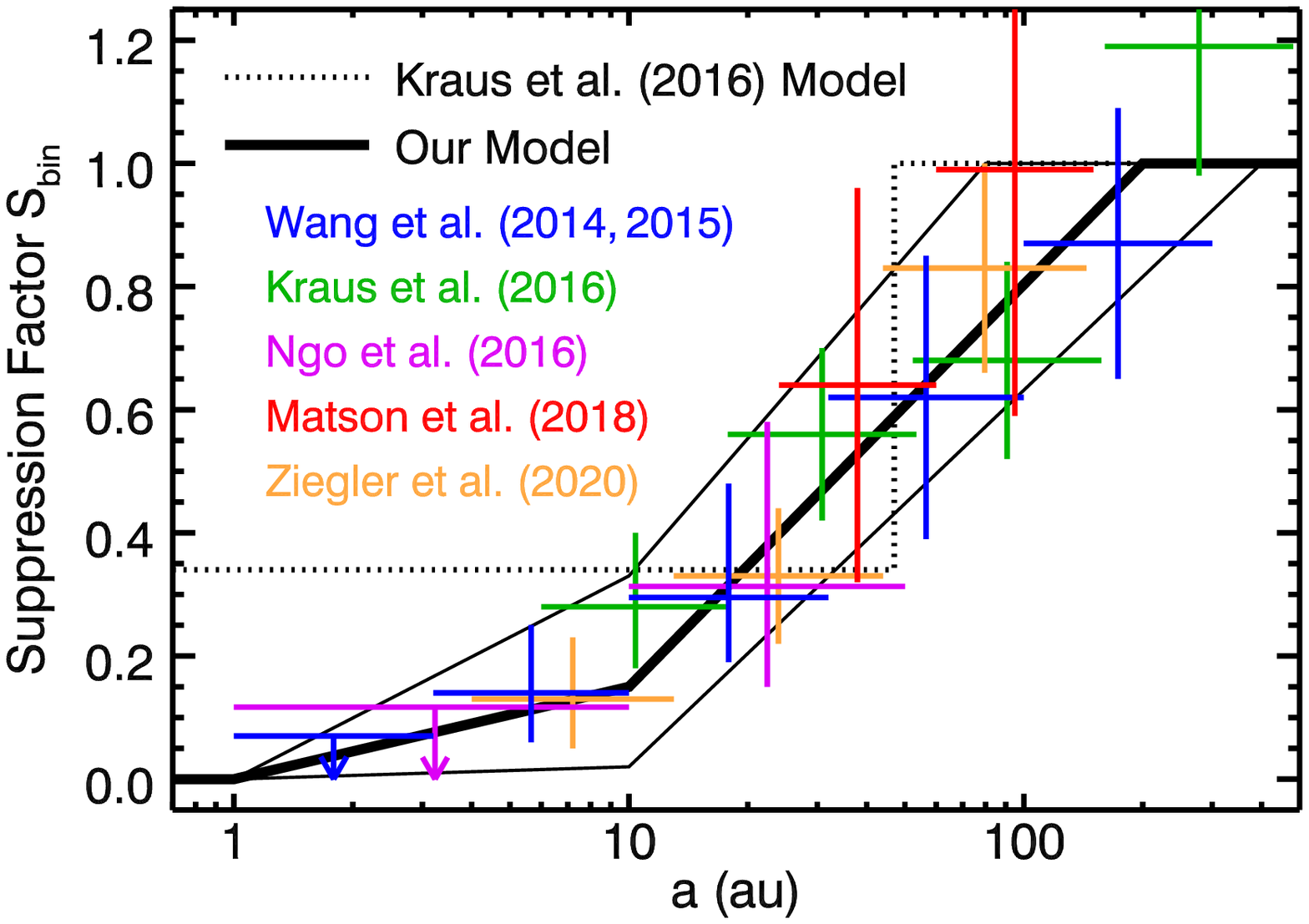}}
\caption{The suppression factor $S_{\rm bin}$, i.e., the ratio of the stellar companion fraction in planet hosts versus field stars, as a function of binary separation. We display the measurements from RV monitoring of hot Jupiter hosts \citep[][magenta]{Ngo2016}, AO observations of KOIs \citep[][green]{Kraus2016}, speckle imaging of K2 planet candidate hosts \citep[][red]{Matson2018}, speckle imaging of TOIs \citep[][orange]{Ziegler2019}, and the combined RV and AO observations of transiting KOIs, giant planets, and multi-planet systems \citep[][blue]{Wang2014b,Wang2015c,Wang2015d}. \citet{Kraus2016} fitted a step-function model for suppression such that only $S_{\rm bin}$~=~34\% of binaries within $a$~$<$~47~au could potentially host planets (dotted line). The RV observations instead demonstrate that the suppression factor is even smaller at closer separations, and so we adopt full suppression ($S_{\rm bin}$ = 0\%) of S-type planets when binary separations are within $a$~$<$~1~au, nearly complete suppression ($S_{\rm bin}$ = 15$_{-12}^{+17}$\%) at $a$~=~10~au, and no suppression ($S_{\rm bin}$ = 100\%) beyond $a$~$>$~200$_{-120}^{+200}$~au (solid lines).  }
\label{suppressfactor}
\end{figure}

With our binary statistics in hand, we now compute the planet suppression rate as a function of binary orbital separation. Utilizing AO imaging of KOIs, \citet{Kraus2016} discovered an intrinsic deficit of stellar companions below $a$~$<$~50~au relative to the field population.  Accounting for the resolution and sensitivity of their AO observations, they fit a two-parameter step function to model planet suppression.  \citet{Kraus2016} reported that only $S_{\rm bin}$~=~34$_{-15}^{+14}$\% of binaries inside of $a_{\rm cut}$~$<$~47$_{-23}^{+59}$~au are capable of hosting interior S-type planets compared to wider binaries and single stars (dotted line in our Fig.~\ref{suppressfactor}). The suppression factor $S_{\rm bin}$ is therefore the ratio of the stellar companion fraction in planet hosts versus field stars, e.g.,  $S_{\rm bin}$~=~0\% corresponds to no planets due to complete suppression by binaries whereas $S_{\rm bin}$ = 100\% implies the stellar companion rate of planet hosts is equal to that of field stars. We also show in Fig.~\ref{suppressfactor} the actual \citet{Kraus2016} measurements for $S_{\rm bin}$ across $\rho$~=~6\,-\,500~AU (red histogram divided by blue curve in right panel of their Fig.~7). \citet{Kraus2016} identified a few additional stellar companions to KOIs with $\rho$~=~2\,-\,6 au, but they argued they are seen in projection and therefore likely have true orbital separations beyond $a$~$>$~6~au. 

A non-zero suppression factor  cannot arbitrarily extend to very small separations. For S-type planets with $P_{\rm p}$~=~10~days, dynamical stability alone requires $S_{\rm bin}$~=~0\% for $a$~$<$~0.4~au \citep{Holman1999}. Although AO imaging cannot readily detect faint WD or low-mass companions within $a$~$<$~10~au, high-precision spectroscopic RV monitoring is sensitive to most stellar-mass companions inside this limit. Based on long-term RV monitoring of stars hosting close giant planets, \citet{Knutson2014} and \citet{Bryan2016} showed that the close binary fraction is extremely small, consistent with zero within $a$~$<$~10~au. Combining their results, \citet{Ngo2016} reported that the binary fraction of hot Jupiter hosts is 4\%$_{-2\%}^{+4\%}$ inside of $a$~$<$~50~au. The observed solar-type field binary fraction is 34\%\,$\pm$\,4\% within the same separation limit (see Fig.~\ref{solarperiod}), providing $S_{\rm bin}$~=~0.04/0.34~=~12$_{-6}^{+12}$\%. After dividing their observations into two separation intervals and comparing to the field solar-type binary period distribution (Fig.~\ref{solarperiod}), we estimate $S_{\rm bin}$~$<$~12\% across $a$~=~1\,-\,10~au and $S_{\rm bin}$~$=$~31$_{-16}^{+27}$\% across $a$~=~10\,-\,50~au, which we display in Fig.~\ref{suppressfactor}.

In a series of papers, Wang et al. combined both RV and AO observations to identify binary star companions in different types of planetary systems, including KOIs, which are dominated by smaller planets \citep{Wang2014a,Wang2014b}, transiting giant planets \citep{Wang2015c}, and transiting multi-planet systems \citep{Wang2015d}.  The details vary from sample to sample, but they concluded that there is nearly complete suppression of S-type planets when binary separations are below $a$~$<$~10~au, moderate suppression for binaries with $a$~=~10\,-\,100~au, and little to no planet suppression in wide binaries with $a$~$>$~100~au. \citet{Wang2014b} speculated that small planet KOIs may be slightly suppressed ($S_{\rm bin}$~=~60\%) in wide binaries with $a$~=~200\,-\,1,500~au, but we attribute this minor effect to photometric dilution whereby bright stellar companions decrease the probability of detecting small transiting planets (see below). To increase the sample size to 278 total systems, we combine their AO observations of 56 KOIs \citep{Wang2014b}, 84 transiting giant planets \citep{Wang2015c} and 138 transiting multi-planet systems \citep{Wang2015d}, yielding 7 (2.5\%\,$\pm$\,0.9\%) and 13 (4.7\%\,$\pm$\,1.3\%) detected stellar companions across $a$~=~32\,-\,100 au and $a$~=~100\,-\,320~au, respectively. Considering their AO sensitivity limits, we estimate corrected MS companion fractions of 4.5\%\,$\pm$\,1.6\% across $a$~=~32\,-\,100~au and 6.0\%\,$\pm$\,1.5\% across $a$~=~100\,-\,320~au. Based on our analysis of field binaries (see \S\ref{closebinfrac} and Fig.~\ref{solarperiod}), 7.2\%\,$\pm$\,1.5\% and 6.9\%\,$\pm$\,1.4\% of solar-type stars have MS (non-WD) stellar companions across $a$~=~32\,-\,100 au and $a$~=~100\,-\,320~au, respectively. The inferred suppression factors are therefore $S_{\rm bin}$~=~62\%\,$\pm$\,23\% for $a$~=~32\,-\,100~au and $S_{\rm bin}$ = 87\%\,$\pm$\,22\% for $a$~=~100\,-\,320~au, which we display in Fig.~\ref{suppressfactor}. The RV observations from  \citet{Wang2014b} yield stringent suppression factors at closer separations (see their Fig.~6 and Table~5): $S_{\rm bin}$~<~7\% across $a$~=~1.0\,-\,3.2~au, $S_{\rm bin}$~=~14$_{-8}^{+11}$\% across $a$~=~3.2\,-\,10~au, and $S_{\rm bin}$~=~29$_{-10}^{+19}$\% across $a$~=~10\,-\,32~au (see Fig.~\ref{suppressfactor}). 

\citet{Matson2018} performed a speckle imaging survey with WIYN and Gemini for stellar companions to K2 planet candidate host stars. After accounting for the sensitivity and resolution of their observations, they reported a binary star fraction across projected separations of 15\,-\,300~au that is consistent with the field. Based on their observed sample of 10 binaries with projected separations within $<$\,50~au, they concluded that close binaries do not suppress planet formation, opposite the conclusions of the previously cited studies. However, \citet{Matson2018} detected most of their stellar companions beyond $a$~$>$~50~au, and therefore had minimal leverage in constraining the suppression factor $S_{\rm bin}$ inside of $a$~$<$~50~au. We consider their higher resolution Gemini observations, which were sensitive to nearly all stellar MS companions beyond $>$\,0.08$''$ (see top panel of their Fig.~7).  Within this sample of 102 K2 planet hosts (median distance $d$~=~300\,pc), \citet{Matson2018} identified 4 (3.9\%\,$\pm$\,1.9\%) and 6 (5.9\%\,$\pm$\,2.4\%) companions across projected separations of $\rho$~=~0.08''\,-\,0.2'' ($a$~=~24\,-\,60~au assuming median $d$~=~300\,pc) and $\rho$~=~0.2''\,-\,0.5'' ($a$~=~60\,-\,150~au), respectively. We estimate that 6.1\%\,$\pm$\,1.3\% and 5.9\%\,$\pm$\,1.2\% of solar-type primaries have MS companions across $a$~=~24\,-\,60~au and $a$~=~60\,-\,150~au, respectively, yielding $S_{\rm bin}$~=~64\%\,$\pm$\,32\% and 99\%\,$\pm$\,40\% for the same separation intervals (see Fig.~\ref{suppressfactor}). The suppression factor of $S_{\rm bin}$~=~64\%\,$\pm$\,32\% across $a$~=~24\,-\,60~au is marginally consistent with no suppression, as concluded in \citet{Matson2018}, but is also consistent with the \citet{Kraus2016} value of $S_{\rm bin}$~=~34\% for $a$~$<$~50~au. All the surveys compiled in Fig.~3 suggest $S_{\rm bin}$~=~55\% across $a$~=~24\,-\,60~au. To rule out such an intermediate suppression factor with $>$2$\sigma$ confidence,  \citet{Matson2018} needed to detect nine companions  across $a$~=~24\,-\,60~au, which is roughly double the observed value. The \citet{Matson2018} sample of binaries is simply too small, especially at close separations within $a$~$<$~50~au, to reliably measure the influence of close binaries on planet statistics.

\citet{Ziegler2019} recently obtained SOAR speckle images of 542 {\it TESS} objects of interest (TOIs). They measured a deficit of close binaries ($a$ $<$ 150 au) and a slight excess of wide binaries ($a$ = 150\,-\,3,000 au) compared to the field.  For the three bins across 4\,-\,150~au in their Fig.~8, we compute $S_{\rm bin}$  by dividing their number of detected binaries (histogram with error bars) by their expectations from field binaries (thick black continuous distribution). We present the results in Fig.~\ref{suppressfactor}. \citet{Ziegler2019} also adopted a step-function model for planet suppression as done in \citet{Kraus2016},  and they fitted $S_{\rm bin}$~=~24\% below $a$~$<$~46~au, consistent with the values reported in \citet{Kraus2016}. \citet{Ziegler2019} showed that solar-type hosts of large planet candidates, solar-type hosts of small planet candidates, and M-dwarf hosts of small planet candidates all exhibit a similar deficit of close stellar companions.

The error bars displayed in Fig.~\ref{suppressfactor} represent measurement uncertainties combined with the associated systematic uncertainties from completeness corrections, but we expect other sources of systematic error should also contribute. For example, some KOIs and TOIs are EB false positives \citep{Fressin2013,Sullivan2015}, and very close binaries are known to exhibit an excess of tertiary companions \citep{Tokovinin2006}. We show in Section~\ref{HJvBin} that a significant fraction of {\it TESS} giant planet candidates in the \citet{Ziegler2019} sample are EB false positives, and so their inferred excess of wide stellar companions beyond $a$~$>$~100~au is spurious, i.e., most of their detected wide companions are tertiaries in hierarchical star systems. Conversely, it is more difficult to detect small transiting planets if their hosts have bright stellar companions that dilute the photometric signal. The slight deficit of wide stellar companions beyond $a$~$>$~200~au to small planet KOIs \citep{Wang2014b} and small planet TOIs \citep{Ziegler2019} are likely due to this transit dilution selection bias (see Section~\ref{SmallPlanet}). Fortuitously, the effects of EB false positives and transit dilution roughly cancel. Moreover, surveys of transiting multi-planet systems \citep{Wang2015d} and dynamically confirmed hot Jupiters \citep{Knutson2014,Ngo2016}, which are relatively immune to transit dilution effects and EB false positives, exhibit the same suppression factors. Most important, the {\it relative} change in $S_{\rm bin}$ as a function of binary separation is rather robust to these systematic biases.  The ratio of suppression factors $S_{\rm bin}$($a$=6\,au)/$S_{\rm bin}$($a$=200\,au) = (0.13\,$\pm$\,0.10)/(1.0\,$\pm$\,0.1) = 0.13\,$\pm$\,0.10 inferred from Fig.~\ref{suppressfactor} is therefore accurate. 

The suppression factor likely varies with binary properties other than orbital separation, e.g., spectral type, mass ratio, and eccentricity, as well as planet characteristics such as period, size, and eccentricity. Eccentric warm Jupiters in particular are known to reside in stellar binaries with $a$~$<$~10~au, including Kepler-420b \citep{Santerne2014} and Kepler-693b \citep{Masuda2017}, indicating they formed via dynamical interactions \citep{Gong2018,Fragione2019}. Similarly, the host of the eccentric hot Jupiter CoRoT-20b ($e_{\rm P}$~=~0.59) has an eccentric brown dwarf companion ($e$~=~0.60) with $M$\,sin\,$i$ = 17\,M$_{\rm J}$ at $a$~=~2.9~au, providing strong evidence that the hot Jupiter migrated via the eccentric Kozai-Lidov mechanism and tidal friction \citep{Rey2018}. \citet{Triaud2017} investigated two other hot Jupiter hosts, WASP-53 and WASP-81, which also have brown dwarf companions closely orbiting at $a$~=~3.7\,au and 2.4\,au, respectively. The existence of CoRoT-20, WASP-53, and WASP-81 suggests that close brown dwarf companions do not significantly suppress planet formation, consistent with our measurements for the stellar binary fractions where we corrected for incompleteness only down to the hydrogen-burning limit (see Section~\ref{closebinfrac}). Kepler-444 harbors five very small planets spanning $R_{\rm p}$ = 0.4\,-\,0.8\,\Rearth\ within $P_{\rm p}$ $<$\,10 days and a stellar companion, which happens to be a tight pair of M-dwarfs, in a highly eccentric orbit with $e$ = 0.86, $a$~=~37~au, and $r_{\rm peri}$~=~5~au \citep{Campante2015, Dupuy2016}. Such an eccentric binary companion likely truncated the inner circumprimary disk and sculpted planet formation differently than binaries with circular orbits. Nevertheless, various types of small planets \citep{Wang2014a,Wang2014b,Kraus2016,Ziegler2019} and giant planets \citep{Knutson2014,Wang2015c,Bryan2016,Ngo2016,Ziegler2019} all exhibit a consistent trend whereby the binary star separation largely dictates the suppression factor. Moreover, \citet{Ziegler2019} provided the first observational evidence that close M-dwarf binaries suppress planets in a fashion similar to close solar-type binaries. 

We therefore adopt a simple functional form for $S_{\rm bin}$($a$) that matches the current observations: very close binaries with $a$~$<$~1~au fully suppress S-type planets ($S_{\rm bin}$~=~0\%), binaries with $a$~$=$~10~au host close planets at $S_{\rm bin}$~=~15$_{-12}^{+17}$\% the occurrence rate of single stars, and wide binaries with $a$~$>$~200$_{-120}^{+200}$~au have no effect  on close planet statistics ($S_{\rm bin}$~=~100\%). We interpolate $S_{\rm bin}$ with respect to log~$a$, as shown in Fig.~\ref{solarperiod} and Fig.~\ref{suppressfactor}. We more thoroughly examine the occurrence rates of wide stellar companions to hosts of small and giant planets in Sections 3.3.1 and 4, respectively. We find that the wide companion fraction of planet hosts is consistent with expectations, i.e., wide binaries neither enhance nor suppress planet formation at a statistically significant level, which motivates our choice of a model that plateaus to $S_{\rm bin}$~=~100\% toward wide separations. The errors in our model parameters account for both measurement and systematic uncertainties. The true shape of $S_{\rm bin}$($a$) may be shallower or steeper than our best model fit. For example, a suppression factor that quickly increases from $S_{\rm bin}$~=~5\% at $a$~=~10~au to $S_{\rm bin}$~=~100\% at $a$~=~80~au is fully consistent with the data. The main advantage of our segmented model (versus a step-function model) is that the suppression factor can gradually decrease to $S_{\rm bin}$~=~0\% within $a$~$<$~1~au as constrained by the various RV surveys of transiting planet hosts. Our model also assumes that the suppression factor is independent of host mass and planet size. Although we expect the suppression factors to depend on these secondary parameters, the variations cannot be too significant. The currently measured suppression of giant planets in solar-type binaries, small planets in solar-type binaries, and small planets in M-dwarf binaries are all consistent with our model fit. 

\subsection{Overall Impact of Close Binaries}
\label{CloseBinary}

Given our detailed statistical accounting of both field binaries and companions to planet hosts, we can now evaluate the total impact of close binaries on planet occurrence rates. We define $F_{\rm no\,planet}$($M_1$) as the fraction of primaries that do not have close planets due to suppression by close stellar companions. This fraction therefore depends on $S_{\rm bin}$ and the frequency $f_{\rm dlogP}$ of stellar companions per decade of orbital period according to:

\begin{equation}
 F_{\rm no\,planet} = \int f_{\rm dlogP} (1 - S_{\rm bin}) d{\rm log}P.
\end{equation}

\noindent Given the observed binary period distributions and our model for $S_{\rm bin}$, then $F_{\rm no\,planet}$ is about halfway between $F_{\rm a<10au}$ and $F_{\rm a<100au}$. For the mean field metallicity, we set:

\begin{equation}
 F_{\rm no\,planet} = 0.33 + 0.36\,{\rm log}(M_1/{\rm M}_{\odot})+0.14[{\rm log}(M_1/{\rm M}_{\odot})]^2,
\label{Fclose}
\end{equation}

\noindent which is shown as the dashed magenta line in Fig.~\ref{closebin}. Similar to our model fits for $F_{\rm a<10au}$ and $F_{\rm a<100au}$, the quadratic fit yields $F_{\rm no\,planet}$~=~33\% for $M_1$~=~1\,\Msun\ primaries, which is 1\% larger than the value of $F_{\rm no\,planet}$~=~32\% measured for solar-type stars (see Fig.~\ref{solarperiod}).

There are two main sources of uncertainty for $F_{\rm no\,planet}$. First is the uncertainty in the multiplicity statistics of field stars. In Section~\ref{closebinfrac}, we estimated $\delta F_{\rm a<10au}$/$F_{\rm a<10au}$~=~15\% and $\delta F_{\rm a<100au}$/$F_{\rm a<100au}$~=~10\% across $M_1$~=~0.3\,-\,2.4\,\Msun. For this source of uncertainty, we adopt an intermediate value of $\delta F_{\rm no\,planet}$/$F_{\rm no\,planet}$~=~13\% considering $F_{\rm no\,planet}$ is about halfway between $F_{\rm a<10au}$ and $F_{\rm a<100au}$. For $M_1$~=~1\,\Msun\ primaries, the uncertainties in the multiplicity statistics contribute $\delta F_{\rm no\,planet}$~=~4\%. Second, the uncertainty in the suppression factor $S_{\rm bin}$($a$) propagates into the uncertainty in $F_{\rm no\,planet}$. Folding one of the thin solid lines in Fig.~\ref{suppressfactor} (i.e., the 1$\sigma$ upper and lower limits on $S_{\rm bin}$) into the solar-type binary separation distribution in Fig.~\ref{solarperiod} would result in a fraction that is $\delta F_{\rm no\,planet}$~=~4\% higher or lower. These two sources of uncertainty are not independent, e.g., the denominator in the suppression factor $S_{\rm bin}$ is the stellar companion fraction of field stars. For a volume-limited sample of $M_1$ = 1.0\,\Msun\ primaries, we conclude that $F_{\rm no\,planet}$ = 33\%\,$\pm$\,5\% do not host close planets due to suppression by close binaries.

Future RV and imaging surveys of planet hosts will increase the sample size of known S-type planets in binaries. However, simply enlarging surveys of planet hosts will not significantly decrease the uncertainties in planet suppression. As illustrated above, the measurement uncertainties in the multiplicity statistics of normal field stars, errors in modeling the sensitivity and completeness of the observations, and systematic biases such as EB false positives and transit dilution currently dominate the overall uncertainty in $S_{\rm bin}$ and $F_{\rm no\,planet}$.  We therefore advocate that future surveys not only target planet hosts, but also a comparably sized control sample of normal field stars most similar to the planet hosts. Not only will this provide a robust measurement of $S_{\rm bin}$ independent of the sensitivity of the observational methods, but also significantly reduce the uncertainties in $F_{\rm no\,planet}$.

\subsection{Implications for Planet Statistics}
\label{Implications}

We find that the effects of planet suppression by close binaries are much more substantial than previously realized. According to Fig.~\ref{closebin} and Eqn.~\ref{Fclose},  $F_{\rm no\,planet}$~=~18\%\,$\pm$\,4\%, 33\%\,$\pm$\,5\%, and 45\%\,$\pm$\,7\% of M-dwarf ($M_1$~=~0.3\,\Msun), G-dwarf (1.0\,\Msun), and A-dwarf (2.0\,\Msun) primaries in volume-limited samples do not host close planets due to suppression by close stellar companions.  Our value of  $F_{\rm no\,planet}$~$=$~33\%\,$\pm$\,5\% for solar-type primaries is larger than the value of $F_{\rm no\,planet}$~=~19\% reported by \citet{Kraus2016} for two reasons.  First, we corrected the \citet{Raghavan2010} field solar-type binary survey for incompleteness and incorporated other surveys to firmly anchor the field binary statistics.  As discussed in \citet{Moe2017} and Appendix~\ref{Appendix}, \citet{Raghavan2010} missed the majority of WDs and late-M companions across intermediate separations of $a$~=~5\,-\,30~au. Second, \citet{Kraus2016} assumed that the suppression factor of $S_{\rm bin}$~=~34\% extended toward arbitrarily small separations, but spectroscopic RV monitoring instead suggests the suppression factor gradually tapers to $S_{\rm bin}$~=~15\% inside of $a$~$<$~10~au and $S_{\rm bin}$~$=$~0\% within $a$~$<$~1~au (see Fig.~\ref{suppressfactor}).  

In magnitude-limited samples, the close binary fraction is even larger due to Malmquist bias, sometimes referred to as the \citet{Opik1924} effect or \citet{Branch1976} bias in the context of binary stars. In magnitude-limited surveys, twin binaries with equally bright components are over-represented by a factor of 2$^{\nicefrac{3}{2}}$~$\approx$~2.8.  The mass-ratio distribution of solar-type binaries is fairly uniform overall \citep{Duquennoy1991,Raghavan2010}, but close solar-type binaries exhibit an excess fraction of twins with mass ratios $q$~$=$~0.95\,-\,1.00 \citep{Tokovinin2000}.  In volume-limited samples, the excess twin fraction gradually declines from $F_{\rm twin}$~=~30\% inside of $a$~$<$~0.1~au to $F_{\rm twin}$~=~10\%  near $a$~$=$~50~au \citep{Moe2017}.  The close binary fraction of field solar-type stars is therefore $f_{\rm Malmquist}$~=~1.3\,$\pm$\,0.1 times larger in magnitude-limited surveys compared to volume-limited samples due to Malmquist bias (see also \citealt{Moe2019}). Close M-dwarf binaries, especially late-M binaries, are further skewed toward large mass ratios \citep{Burgasser2003,Law2008,Dieterich2012,Duchene2013,Winters2019}.  For example, in the \citet{Law2008} magnitude-limited sample of 77 M4.5-M6.0 stars ($M_1$~$\approx$~0.15\,\Msun), 13 of the 21 detected binaries across $\rho$~=~2\,-\,80\,au have small brightness contrasts within $\Delta i$~$<$~0.8~mag, even though the observations were sensitive to $\Delta i$~=~3.0~mag. As done in \citet{Burgasser2003} and \citet{Law2008}, we adopt $f_{\rm Malmquist}$~=~2.0\,$\pm$\,0.2 for late-M binaries ($M_1$~=~0.1\,\Msun).  Very close A-type and B-type binaries have a smaller twin fraction ($F_{\rm twin}$~=~10\%), and slightly wider companions to intermediate-mass stars are skewed toward small mass ratios  \citep{Rizzuto2013,Gullikson2016,Moe2017,Murphy2018}.  We estimate $f_{\rm Malmquist}$~=~1.15\,$\pm$\,0.05 for $M_1$~$=$~2\,\Msun\ and $f_{\rm Malmquist}$~=~1.05\,$\pm$\,0.03 for $M_1$~$=$~10\,\Msun. We interpolate $f_{\rm Malmquist}$ as a continuous function of log\,$M_1$. 

In Fig.~\ref{closebin}, we display the magnitude-limited fraction of stars that do not host close planets as a function of primary mass (dash-dotted red). We estimate  $f_{\rm Malmquist}$\,$\times$\,$F_{\rm no\,planet}$~=~28\%\,$\pm$\,5\%, 43\%\,$\pm$\,7\%, and 52\%\,$\pm$\,8\% for M-dwarf, G-dwarf, and A-dwarf primaries, respectively. In magnitude-limited samples, nearly half of solar-type field primaries do not host close planets due to suppression by close binaries.

\begin{figure}
\centerline{
\includegraphics[trim=0.5cm 0.2cm 0.3cm 0.3cm, clip=true, width=3.3in]{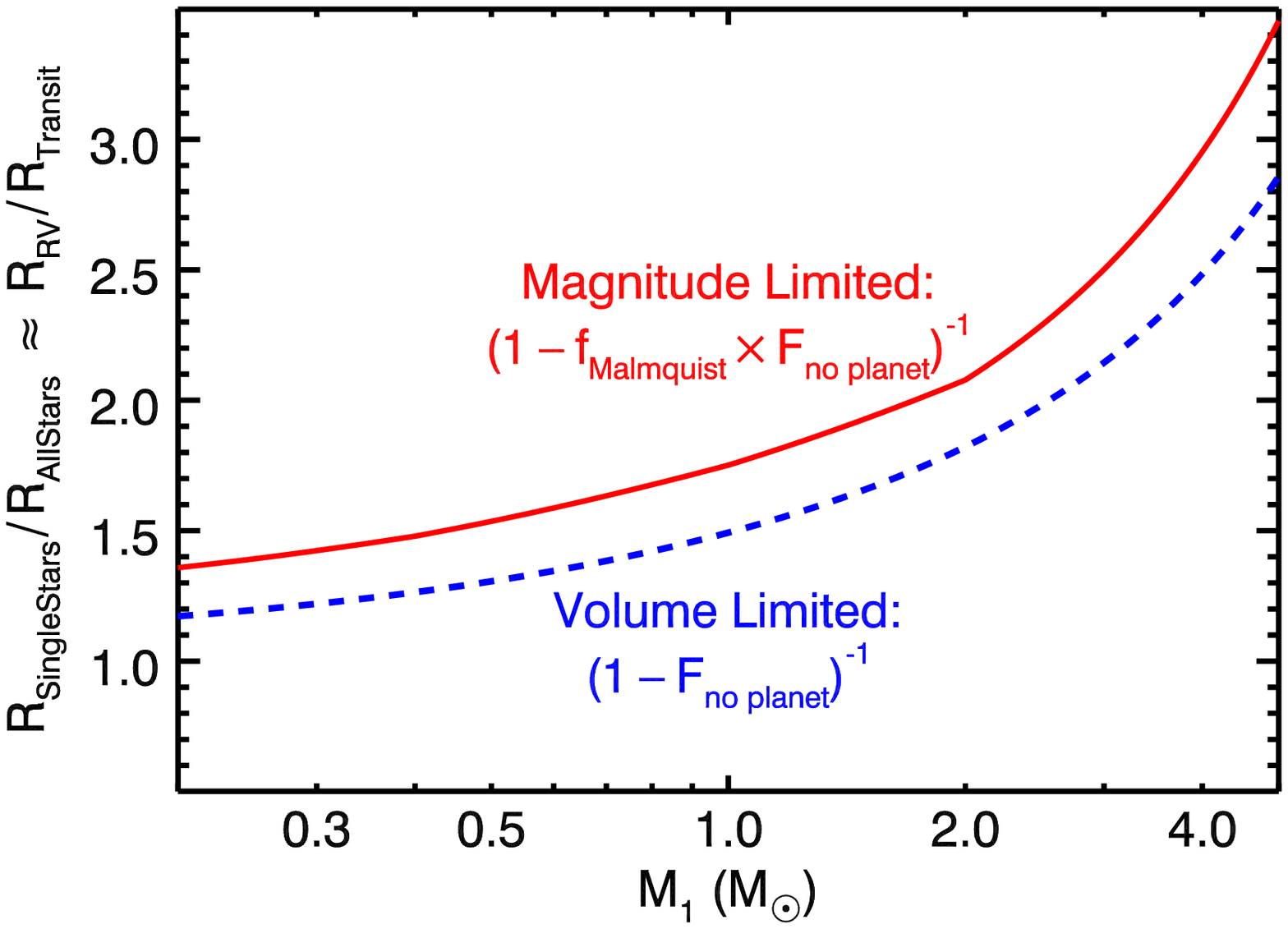}}
\caption{Planet occurrence rate $R_{\rm SingleStars}$ for effectively single stars divided by planet occurrences rate $R_{\rm AllStars}$ for all primary stars, including those in close binaries. We display this ratio for volume-limited (dashed blue line) and magnitude-limited (solid red line) samples. After initial spectroscopic monitoring, RV searches for Jovian planets typically remove close stellar-mass companions from their sample ($R_{\rm RV}$~$\approx$~$R_{\rm SingleStars}$), while transiting surveys make no significant selections against close binaries ($R_{\rm Transit}$~$\approx$~$R_{\rm AllStars}$). In magnitude-limited surveys of M, G, and A primaries, the planet occurrence rates $R_{\rm RV}$ from RV surveys are 1.4, 1.8, and 2.1 times larger, respectively, than planet occurrence rates $R_{\rm Transit}$ inferred from transiting surveys. This selection bias fully accounts for the  3$\sigma$ discrepancy in hot Jupiter occurrence rates measured from RV ($R_{\rm HJ;RV}$~=~0.9\,-\,1.2\,\%) versus transit ($R_{\rm HJ;Transit}$~=~0.4\,-\,0.6\,\%) methods.}
\label{ratio}
\end{figure}

Planet occurrence rates can be reported in two different ways: (1)~the rate $R_{\rm AllStars}$ of planets with respect to all primary stars, including those in close binaries, and (2)~the rate $R_{\rm SingleStars}$ of planets per effectively single stars, i.e., truly single stars and primaries in wide binaries. The former matches observational constraints when there are no selection biases with respect to close binaries, e.g., the {\it Kepler} and {\it TESS} transiting surveys (see below), while the latter should anchor theoretical models of planet formation in single stars. The ratio $R_{\rm SingleStars}$/$R_{\rm AllStars}$ is simply given by (1\,$-$\,$F_{\rm no\,planet}$)$^{-1}$ for volume-limited samples and (1\,$-$\,$f_{\rm Malmquist}$\,$\times$\,$F_{\rm no\,planet}$)$^{-1}$ for magnitude-limited samples, which we display in Fig.~\ref{ratio}. In magnitude-limited surveys of M, G, and A field primaries, the planet occurrence rates for single stars are 1.4\,$\pm$\,0.1, 1.8\,$\pm$\,0.2, and 2.1\,$\pm$\,0.3 times larger, respectively, than the planet occurrence rates for all primary stars within the sample. We discuss the differences between $R_{\rm AllStars}$ and $R_{\rm SingleStars}$ in the context of hot Jupiters, giant planets, and small planets in the following section.

\section{Binaries Bias Planet Occurrence Rates and Trends}
\label{bias}
\subsection{Hot Jupiters}
\label{HJRate}

As introduced in Section~\ref{Introduction}, there is a well-known 3$\sigma$ discrepancy in the measured occurrence rates of hot Jupiters orbiting solar-type stars: various RV surveys find $R_{\rm HJ;RV}$~=~0.9\%\,-\,1.2\% \citep{Marcy2005,Mayor2011,Wright2012} while the {\it Kepler} transiting survey yields $R_{\rm HJ;Transit}$~=~0.4\%\,-\,0.6\% \citep{Howard2012,Fressin2013,Mulders2015a,Santerne2016,Masuda2017,Petigura2018}. \citet{Zhou2019} recently measured $R_{\rm HJ;Transit}$~=~0.45\%\,$\pm$\,0.10\% based on {\it TESS} observations of AFG dwarfs, consistent with the hot Jupiter occurrence rate of {\it Kepler} FGK stars. Previous studies have investigated some of the possible biases between RV and transiting methods, but none have yet accounted for the observed factor of two discrepancy. For example, \citet{Guo2017} suggested that the RV surveys systematically targeted more metal-rich stars, which have higher hot Jupiter occurrence rates \citep{Santos2004,Fischer2005,Johnson2010,Petigura2018,Buchhave2018}. Utilizing follow-up spectroscopy, \citet{Guo2017} indeed confirmed that the RV samples have an average metallicity that is $\Delta\langle$[Fe/H]$\rangle$~=~0.04~dex higher than the {\it Kepler} sample. Nonetheless, \citet{Guo2017} concluded that such a small metallicity difference can account for only a 20\% relative change in the hot Jupiter occurrence rate, well short of the observed factor of two difference. In a different study, \citet{Wang2015b} investigated how subgiants and binaries within the {\it Kepler} sample dilute the depths of planetary transits compared to hosts that are single dwarfs. They estimated that 13\% of hot Jupiters were misclassified as smaller planets in the {\it Kepler} pipeline due to this transit dilution effect, which again cannot account for the observed discrepancy. \citet{Guo2017} and \citet{Bouma2018} also confirmed that photometric dilution due to subgiants and unresolved binaries cannot explain the discrepancy in hot Jupiter occurrence rates.  

Our proposed solution lies in the fact that RV surveys for giant planets are substantially biased against binaries, especially close spectroscopic binaries \citep[][references therein]{Marcy2005,Mayor2011,Wright2012}. When selecting their initial targets, they excluded previously known spectroscopic binaries, mainly SB2s and short-period SB1s. After obtaining a few RV epochs, they further removed stars that exhibited large-amplitude RV variability due to either strong atmospheric activity or orbital motion with a binary companion.  They may not have had the necessary number of epochs or temporal baseline to fit a binary orbit, but had sufficient observations to infer the reflex motion from a stellar-mass companion. The RV surveys continually removed spectroscopic binaries from their samples as they were discovered so that they could concentrate their remaining observing efforts on stars in which the dominant RV variability could potentially be due to orbiting Jovian planets. Hence, the final RV sample that was continuously monitored for $>$\,5 years contained hosts of giant planets and stars that did not exhibit detectable RV variability from stellar-mass companions.

Given only three RV epochs across a three-year timespan, \citet{Moe2019} showed that 80\% of solar-type binaries within $a$~$<$~10~au produce RV variations above $>$\,1~km\,s$^{-1}$ (see their Fig.~10). Surveys for giant planets have substantially better RV precision and cover a longer timespan, and therefore are sensitive to spectroscopic binaries across a broader parameter space. In fact, the removal of spectroscopic binaries within RV surveys for giant planets closely mimics the suppression factor $S_{\rm bin}$. In other words, RV planet searches removed 100\% of binaries within $a$~$<$~1\,au, 85\% of binaries with $a$~=~10\,au, and no binaries with $a$~=~200\,au. We reiterate that the deficit of close binaries orbiting hosts of close planets is a real suppression effect, as is demonstrated by AO imaging and RV monitoring of planets that were originally discovered via the transit method \citep{Wang2014a,Wang2014b,Wang2015c,Kraus2016,Ngo2016,Ziegler2019}. It just so happens that the close binary population removed by RV planet surveys nearly coincides with the population of close binaries that do not host close planets. We therefore conclude that $R_{\rm HJ; RV}$ is approximately $R_{\rm SingleStars}$. Magnitude-limited RV surveys excluded $f_{\rm Malmquist}$\,$\times$\,$F_{\rm no\,planet}$~=~43\%\,$\pm$\,7\%  of solar-type primaries,  i.e., those that are close spectroscopic binaries that do not host close planets.

Meanwhile, transiting surveys do not make any significant selection biases against binaries. {\it Kepler} may have excluded stars with previously known bright visual companions across $\rho$~$\approx$~1$''$\,-\,3$''$ that would have landed on the same pixel \citep{Batalha2010}, but such a bias across a narrow interval of binary separations and companion masses is negligible. Most important, a bias against wide companions would not affect the inferred hot Jupiter occurrence rate, as only close binaries suppress giant planet formation. The hot Jupiter occurrence rate $R_{\rm HJ;Transit}$ inferred from {\it Kepler} is representative of $R_{\rm AllStars}$. 

\begin{figure}
\centerline{
\includegraphics[trim=0.5cm 0.2cm 0.3cm 0.3cm, clip=true, width=3.3in]{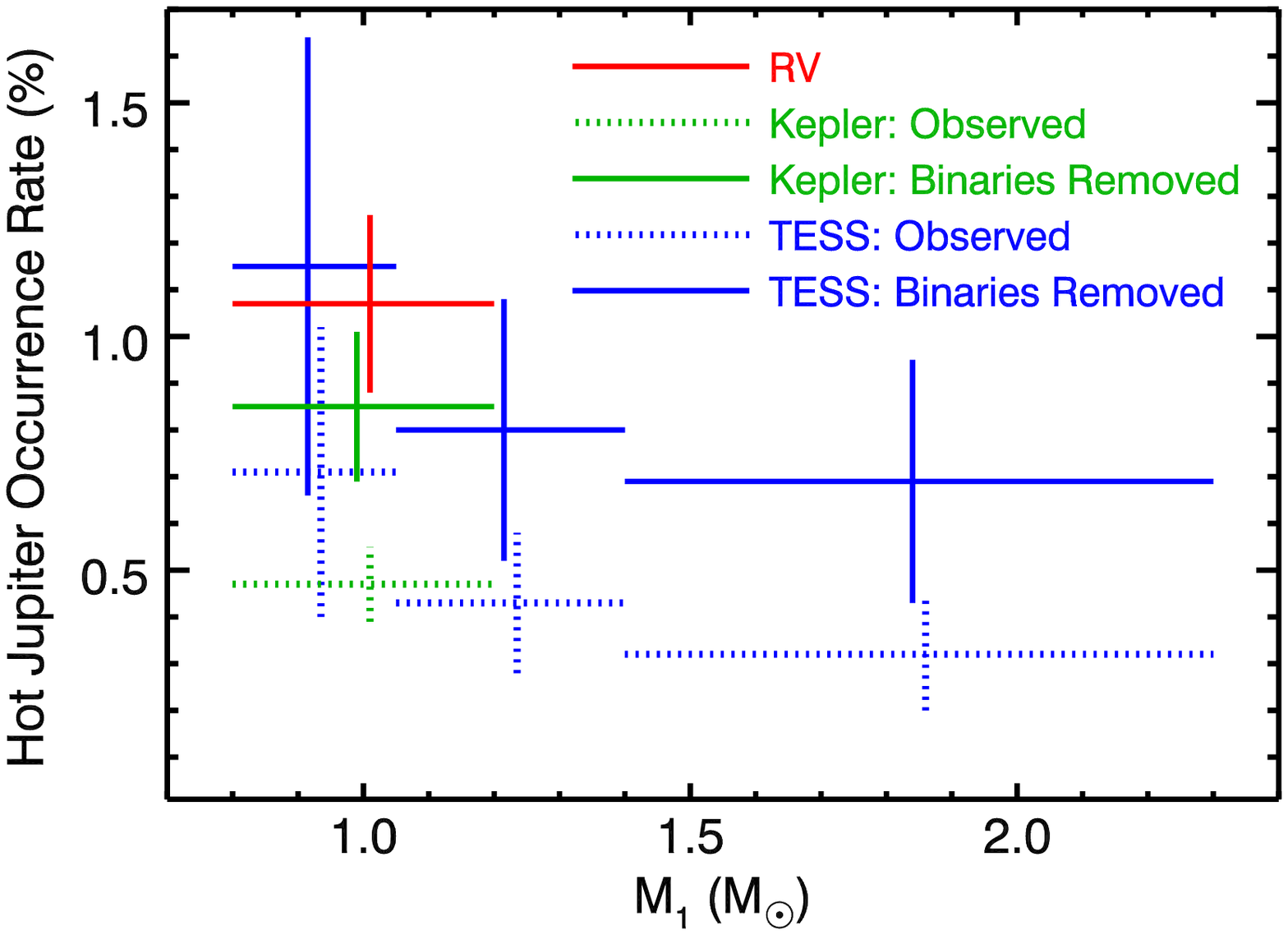}}
\caption{The observed hot Jupiter occurrence rate based on {\it Kepler} (dotted green) is two times lower than the RV value (solid red) at the 3$\sigma$ significant level. {\it TESS} observations (dotted blue) indicate that the hot Jupiter occurrence rate for AF dwarfs is even lower. After statistically removing close binaries within the {\it Kepler} and {\it TESS} transiting surveys (solid green and blue, respectively), the corrected hot Jupiter occurrence rates in effectively single stars are all consistent with $R_{\rm HJ;SingleStars}$ = 0.9\%\,$\pm$\,0.2\% across $M_1$ = 0.8\,-\,2.3\,\Msun.}
\label{HJ}
\end{figure}

By removing close spectroscopic binaries from their magnitude-limited samples of solar-type stars, RV searches for hot Jupiters boost their detection rates by a factor of $R_{\rm HJ;RV}$/$R_{\rm HJ;Transit}$ $\approx$ $R_{\rm SingleStars}$/$R_{\rm AllStars}$ = 1/(1$-$0.43) = 1.8\,$\pm$\,0.2 compared to transiting methods (see Fig.~\ref{ratio} and Fig.~\ref{HJ}). This ratio is fully consistent with the observed factor of two discrepancy in hot Jupiter occurrence rates between the transit and RV samples. The larger frequency of close binaries, which do not host close planets, within the {\it Kepler} and {\it TESS} samples fully explains their lower hot Jupiter occurrence rates compared to RV surveys.

\subsection{Giant Planets}
\label{GiantPlanet}

\subsubsection{Occurrence Rates}

 Like hot Jupiters, the removal of close binaries should also increase the RV occurrence rate of giant planets with longer periods. However, the measured occurrence rates of such temperate giant planets in RV and {\it Kepler} samples currently appear to be consistent with each other \citep{Santerne2016,Fernandes2019}. We argue this similarity is due to three effects. First, their samples of long-period giant planets are too small to distinguish a factor of 1.8\,$\pm$\,0.2 difference between the two methods. For example, 4.5\%\,$\pm$\,0.9\% and 4.1\,$\pm$\,0.8\% of solar-type stars have giant planets with $P$~=~10\,-\,400 days based on RV and transit techniques, respectively \citep{Mayor2011,Santerne2016}. Although these values are consistent with each other, the hypothesis that the RV occurrence rate is $>$1.8 times the transit occurrence rate cannot be ruled out with strong confidence (1.6$\sigma$).  Second, many of the transiting warm and cool Jupiters have not yet been validated with RV measurements, and some may in fact be EB false positives \citep{Fressin2013,Santerne2016}. \citet{Santerne2016} noted that 80\% of their hot Jupiters have secure RV classifications while only 50\% of the temperate giant planets utilized in their statistical analysis are well established. Eccentric eclipsing binaries, which tend to have $P$~$>$~10~days and can produce only one eclipse per orbit \citep{Moe2015b,Kirk2016}, can closely mimic the light curves of long-period transiting giant planets.

Third and perhaps most important, RV surveys identified giant planets according to some minimum mass threshold, typically $M_{\rm p}$\,sin\,$i$ $>$ 0.1\,M$_{\rm J}$  \citep{Johnson2007,Johnson2010,Santerne2016,Fernandes2019}, whereas {\it Kepler} classified giant planets according to their deep transits and large radii, i.e., $R_{\rm p}$~$>$~5\,\Rearth. There is a dearth of hot sub-Saturns and large Neptunes within $P_{\rm p}$~$<$~10~days \citep{Szabo2011,Mazeh2016,Morton2016,Owen2018,Szabo2019}. Thus the populations of transiting and RV hot Jupiters can easily be compared with little probability of contamination by smaller or less massive planets. At longer orbital periods, however, there is a continuous distribution of planet types, and so linking giant planet radii and masses is more ambiguous. \citet{Fernandes2019} found that the occurrence rates of long-period RV and transiting giant planets match each other by assuming that planets with $M_{\rm p}$~=~0.1\,-\,20\,M$_{\rm J}$ correspond to those with $R_{\rm p}$~=~5\,-\,20\,\Rearth\ (see their Fig.~2).  However, a non-negligible fraction of transiting {\it Kepler} planets with $P$~=~10\,-\,400~days are ``super-puffs'' \citep{Masuda2014,Lopez2014,Lee2016}, which have large radii ($R_{\rm p}$~=~5\,-\,8\,\Rearth) but very small masses ($M_{\rm p}$~=~3\,-\,10\,\Mearth~$=$~0.01\,-\,0.03\,M$_{\rm J}$).
 Kepler-51\,b (7.0\,\Rearth, 0.01\,M$_{\rm J}$, $P_{\rm P}$~=~45~days), Kepler-79\,d (7.2\,\Rearth, 0.02\,M$_{\rm J}$, 52~days), and Kepler-87\,c (6.3\,\Rearth, 0.02\,M$_{\rm J}$, 192~days) are a few case examples of super-puff planets \citep{Masuda2014,JontofHutter2014,Ofir2014,Hatzes2015,Santerne2016}. Such super-puffs would generally be classified as giant planets in transiting surveys, but missed in RV searches for Jovian-mass planets. We predict that with larger samples, more complete vetting of false positives, and robust mapping between giant planet radii and masses, a statistically significant factor of 1.8\,$\pm$\,0.2 discrepancy in the occurrence rate of temperate giant planets will become apparent between RV surveys (which exclude close binaries) and transit surveys (which do not).

\subsubsection{Trends with Stellar Mass}

Close binaries also bias inferred planet trends with respect to host mass. Namely, transiting surveys of AF stars severely underestimate $R_{\rm SingleStars}$ due to their larger close binary fraction. For example, RV surveys have revealed that the giant planet occurrence rate monotonically increases with primary mass across $M_1$~=~0.2\,-\,2.0\,\Msun\  \citep{Johnson2010,Bowler2010,Bonfils2013}. Meanwhile, {\it Kepler} found that the giant planet occurrence rate within $a$~$<$~1~au initially increases from 3.6\%\,$\pm$\,1.7\% for M-dwarfs to 6.1\%\,$\pm$\,0.9\% for GK-dwarfs, but then decreases to 4.3\%\,$\pm$\,1.0\% for AF-dwarfs \citep{Fressin2013}. Early {\it TESS} observations also indicated that the hot Jupiter occurrence rate decreases with primary mass across $M_1$~=~0.8\,-\,2.3\,\Msun\ (\citealt{Zhou2019}; see our Fig.~\ref{HJ}). One possible explanation for the discrepancy is that Jovian planets orbiting AF stars are at systematically wider separations ($a$~$>$~1~au) and therefore missed by {\it Kepler} and {\it TESS} but detected in RV surveys \citep{Fressin2013}. A deficit of Jovian planets orbiting 2\,-\,5\,\Msun\ stars inside of $a$~$<$~1~au at least partially accounts for the observed discrepancy \citep{Bowler2010,Reffert2015}. Another possibility is that retired A-stars, i.e., $\approx$\,2\,\Msun\ sub-giants that are utilized by RV surveys to measure giant planet occurrence rates for more massive stars, actually evolved from lower mass F-dwarfs \citep{Lloyd2011,North2017}.

A third contributing factor is that the {\it Kepler} and {\it TESS} samples of AF-dwarfs contain a substantially larger population of close binaries. In a magnitude-limited sample of AF stars, we estimate that the giant planet occurrence rate is 2.0\,$\pm$\,0.3 times larger in RV surveys compared to transiting surveys. Based on {\it Kepler} observations, the giant planet occurrence rate within $a$~$<$~1\,au of all AF primaries is $R_{\rm AllStars}$~=~4.3\%\,$\pm$\,1.0\% \citep{Fressin2013}. We therefore estimate a bias-corrected rate of $R_{\rm SingleStars}$~=~8.6\%\,$\pm$\,2.3\% within $a$~$\lesssim$~1\,au of single AF primaries, which is consistent with the RV measurements. Similarly, although the observed {\it TESS} hot Jupiter occurrence rate may decrease with stellar mass,  the bias-corrected rate in effectively single stars is nearly independent of host mass, at least within the statistical uncertainties, i.e., $R_{\rm HJ;SingleStars}$ = 0.9\%\,$\pm$\,0.2\% across $M_1$ = 0.8\,-\,2.3\,\Msun (see Fig.~\ref{HJ}). Future studies need to consider the effects of binaries when comparing RV and transit surveys, especially for AF primaries where $f_{\rm Malmquist}$\,$\times$\,$F_{\rm no\,planets}$~=~50\%\,$\pm$\,8\% do not host close planets due to close binary suppression.

\subsection{Small Planets}
\label{SmallPlanet}

\subsubsection{Trends with Stellar Mass}

{\it Kepler} showed that the occurrence rate of sub-Neptunes and super-Earths with short orbital periods decreases with respect to stellar mass \citep{Howard2012,Petigura2013,Morton2014,Dressing2015,Mulders2015a,Mulders2015b,Gaidos2016}. In particular, \citet{Mulders2015a} measured the frequency of planets with $P_{\rm p}$~$<$~50~days and $R_{\rm p}$ = 1\,-\,4\,\Rearth\ orbiting M-dwarfs to be twice the frequency compared to G-dwarfs and triple that of F-dwarfs. \citet{Mulders2015b} subsequently showed that M-dwarfs host close, small planets with $R_{\rm p}$ = 1.0\,-\,2.8\,\Rearth\ within $P_{\rm p}$~$<$~50~days at 3.5 times the occurrence rate of FGK stars. They argued that the correlation is intrinsic to the formation process of close, small planets, and that the change in the binary fraction with respect to stellar mass only mildly biases the observed trend. In the following, we show that binaries can explain half (but not all) of the observed variation.

\begin{figure}
\centerline{
\includegraphics[trim=3.6cm 12.1cm 7.9cm 0.9cm, clip=true, width=3.3in]{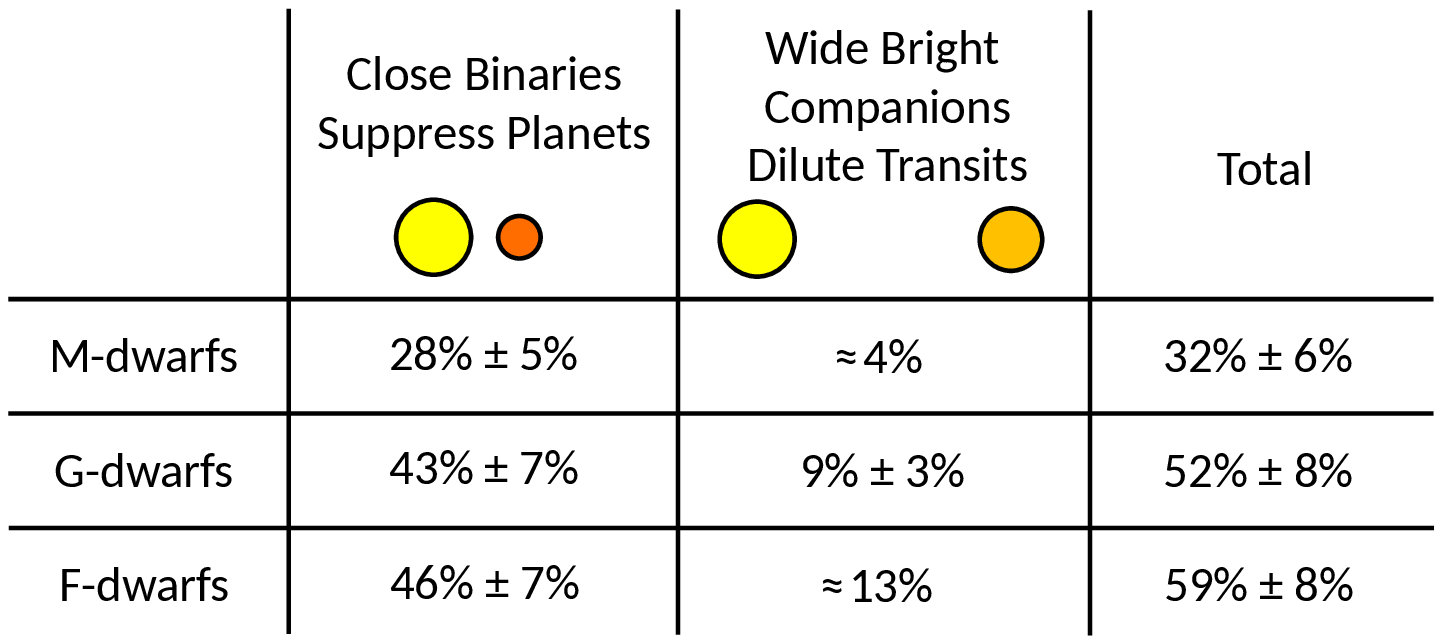}}
\caption{Impact of binary stars on the occurrence rates of small planets as a function of stellar mass. In magnitude-limited samples, $F_{\rm no\,planet}$\,$\times$\,$f_{\rm Malmquist}$ = 28\%, 43\% and 46\% of M-dwarf, G-dwarf, and F-dwarf primaries, respectively, do not host planets (small or large) due to suppression by close binaries (first column). In addition, a non-negligible fraction of {\it Kepler} targets have wide (but unresolved) bright stellar companions that inhibit the detection of shallow transits from small planets (second column). The frequency of small planets orbiting single G-dwarfs is 1/(1$-$0.52) = 2.1 \,$\pm$\,0.3 times larger than the rate inferred from all {\it Kepler} G-dwarfs. }
\label{smallplanetFig}
\end{figure}

We consider two selection effects whereby binaries considerably decrease the inferred occurrence rate of small transiting planets orbiting solar-type stars. First, close binaries suppress planet formation, and the close binary fraction increases with stellar mass.  Specifically, in volume-limited samples, the fraction of stars that do not host close planets due to close binary suppression increases from  $F_{\rm no\,planet}$~=~18\%\,$\pm$\,4\% for M-dwarfs ($M_1$ = 0.3\,\Msun) to 37\%\,$\pm$\,6\% for F-dwarfs (1.3\,\Msun) (Eqn.~\ref{Fclose}). Similarly, in magnitude-limited samples, the fraction of stars that do not host close planets increases from $F_{\rm no\,planet}$\,$\times$\,$f_{\rm Malmquist}$ = 28\%\,$\pm$\,5\% for M-dwarfs to to 46\%\,$\pm$\,7\% for F-dwarfs (see first column in Fig.~\ref{smallplanetFig}).

Second, even wide bright stellar companions dilute the photometric signal, decreasing the probability of detecting small transiting planets. Solar-type hosts of small {\it Kepler} planets \citep{Wang2014b} and small {\it TESS} planets \citep{Ziegler2019} both exhibit a slight deficit of wide stellar companions, $S_{\rm bin}$~=~60\% and 40\%, respectively, most likely due to this transit dilution selection bias (see Section~\ref{Suppression}). To illustrate more conclusively, \citet{Ziegler2019} discovered 65 wide companions with $a$~=~100\,-\,2,000~au and brightness contrasts $\Delta I$~$<$~5.1~mag to hosts of {\it TESS} planet candidates. The majority (53/65~=~82\%\,$\pm$\,5\%) of these wide binaries host large planet candidates with $R_{\rm p}$~$>$~4.0\,\Rearth. Large planet hosts include both bright and faint wide companions, i.e., 15/53~=~28\%\,$\pm$\,6\% with $\Delta I$~$<$~1.5~mag and  38/53~=~72\%\,$\pm$\,6\% with $\Delta I$~=~1.5\,-\,5.1~mag. Meanwhile, all 12 wide companions to hosts of small planets are relatively faint with $\Delta I$~=~1.7\,-\,5.1~mag. The probability of not detecting any companions brighter than $\Delta I$~$<$~1.5~mag to the small planet hosts when we expected  0.28\,$\times$\,12~=~3.4 is $p$~=~0.03. The known sample of small planet hosts are significantly biased against bright wide companions due to transit dilution.

For hosts of small {\it Kepler} planets, \citet{Wang2014b} measured a non-negligible suppression factor of  $S_{\rm bin}$~=~60\%\,$\pm$\,10\% across $a$~=~50\,-\,2,000~au, which we attribute to the transit dilution selection bias. We show in Section~\ref{HJcomprate} that $F_{\rm wide}$~=~37\%\,$\pm$\,7\% of solar-type planet hosts in a magnitude-limited sample have wide stellar companions across $a$~=~50\,-\,2,000~au. We therefore empirically compute that (1$-S_{\rm bin}$)\,$\times$\,$F_{\rm wide}$ = 0.40\,$\times$\,0.37 = 15\%\,$\pm$\,5\% of small planet hosts do not have transits detectable by {\it Kepler} due to photometric dilution by wide stellar companions. In other words, $F_{\rm dilute}$ =  (0.15\,$\pm$\,0.05)\,$\times$\,(1$-$0.43) = 9\%\,$\pm$\,3\% of all G-dwarfs have bright, wide companions that inhibit the detection of small transiting planets (see second column in Fig.~\ref{smallplanetFig}). 

The probability of detecting small planets transiting larger F-dwarfs is even smaller, especially if their hosts have bright stellar companions. Moreover, the wide binary fraction also increases with stellar mass, especially across the interval $M_1$ = 0.3\,-\,1.3\,\Msun\ (see Appendix~\ref{Appendix}). We estimate that $F_{\rm dilute}$~$\approx$~13\% of F-dwarfs have bright wide companions that inhibit the detection of small transiting planets. Conversely, perhaps only $F_{\rm dilute}$~$\approx$~4\% of M-dwarfs within the {\it Kepler} sample have wide companions that dilute the transits of small planets below the detection threshold.  The fractions of M-dwarfs and F-dwarfs with wide companions that inhibit transit detection are more uncertain than our empirical estimate for G-dwarfs. Fortunately, transit dilution by wide binaries only slightly biases the inferred planet occurrence rates compared to the main effect of planet suppression by close binaries, where the measurements and uncertainties are more robust (see Fig.~\ref{smallplanetFig}).

Under the hypothesis that the small planet occurrence rate in single stars is independent of stellar mass, then we would predict that the observed {\it Kepler} sample of M-dwarfs should have (1$-$0.32)/(1$-$0.59)~=~1.7\,$\pm$\,0.4 times the occurrence rate of small planets than {\it Kepler} F-dwarfs. This is only half of the observed factor of 3.0\,-\,3.5 variation measured by \citet{Mulders2015a} and \citet{Mulders2015b}. Even after correcting for binaries, the occurrence rate of small, close planets intrinsically decreases with increasing stellar mass, consistent with their main conclusion. 

To more conclusively demonstrate that binaries cannot fully account for the observed trend in small planet occurrence rates, we consider the extremes in $F_{\rm no\,planet}$+$F_{\rm dilute}$ allowed by the observations. In a magnitude-limited sample, the overall binary fraction of F-dwarfs is 70\%\,$\pm$\,7\%, a non-negligible fraction of which have faint, wide secondaries that can neither suppress planet formation nor significantly dilute transits. The 1.4$\sigma$ upper limit of $F_{\rm no\,planet}$+$F_{\rm dilute}$ = 70\% for F-dwarfs inferred from Fig.~\ref{smallplanetFig} is therefore quite conservative. Meanwhile, the magnitude-limited binary fraction of M-dwarfs ($M_1$~=~0.3\Msun) within $a$~$<$~10~au is 21\%\,$\pm$\,4\%, a significant majority of which suppress planet formation \citep{Ziegler2019}. An additional 14\%\,$\pm$\,3\% of M-dwarfs in magnitude-limited samples have stellar companions across $a$~=~10\,-\,100 AU, a large fraction of which have large mass ratios above $q$~$>$~0.7. Even if these M-dwarf binaries with intermediate separations do not suppress planet formation, the non-negligible fraction with large mass ratios will at least inhibit the detection of small transiting planets. The 1.4$\sigma$ lower-limit on the sum of $F_{\rm no\,planet}$+$F_{\rm dilute}$~=~23\% for M-dwarfs is robust (see Fig.~\ref{smallplanetFig}).  The 1.4$\sigma$ upper and lower-limits on $F_{\rm no\,planet}$+$F_{\rm dilute}$ for F-dwarfs and M-dwarfs, respectively, yield a 2.0$\sigma$ spread. The maximum  (2.0$\sigma$) variation in small planet occurrence rates between M-dwarfs and F-dwarfs due to binaries is therefore a factor of (1.0$-$0.23)/(1.0$-$0.70) = 2.6, less than the observed factor of 3.0\,-\,3.5 variation. We can rule out the hypothesis that binaries fully account for the observed trend between stellar mass and occurrence rate of small, close planets with 2.9$\sigma$ confidence.

Nonetheless, binaries account for roughly half of the observed trend. Thus the bias-corrected occurrence rate of small planets closely orbiting single M-dwarfs is (3.0\,-\,3.5)/(1.7\,$\pm$\,0.4) = 1.9\,$\pm$\,0.4 times larger than the rate for single F-dwarfs. Future studies of planet formation in single stars should anchor their models to this bias-corrected result.

\subsubsection{Impact on $\eta_{\oplus}$}

A total of 43\%\,+\,9\% = 52\%\,$\pm$\,8\% of G-dwarf primaries do not have small planets detectable by {\it Kepler} due to the combined effects of close binary suppression and wide binary transit dilution (third column in Fig.~\ref{smallplanetFig}). The occurrence rate of small planets orbiting single G-dwarfs is therefore 1/(1$-$0.52) = 2.1\,$\pm$\,0.3 times larger than the rate inferred from all {\it Kepler} G-dwarfs. The frequency $\eta_{\oplus}$ of Earth-sized planets in the habitable zone of single solar-type stars is also 2.1\,$\pm$\,0.3 times larger than the frequency previously measured for all solar-type stars.

The strong dependence of $\eta_{\oplus}$ on binary status has profound implications for the expected yields and target prioritization of directing planet imaging surveys. For example, the nearest G-type star to our solar system, $\alpha$\,Centauri~A, contains a K1V companion at $a$~=~18\,au \citep{Pourbaix2002}. Considering $S_{\rm bin}$~=~30\% near $a$~=~18~au (Fig.~\ref{suppressfactor}), then an earth-sized planet in the habitable zone is highly unlikely to orbit $\alpha$~Centauri~A. Meanwhile, the next closest G-type star, $\tau$\,Ceti, is single, exhibits a debris disk, and hosts several small planet candidates \citep{Lawler2014,MacGregor2016,Feng2017}. In Paper~II, we discuss $\eta_{\oplus}$ as a function of both binary status and host star metallicity. 

\subsubsection{Close Neptunes}

Like hot Jupiters and giant Jovian planets, the occurrence rate of close Neptunes inferred from RV surveys is larger than the rate measured from transit surveys. After correcting for geometric transit probabilities, 16\%\,$\pm$\,2\% of all {\it Kepler} solar-type stars have $R_{\rm p}$~=~2\,-\,6\,\Rearth\ planets within $P_{\rm p}$~$<$~50~days \citep{Howard2012,Fressin2013,Mulders2015a}. The majority of transiting Neptunes within $P_{\rm p}$~$<$~50~days are detectable by {\it Kepler}, even if they have bright twin companions that dilute the transit depth by a factor of two. Nonetheless, Neptunes are skewed significantly toward small radii, i.e., $f$ $\propto$ $R_{\rm p}^{-3}$ \citep{Howard2012,Mulders2015b}, and so most Neptunes transiting hosts with bright twin stellar companions are misclassified as super-Earths below $R_{\rm p}$~$<$~2\,\Rearth. We must still account for both suppression by close binaries and transit dilution by wide binaries, and so we expect that (16\%\,$\pm$\,2\%)(2.1\,$\pm$\,0.3) = 33\%\,$\pm$\,5\% of single G-dwarfs have $R_{\rm p}$~=~2\,-\,6\,\Rearth\ planets within $P_{\rm p}$~$<$~50~days. 

Meanwhile, high-precision RV surveys find that 28\%\,$\pm$\,5\% of solar-type stars have Neptunes with $M_{\rm p}$\,sin\,$i$ = 3\,-\,30\,\Mearth\ within the same period range (Table~2 in \citealt{Mayor2011}). Assuming $M_{\rm p}$ = 3\Mearth\ roughly corresponds to $R_{\rm p}$ = 2\,\Rearth\ (see \citealt{Lopez2014}), then the RV surveys find a factor of  (0.28\,$\pm$\,0.05)/(0.16\,$\pm$\,0.02) = 1.8\,$\pm$\,0.3 times more close Neptunes than {\it Kepler}, consistent with our predicted factor of 2.1\,$\pm$\,0.3 discrepancy. The fact that the occurrence rates of both hot Jupiters and close Neptunes are two times higher in the RV surveys compared to transiting surveys confirms our  conclusion that binaries within the {\it Kepler} sample measurably reduce the inferred planet occurrence rates. 

\section{Wide Companions to Hot Jupiter Hosts}
\label{HJwide}
\subsection{Mass-ratio Distribution}
\label{HJq}

Now that we have carefully considered how close binaries affect the formation and occurrence rates of hot Jupiters, we next investigate how wide stellar companions influence hot Jupiters. In order to measure the intrinsic wide binary fraction of hot Jupiter hosts (Section~\ref{HJcomprate}), we must first analyze their binary mass-ratio distribution. Both \citet{Ngo2016} and \citet{Evans2018} noted that the observed wide companions to known hot Jupiter hosts are weighted toward small mass ratios, inconsistent with the field solar-type binary mass-ratio distribution. They argued that the deficit of wide binaries with $q$~$>$~0.7 to known hot Jupiter hosts is due to a selection bias whereby bright companions within the same pixel dilute the photometric transit depths of hot Jupiters, making it more difficult to discover them. Indeed, the majority of hot Jupiters in their samples were discovered by the HAT and WASP transiting surveys \citep{Bakos2004,Pollacco2006}, which have very large 14$''$ pixels ($\approx$\,4,000~au at the average distance $\approx$\,300\,pc to their targets). \citet{Ngo2016} and \citet{Evans2018} speculated that the wide binary fraction of hot Jupiter hosts might be even larger than their reported values after accounting for the possible bias against bright companions. However, in the following, we show that the mass-ratio distributions of wide companions to hot Jupiter hosts and wide solar-type binaries are actually consistent with each other. The transit dilution bias against bright companions to hot Jupiter hosts is therefore negligible. 

\citet{Ngo2015} and \citet{Ngo2016} listed the primary and secondary masses of all their AO-detected binaries. Corrections for incompleteness become non-negligible and uncertain below $q$~$<$~0.2 for both nearby solar-type field binaries \citep{Chini2014,Moe2017} and especially more distant hosts of hot Jupiters \citep{Ngo2016}. We therefore select the 19 wide companions to hot Jupiter hosts with $q$~$>$~0.2 and $a$ = 50\,-\,2,000 au in their sample. 

\citet{Evans2016} and \citet{Evans2018} utilized lucky imaging with the Two Colour Instrument (TCI) to detect companions primarily on the red side, which has a broad spectral response across 660\,-\,1050~nm. \citet{Evans2018} reported the brightness contrasts $\Delta r_{\rm TCI}$ of their binary companions, but did not fit isochrones to estimate the mass ratios. Nonetheless, there are five binaries common to both surveys (HAT-P-30, HAT-P-35, HAT-P-41, WASP-8, WASP-36), where \citet{Ngo2016} estimated the mass ratios $q$~=~0.35\,-\,0.48 while \citet{Evans2018} measured the brightness contrasts $\Delta r _{\rm TCI}$~=~3.4\,-\,4.6 mag. Averaging these five systems provides a firm empirical relation such that  $\Delta r _{\rm TCI}$~=~4.0~mag maps to $q$~=~0.43. \citet{Evans2018} also estimated that $\Delta r _{\rm TCI}$~=~2.0~mag corresponds to $q$~=~0.75. To anchor smaller mass ratios, we rely on the similarity between the average wavelength of $r_{\rm TCI}$ and $I$-band. E.~Mamajek compiled empirical absolute $I$-band magnitudes and stellar masses as a function of spectral type\footnote{http://www.pas.rochester.edu/~emamajek/spt/}. Interpolating his tables, a 0.43\,\Msun\ dwarf is $\Delta I$~=~4.0~mag fainter than the sun, identical to our estimate of $\Delta r _{\rm TCI}$~=~4.0~mag for $q$~=~0.43. Meanwhile, a 0.20\,\Msun\ dwarf is $\Delta I$~=~6.0~mag fainter than the sun. We therefore adopt $q$ = 0.20, 0.43, 0.75, and 1.0 for $\Delta r_{\rm TCI}$~=~6.0, 4.0, 2.0, and 0.0 mag, respectively, and interpolate to estimate the mass ratios of the wide companions to hot Jupiter hosts in \citet{Evans2018}. There are 10 additional companions to hot Jupiter hosts with $q$~$>$~0.2 ($\Delta r _{\rm TCI}$~$<$~6.0~mag) and $a$~=~50\,-\,2,000~au in Table~6 of \citet{Evans2018}, bringing the total to 29 unique systems.

\begin{figure}
\centerline{
\includegraphics[trim=0.4cm 0.2cm 0.3cm 0.3cm, clip=true, width=3.3in]{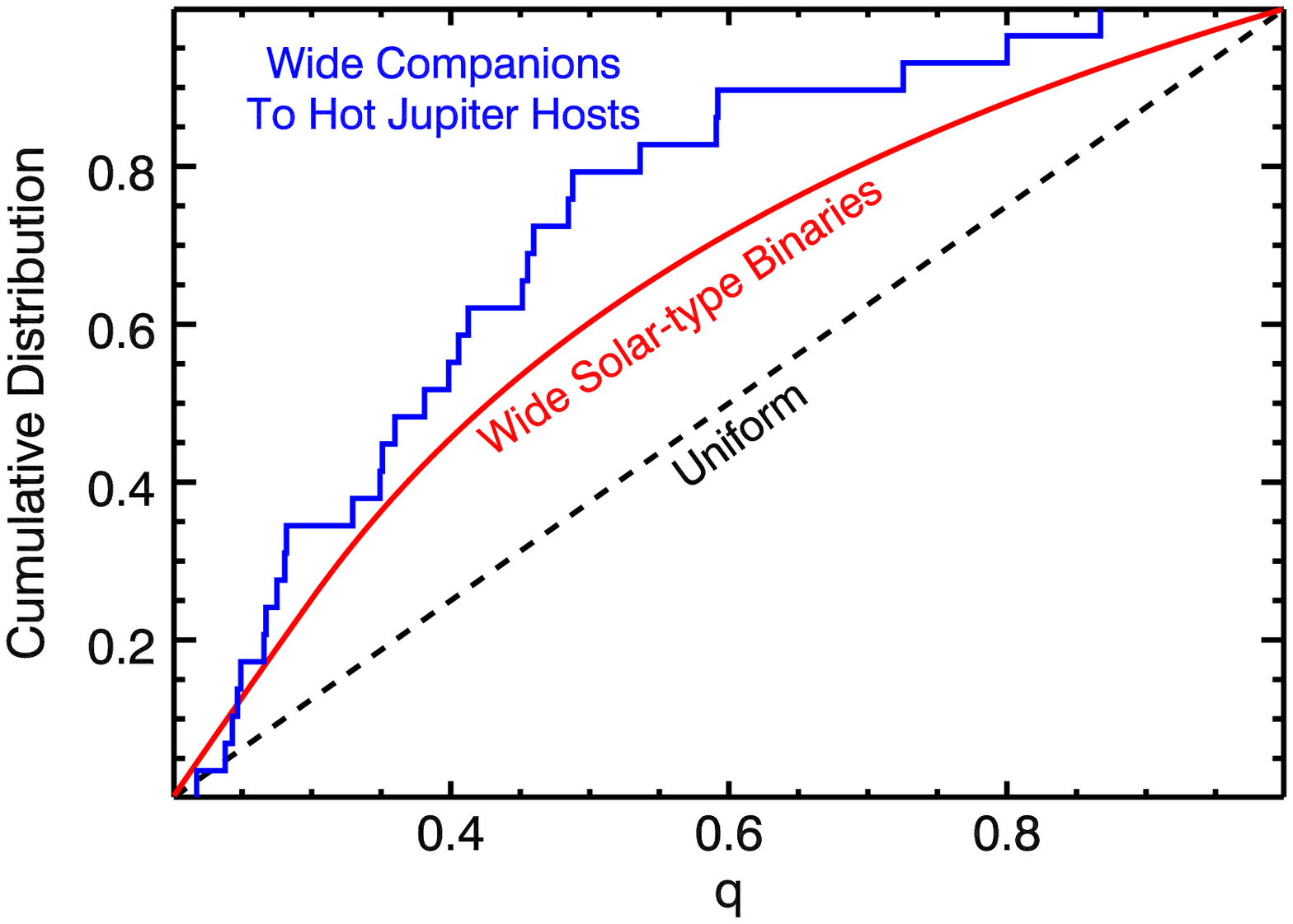}}
\caption{Cumulative mass-ratio distribution of the 29 wide companions to hot Jupiter hosts with $a$~=~50\,-\,2,000~au and $q$~$>$~0.2 within the combined \citet{Ngo2016} and \citet{Evans2018} samples (blue), which is bottom-heavy compared to a uniform mass-ratio distribution (dashed black). Although the overall mass-ratio distribution of solar-type binaries is roughly uniform \citep{Raghavan2010}, wide solar-type binaries are weighted toward small mass ratios \citep[red;][]{Moe2017,ElBadry2019b}. Wide companions to hot Jupiter hosts and wide solar-type binaries therefore have mutually consistent mass-ratio distributions ($p_{\rm KS}$ = 0.16), and so the transit dilution bias against detecting hot Jupiters in wide twin binaries is statistically insignificant.}
\label{q_HJ}
\end{figure}

In Fig.~\ref{q_HJ}, we plot the cumulative mass-ratio distribution of the 29 companions to hot Jupiter hosts with $q$~$>$~0.2 and $a$~=~50\,-\,2,000 au. We confirm the conclusions of \citet{Ngo2016} and \citet{Evans2018} that wide companions to hot Jupiter hosts are bottom-heavy compared to a uniform mass-ratio distribution, discrepant at the $p_{\rm KS}$~=~2$\times$10$^{-5}$ level according to a Kolmogorov-Smirnov test. However, although the overall solar-type binary mass-ratio distribution is roughly uniform with a small excess of twins \citep{Raghavan2010}, wide solar-type binaries are weighted toward small mass ratios with a negligible excess twin fraction \citep{Moe2017,ElBadry2019b}. Utilizing the \citet{Raghavan2010} sample, \citet{Moe2017} demonstrated that wide solar-type binaries with $a$~=~200\,-\,5,000~au are accurately modeled by a two-component power-law distribution $f$~$\propto$~$q^{\gamma}$ with a bottom-heavy slope of $\gamma_{\rm largeq}$~=~$-$1.1 across large mass ratios $q$~=~0.3\,-\,1.0 and a turn-over toward $\gamma_{\rm smallq}$~=~0.3 across small mass ratios $q$~=~0.1\,-\,0.3. \citet{ElBadry2019b} examined {\it Gaia} common-proper-motion binaries, and confirmed that wide solar-type binaries with $a$~$>$~200~au are weighted toward small mass ratios. They fitted a slightly different break, i.e., $\gamma_{\rm largeq}$~=~$-$1.4 across $q$~=~0.5\,-\,1.0 and a flattening toward $\gamma_{\rm smallq}$~=~0.0 across $q$ = 0.1\,-\,0.5.  Based on the \citet{DeRosa2014} AO survey of A-type primaries, \citet{Moe2017} showed that wide companions near $a$~=~400\,au are even further skewed toward small mass ratios, i.e., $\gamma_{\rm largeq}$~=~$-$2.0 across $q$~=~0.3\,-\,1.0 and  $\gamma_{\rm smallq}$~=~$-$1.0 across $q$~=~0.1\,-\,0.3. Similarly, \citet{ElBadry2019b} found that wide {\it Gaia} binaries with more massive primaries are weighted toward smaller mass ratios, also with a break near $q$~=~0.3. The average primary mass of the 77 hosts of hot Jupiters examined in \citet{Ngo2016}  is $\langle M_1 \rangle$~=~1.2\,\Msun. Interpolating the measurements, wide companions to $M_1$~=~1.2\,\Msun\ primaries and hot Jupiter hosts (assuming there is no transit dilution selection bias) should follow a bottom-heavy mass-ratio distribution with $\gamma_{\rm largeq}$~=~$-$1.3\,$\pm$\,0.3 across $q$~=~0.3\,-\,1.0 and $\gamma_{\rm smallq}$~=~$-$0.1\,$\pm$\,0.5 across $q$ = 0.1\,-\,0.3 (red line in Fig.~\ref{q_HJ}).

As displayed in Fig.~\ref{q_HJ}, the mass-ratio distributions of wide companions to hot Jupiter hosts and our best-fit model for  wide solar-type binaries are mutually consistent ($p_{\rm KS}$~=~0.16 according to a one-sample KS test). Our model for the mass-ratio distribution of wide solar-type binaries also has uncertainties (see above). Hence, the statistical significance of disagreement between the two populations is even smaller, which strengthens our conclusion that wide solar-type binaries and wide companions to known hot Jupiter hosts do not have statistically different mass-ratio distributions. Transit surveys may fail to detect a very small fraction of hot Jupiters in wide binaries with $q$~$>$~0.7, but the effect is minor and statistically insignificant. Although {\it Kepler} and {\it TESS} missed a significant fraction of transiting small planets in wide binaries due to photometric dilution (Section~\ref{SmallPlanet}), the deeper and more frequent transits of hot Jupiters are more immune to wide binary dilution, even in ground-based surveys. We conclude that the completeness-corrected wide binary fraction of hot Jupiter hosts reported in \citet{Ngo2016} and \citet{Evans2018} are not lower limits but instead unbiased accurate measurements. 

\subsection{Wide Binary Fraction}
\label{HJcomprate}

\begin{figure}
\centerline{
\includegraphics[trim=1.7cm 8.3cm 5.3cm 0.5cm, clip=true, width=3.3in]{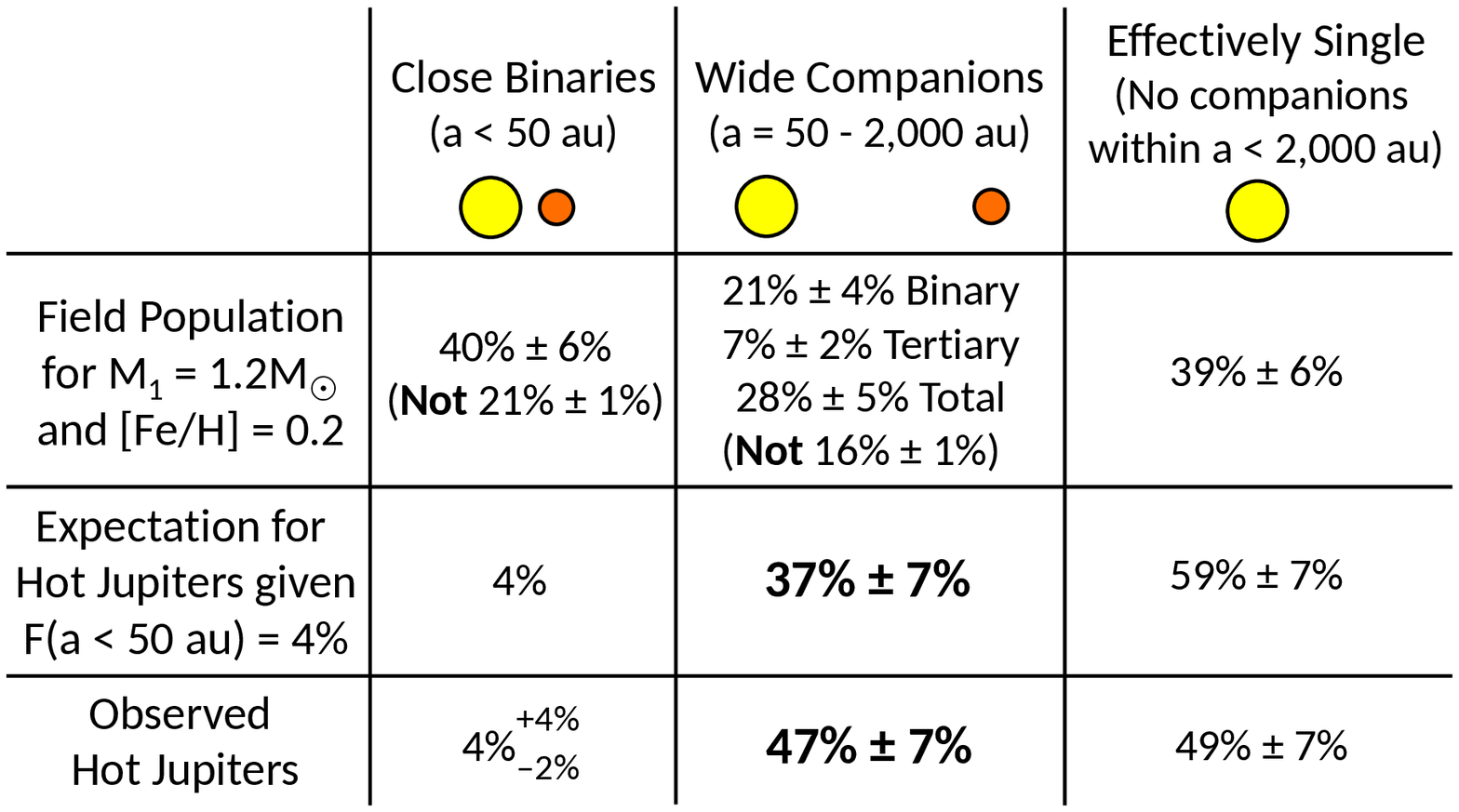}}
\caption{The multiplicity fractions of field stars (top row) versus those expected (middle) and measured (bottom) for hot Jupiter hosts. For a magnitude-limited sample of $M_1$~=~1.2\Msun\ field primaries with [Fe/H]~=~0.2, 40\%\,$\pm$\,6\% have close stellar companions within $a$~$<$~50~au (higher than the 21\%\,$\pm$\,1\% estimated in \citealt{Ngo2016}) and 28\%\,$\pm$\,5\% have wide MS companions across $a$~=~50\,-\,2,000~au (also higher than the 16\%\,$\pm$\,1\% estimated in \citealt{Ngo2016}). After correcting for incompleteness, \citet{Ngo2016} reported that 4\%$_{-2\%}^{+4\%}$ and 47\%\,$\pm$\,7\% of hot Jupiter hosts have stellar companions within $a$~$<$~50~au and across $a$~=~50\,-\,2,000~au, respectively. Because close binaries suppress hot Jupiters, we actually expect the wide binary fraction of hot Jupiter hosts to be larger than the field wide binary fraction, even if wide binaries do not influence planet formation. Given $F_{\rm HJ,close}$~=~4\% and a fixed ratio $F_{\rm field,wide}$/$F_{\rm field,single}$ = $F_{\rm HJ,wide}$/$F_{\rm HJ,single}$ = 0.62, then we expect $F_{\rm HJ,wide}$ = 37\%\,$\pm$\,7\%, which is consistent with the measured value of 47\%\,$\pm$\,7\% at the 1.0$\sigma$ level. Wide binaries do not enhance hot Jupiter formation at a statistically significant level, i.e., $f_{\rm enhance}$ = 1.27\,$\pm$\,0.28.}
\label{HJ_Fig}
\end{figure}

After correcting for incompleteness, \citet{Ngo2016} reported that $F_{\rm HJ,close}$ = 4$_{-2}^{+4}$\% and $F_{\rm HJ,wide}$ = 47\,$\pm$\,7\% of hot Jupiter hosts have stellar companions within $a$~$<$~50~au and across $a$ =~50\,-\,2,000~au, respectively. This implies that $F_{\rm HJ,single}$ = 49\,$\pm$\,7\% of hot Jupiter hosts are effectively single, i.e., truly single stars or in very wide binaries beyond $a$~$>$~2,000~au. We list these three fractions in the bottom row of Fig.~\ref{HJ_Fig}.

\citet{Ngo2016} then compared to the \citet{Raghavan2010} sample of field solar-type binaries. Specifically, they adopted a log-normal period distribution with mean of $\mu_{\rm logP}$ = 5.0 and dispersion of $\sigma_{\rm logP}$ = 2.3 scaled to an overall binary fraction of 44\%\,$\pm$\,3\%. By integrating below $a$~$<$~50~au and across $a$~=~50\,-\,2,000~au, they computed $F_{\rm field,close}$~=~21\%\,$\pm$\,1\% and $F_{\rm field,wide}$~=~16\,$\pm$\,1\%, respectively. They concluded that close binaries suppress hot Jupiters, i.e., $S_{\rm bin}$ = $F_{\rm HJ,close}$/$F_{\rm field,close}$ = 0.04/0.21~=~0.2, whereas wide binaries enhance the formation of hot Jupiters, i.e., $f_{\rm enhance}$ = $F_{\rm HJ,wide}$/$F_{\rm field,wide}$ = 0.47/0.16~=~2.9. However, the \citet{Raghavan2010} sample is also incomplete, contains both inner binaries and outer tertiaries in triples, has an average primary mass and metallicity slightly different than hot Jupiter hosts, and is volume-limited whereas surveys for transiting hot Jupiters are magnitude-limited. Most important, close binaries suppress planets, and so an imaging survey of planet hosts will actually be biased toward a higher fraction of single stars and wide binaries compared to field stars. This bias toward wide binaries not only affects the \citet{Ngo2016} sample of hot Jupiters, but all imaging surveys of planet hosts \citep{Wang2015c,Kraus2016,Matson2018,Ziegler2019}. In the following, we carefully account for these various selection effects and show that the observed wide binary fraction of hot Jupiter hosts is actually fully consistent with expectations.  

For a volume-limited sample of $M_1$ = 1.0\,\Msun\ primaries with [Fe/H]~$=$~$-$0.1, the binary fraction below $a$~$<$~50~au is $F_{\rm field,close}$~=~34\%\,$\pm$\,4\%, which is the integral of the red curve in Fig.~\ref{solarperiod} below log\,$P$\,(days)~$<$~5.0. This is already larger than the value of $F_{\rm field,close}$~=~21\%\,$\pm$\,1\% computed by \citet{Ngo2016} for two reasons. First, although the separation distribution of all companions peaks near $a$~=~50~au, the distribution of inner binaries peaks at closer separations, resulting in a larger close binary fraction (see Fig.~\ref{solarperiod}). It is inconsistent to scale the period distribution of all companions, including inner binaries and outer tertiaries, to the binary fraction (see Section~\ref{Solar} and Fig.~\ref{TokFig}). Second, the \citet{Raghavan2010} sample is incomplete toward WD and late-M companions with intermediate separations of $a$~=~5\,-\,30~au (see Appendix~\ref{Appendix}). Meanwhile, the long-term RV monitoring of hot Jupiter hosts were considerably more precise \citep{Knutson2014,Bryan2016,Ngo2016}, and were sensitive to nearly all $\approx$\,0.6\,\Msun\ WD companions within $a$~$<$~50~au (see Fig.~7 in \citealt{Ngo2016}). The close binary fraction $F_{\rm HJ,close}$ = 4$_{-2}^{+4}$\% of hot Jupiter hosts reported by \citet{Ngo2016} therefore potentially includes WD companions, and is completeness-corrected for missing late-M companions. The corresponding field close binary fraction must therefore also incorporate all WD and late-M companions within $a$~$<$~50~au.

For a magnitude-limited sample, Malmquist bias boosts the solar-type close binary fraction by a factor of $f_{\rm Malmquist}$~=~1.3\,$\pm$\,0.1 (see Section \ref{Implications}). The average primary mass of hot Jupiter hosts in the \citet{Ngo2016} sample is $\langle M_1 \rangle$~=~1.2\,\Msun, and such F-type stars have a  $\Delta F_{\rm close}$/$F_{\rm close}$ = 9\%\,$\pm$\,3\% higher close binary fraction than $M_1$~=~1.0\,\Msun\ primaries (Eqn.~\ref{Fclose}). Their hot Jupiter hosts are also metal-rich with average $\langle$[Fe/H]$\rangle$~=~+0.2, and the close binary fraction of solar-type stars decreases with metallicity \citep{Badenes2018,Moe2019,ElBadry2019a}. \citet{Moe2019} showed that the binary fraction within $a$~$<$~10~au decreases by $\Delta F_{\rm a<10au}$/$F_{\rm a<10au}$~=~$-$27\%\,$\pm$\,7\% across $-$0.1~$<$~[Fe/H]~$<$~0.2. The trend is slightly weaker for wider binaries with $a$~=~50~au \citep{ElBadry2019a}, i.e.,  $\Delta F_{\rm a=50au}$/$F_{\rm a=50au}$ = $-$20\%\,$\pm$\,5\% across the same metallicity interval. See Paper~II for a more detailed discussion of how binaries bias inferred planet trends with respect to metallicity.  Combining these various effects and their uncertainties, a magnitude-limited sample of $M_1$~=~1.2\,\Msun\ primaries with [Fe/H]~=~0.2 yields a close binary fraction of $F_{\rm field,close}$~=~40\%\,$\pm$\,6\% within $a$~$<$~50~au. We list our result in the top-left corner of Fig.~\ref{HJ_Fig}, nearly double the \citet{Ngo2016} estimate of $F_{\rm field,close}$ = 21\%\,$\pm$\,1\%.

At wider separations, 25\%\,$\pm$\,4\% of $M_1$~=~1.0\,\Msun\ primaries with [Fe/H]~=~$-$0.1 have companions across $a$~=~50\,-\,2,000~au (integral of blue curve in Fig.~\ref{solarperiod} across log\,$P$~=~5.0\,-\,7.5). About three-quarters of these wide companions, i.e., 19\%\,$\pm$\,3\% of all systems, are inner binaries (integral of red curve). Although 20\%\,$\pm$\,6\% of close companions to field solar-type primaries are WDs \citep[][see Appendix~\ref{Appendix}]{Moe2017,Murphy2018}, only  7\%\,$\pm$\,3\% of wide companions are WDs \citep{Holberg2016,Toonen2017,ElBadry2018}. The near-IR AO observations of hot Jupiter hosts are insensitive to faint companions, including old WDs, and so the wide binary fraction of $F_{\rm HJ,wide}$ = 47\%\,$\pm$\,7\% to hot Jupiter hosts reported by \citet{Ngo2016} likely excludes WD companions. After removing wide WD companions, we estimate that 18\%\,$\pm$\,3\% of $M_1$~=~1.0\,\Msun\ field primaries in a volume-limited sample have MS inner binary companions across $a$~=~50\,-\,2,000 au. This is consistent with the value of $F_{\rm wide,field}$ = 16\,$\pm$\,1\% adopted by \citet{Ngo2016} for the same set of assumptions. 

Nevertheless, the wide binary fraction for a comparable sample of field stars is larger. For example, the wide binary fraction of $M_1$~=~1.2\,\Msun\ primaries is $\Delta F_{\rm wide}$/$F_{\rm wide}$~=~8\%\,$\pm$\,3\% higher than that of $M_1$ = 1.0\,\Msun\ primaries (Appendix~\ref{Appendix}). Wide solar-type binaries are weighted toward small mass ratios (Section~\ref{HJq}), and so Malmquist bias in a magnitude-limited sample marginally increases the wide binary fraction by a factor of $f_{\rm Malmquist}$~=~1.06\,$\pm$\,0.03. The solar-type binary fraction beyond $a$~$>$~200~au is independent of metallicity, and a weak metallicity dependence emerges below $a$~$<$~200~au \citep{Moe2019,ElBadry2019a}. By multiplying these small correction factors, we find that 21\%\,$\pm$\,4\% of $M_1$~=~1.2\,\Msun\ primaries with [Fe/H]~=~0.2 in a magnitude-limited sample have inner binary MS companions across $a$~=~50\,-\,2,000~au. We estimate an additional 7\%\,$\pm$\,2\% of systems have tertiary companions across the same separation interval, bringing the total wide companion fraction to 28\%\,$\pm$\,5\%. We display these values in the middle of the top row in Fig.~\ref{HJ_Fig}. 

About 60\% of solar-type triples with tertiary companions across $a_{\rm out}$~=~50\,-\,2,000~au have inner binaries within $a_{\rm in}$~$<$~20~au \citep{Tokovinin2014}. Such triples with close inner binaries likely do not host hot Jupiters. Meanwhile, the majority of triples with $a_{\rm out}$~=~50\,-\,2,000~au and $a_{\rm in}$~$>$~20~au are capable of hosting hot Jupiters. We therefore adopt $F_{\rm field,wide}$ = 0.21\,$\pm$\,0.04 + 0.4(0.07\,$\pm$\,0.02) = 0.24\,$\pm$\,0.04. We estimate that 12\%\,$\pm$\,4\% of the stellar companions across $a$~=~50\,-\,2,000~au to hot Jupiter hosts detected by \citet{Ngo2016} are actually outer tertiaries in hierarchical triples. 

Before we can compare the wide companion fraction of hot Jupiter hosts to field stars, we must account for the most important selection bias as previously indicated, which we illustrate in the middle row of Fig.~\ref{HJ_Fig}. Assuming wide stellar companions do not influence planet formation, then we predict that hot Jupiter hosts will have the same ratio $F_{\rm HJ,wide}$/$F_{\rm HJ,single}$ = $F_{\rm field,wide}$/$F_{\rm field,single}$ = 0.24/0.39 = 0.62\,$\pm$\,0.15 of wide companions to effectively single stars as in the field, {\it not} the same wide binary fraction itself. Given the prior that only 4\% of hot Jupiter hosts have stellar companions within $a$~$<$ 50~au due to close binary suppression, then both the wide binary and single star fractions of hot Jupiter hosts must both increase in parallel so that all three columns in Fig.~\ref{HJ_Fig} add to 100\%. With the constraints $F_{\rm HJ,wide}$ + $F_{\rm HJ,single}$ = 96\% and $F_{\rm HJ,wide}$/$F_{\rm HJ,single}$ = 0.62, then we compute $F_{\rm HJ,wide}$ = 37\%\,$\pm$\,7\% and $F_{\rm HJ,single}$ = 59\%\,$\pm$\,7\%, which we display in the middle row of Fig.~\ref{HJ_Fig}. 

Another way of interpreting this selection effect is via Bayes' theorem for conditional probabilities:

\begin{equation}
p({\rm wide}\,|\,{\rm no~close}) = \frac{p({\rm no~close}\,|\,{\rm wide})p({\rm wide})}{p({\rm no~close})}
\label{Bayes}
\end{equation}

\noindent where $p$(wide) = 28\% is the total wide companion fraction of field stars as shown in the top row of Fig.~\ref{HJ_Fig}. Of these wide companions, 15\% are tertiaries in triples with close inner binaries, and so $p$(no~close\,$|$\,wide)~=~85\% is the fraction of wide companions that do not have close inner binaries that suppress planets. The numerator multiplies to our adopted value $F_{\rm field,wide}$ = $p$(no~close\,$|$\,wide)$p$(wide) = 24\%. Finally, we estimated a binary fraction within $a$~$<$~50~au of $F_{\rm field,close}$~=~40\%\,$\pm$\,6\% for field primaries most similar to hot Jupiter hosts, but only $F_{\rm HJ,close}$~=~4\% of hot Jupiter hosts actually have such close companions, a difference of 36\%. Hence, $p$(no~close) = 1$-$0.36 = 64\% is the probability that a field primary does not have a sufficiently close companion to suppress planet formation. Given a field primary that does not have a close binary that suppresses planets, then the expected wide binary fraction is $p$(wide\,$|$\,no~close) = 0.85$\times$0.28/0.64 = 37\%, identical to our estimate above. 

Assuming wide binaries do not influence planet formation, we predict that $F_{\rm HJ,wide}$ = 37\%\,$\pm$\,7\% of hot Jupiter hosts will have MS companions across $a$~=~50\,-\,2,000~au, which is only slightly smaller than the fraction 47\%\,$\pm$\,7\% measured by \citet{Ngo2016}.  These two values are fully consistent with each other, differing at the 1.0$\sigma$ level. Most important, the enhancement factor is $f_{\rm enhance}$ = (0.47\,$\pm$\,0.07)/(0.37\,$\pm$\,0.07) = 1.27\,$\pm$\,0.28, consistent with unity and well below the estimate of $f_{\rm enhance}$~=~2.9 reported in \citet{Ngo2016}. We conclude that wide stellar companions do not enhance the formation of hot Jupiters at a statistically significant level. Because close binaries suppress hot Jupiters, we actually expect the wide binary fraction of hot Jupiter hosts to be larger than the wide binary fraction of comparable field stars.

\subsection{Hot Jupiters versus Close Binaries}
\label{HJvBin}

\subsubsection{Hot Jupiters}

\begin{figure}
\centerline{
\includegraphics[trim=0.2cm 0.2cm 0.1cm 0.2cm, clip=true, width=3.3in]{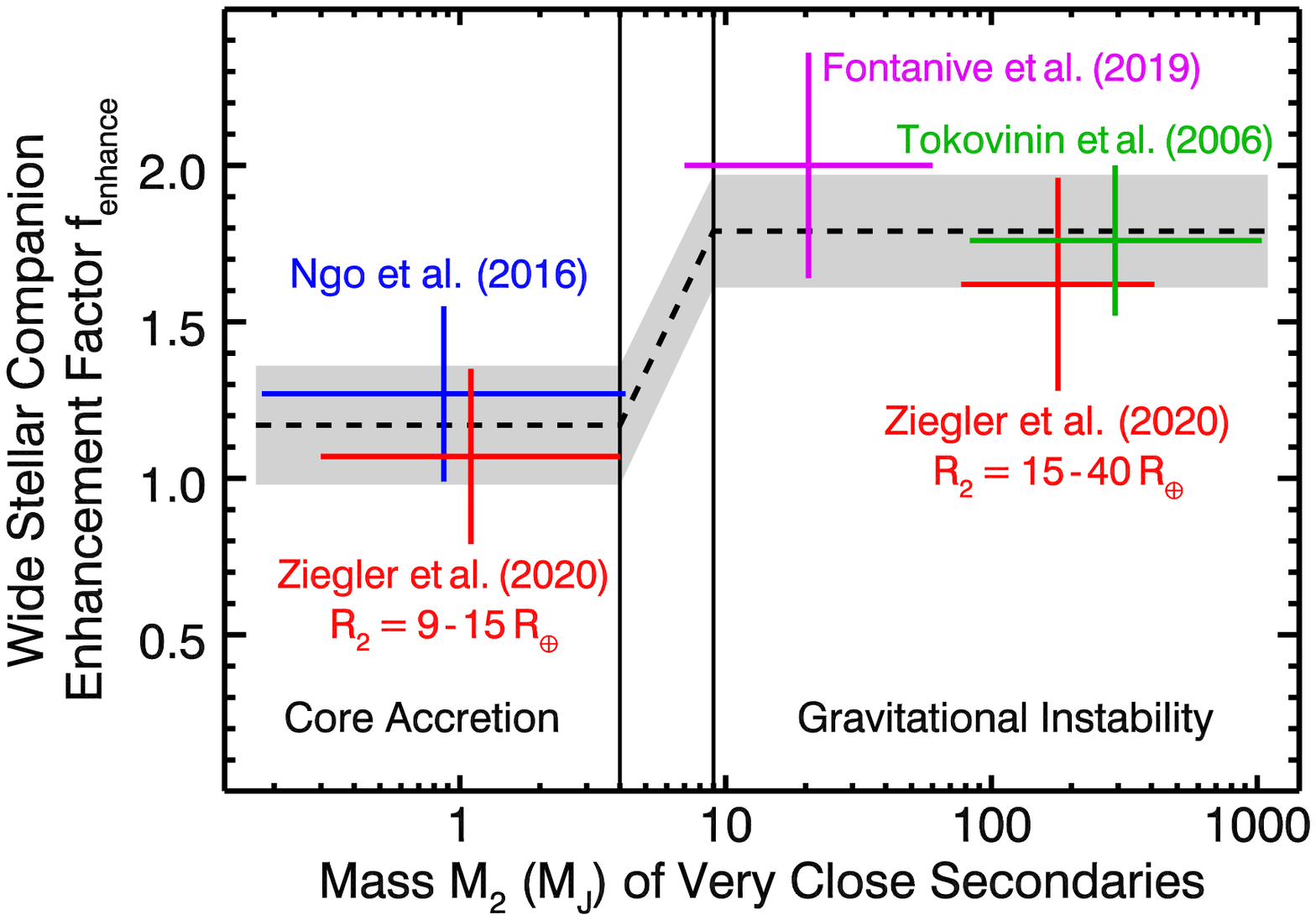}}
\caption{As a function of mass of very close secondaries, the enhancement factor $f_{\rm enhance}$ of wide stellar companions relative to expectations. We display our bias-corrected measurements based on the \citet{Ngo2016} sample of hot Jupiters (blue), close brown dwarf companions and very massive planets (magenta; \citealt{Fontanive2019}), very close binaries with $P$~$<$~7~days in the \citet{Tokovinin2006} sample (green), and TOIs in the \citet{Ziegler2019} survey (red) after separating probable giant planet candidates with $R_{\rm p}$~=~9\,-\,15\,\Rearth\ from EB false positives with M-dwarf secondaries. Hosts of genuine hot Jupiters with $M_{\rm p}$~=~0.2\,-\,4\,M$_{\rm J}$ that formed via core accretion are metal-rich ($\langle$[Fe/H]$\rangle$~=~0.23) and do not exhibit a statistically significant excess of wide stellar companions (weighted average of $\langle f_{\rm enhance} \rangle$~=~1.17\,$\pm$\,0.19; grey). Meanwhile, both very close binaries and brown dwarf companions with $M_2$~$\gtrsim$~9\,M$_{\rm J}$ formed via fragmentation of gravitationally unstable disks, are relatively metal-poor ($\langle$[Fe/H]$\rangle$~=~$-$0.15), and tend to reside in triples, i.e., exhibit an enhanced frequency $\langle f_{\rm enhance} \rangle$~=~1.79\,$\pm$\,0.18 of tertiary companions at the 5.2$\sigma$ significance level. }
\label{comp_enhance}
\end{figure}

The hot Jupiters in the \citet{Ngo2016} sample were originally discovered by the ground-based HAT, TrES,  and WASP transiting surveys \citep{Bakos2004,Alonso2004,Pollacco2006}. They were all spectroscopically confirmed to have dynamical masses below the deuterium-burning limit of 13\,M$_{\rm J}$, and a significant majority (70/77~=~91\%\,$\pm$\,3\%) are contained within the narrow interval of $M_{\rm p}$~=~0.18\,-\,4.2\,M$_{\rm J}$. Hot Jupiters with $M_{\rm p}$~=~0.2\,-\,4\,M$_{\rm J}$ likely formed via core accretion \citep[][see below]{Matsuo2007,Mordasini2008}. We show in Fig.~\ref{comp_enhance} that hosts to hot Jupiters do not exhibit a statistically significant excess of wide stellar companions, i.e., $f_{\rm enhance}$ = 1.27\,$\pm$\,0.28 as determined above.

Although hot Jupiters in general do not exhibit an excess of wide stellar companions, the subset with very tight orbits within $P_{\rm p}$~$<$~3~days may possibly be different (\citealt{Ziegler2018}; see below). \citet{Ngo2016} detected 23 wide companions with $a$~=~100\,-\,2,000~au and $M_{\rm wide}$~$>$~0.15\,\Msun\ in their AO survey of 77 hot Jupiter hosts. This parameter space is relatively complete and unlikely to include background optical doubles. Of their 77 hot Jupiters, only 31 have $P_{\rm p}$~$<$~3~days. Meanwhile, 12 of the 23 wide companions orbit hosts of very hot Jupiters with $P_{\rm p}$~$<$~3~days. The wide companion fractions are therefore $F_{\rm HJ,P<3,wide}$ = 12/31 = 0.39\,$\pm$\,0.09 and $F_{\rm HJ,P>3,wide}$ = 11/46 = 0.24\,$\pm$\,0.06 for hosts of hot Jupiters with $P_{\rm p}$~$<$~3~days and $P_{\rm p}$~=~3\,-\,10~days, respectively. Hence, hosts of very hot Jupiters with $P_{\rm p}$~$<$~3~days have 1.6\,$\pm$\,0.4 times the frequency of wide companions compared to hot Jupiters with $P_{\rm p}$~$=$~3\,-\,10~days, but the difference is only a marginal 1.4$\sigma$ effect. 

Repeating our same procedure as in Section~\ref{HJcomprate} and adopting the mass-ratio distribution of wide companions to $M_1$~= ~1.2\,\Msun\ primaries measured in Section~\ref{HJq}, we expect $p$(wide\,$|$\,no~close)~=~25\%\,$\pm$\,5\% of hot Jupiter hosts to have wide companions across the truncated parameter space of $M_{\rm wide}$~$>$~0.15\,\Msun\ and $a$~=~100\,-\,2,000~au.  The enhancement factor for very hot Jupiters with $P_{\rm p}$~$<$~3~days is $f_{\rm enhance}$ = (0.39\,$\pm$\,0.09)/(0.25\,$\pm$\,0.05) = 1.55\,$\pm$\,0.42. This is larger than the value of $f_{\rm enhance}$ = 1.27\,$\pm$\,0.28 measured for all hot Jupiters, but is still statistically insignificant due to the smaller sample size. Alternatively, we measure $f_{\rm enhance}$ = (0.24\,$\pm$\,0.06)/(0.25\,$\pm$\,0.05) = 0.96\,$\pm$\,0.31 for hosts of hot Jupiters with $P_{\rm p}$~=~3\,-\,10 days. The hosts of giant planets with $M_{\rm p}$~=~0.2\,-\,4\,M$_{\rm J}$ and $P_{\rm p}$~=~3\,-\,10~days, which compose the majority of hot Jupiters, clearly do not exhibit a large excess of wide stellar companions. 

\subsubsection{Very Close Binaries}

Conversely, the majority of very close binaries are in triples. After correcting for incompleteness in their volume-limited sample, \citet{Tokovinin2006} demonstrated that 96\% of binaries with $P$~$<$~3~days have tertiary companions. The triple star fraction steadily decreases with binary period, falling to 34\% for inner binaries with $P_{\rm in}$~=~12\,-\,30~days. It was originally argued that the excess of tertiary companions provided strong evidence that the inner binaries migrated via Kozai-Lidov cycles in misaligned triples coupled to tidal friction over Gyr timescales \citep{Kiseleva1998,Fabrycky2007,Naoz2014}. However, the very close binary fraction of T~Tauri stars matches the field value \citep{Melo2003,Prato2007,Kounkel2019}, and a significant majority of compact triples have small mutual inclinations $i$~$<$~40$^{\circ}$ that cannot induce Kozai-Lidov oscillations \citep{Borkovits2016}. \citet{Moe2018} and \citet{Tokovinin2019} therefore concluded that the majority of very close binaries formed via fragmentation, accretion, and migration within the disk during the embedded Class 0/I phase. Driving a gravitational instability in the disk near $a$~=~30\,-\,300~au followed by significant inward migration to $a$~$<$~0.1~au via energy dissipation in the disk requires substantial mass.  This surplus mass can also form additional (tertiary) stellar companions.  Hence, massive cores and disks are more likely to simultaneously form triple stars and harden the inner binary to $P_{\rm in}$~$<$~10~days, leading to the observed anti-correlation between inner binary period and triple star fraction.

In any case, very close binaries exhibit a statistically significant excess of wide tertiaries. Unlike hot Jupiters, which are suppressed by close binaries within $P$~$<$~300~yr ($a$~$<$~50~au), very close binaries exhibit a deficit of tertiary companions only within the dynamical stability limit of $P_{\rm out}$~$<$~1~yr \citep{Tokovinin2006,Tokovinin2014}. Based on the bias-corrected Fig.~13 in \citet{Tokovinin2006}, 67\%\,$\pm$\,5\% of very close binaries with $P_{\rm in}$~$<$~7~days have tertiary MS companions across intermediate periods of $P_{\rm out}$~=~1\,-\,10$^4$~yr. Meanwhile, a volume-limited sample of field solar-type stars has a multiplicity frequency of 0.42\,$\pm$\,0.04 companions per primary across this period interval (integral of blue curve in Fig.~\ref{solarperiod}), 85\%\,$\pm$\,5\% of which are MS (non-WD) companions.  The median primary mass of the \citet{Tokovinin2006} sample is $M_1$~=~1.1\,\Msun, and so we expect the companion fraction across intermediate periods to be  $\Delta F$/$F$~=~4\%\,$\pm$\,2\% larger. We adopt a companion fraction of $p_{\rm wide}$~=~0.37\,$\pm$\,0.05 across $P$~=~1\,-\,10$^4$~yr for a comparable population of field stars. Of these wide companions, 2\%\,$\pm$\,1\% are tertiaries to inner binaries with $P_{\rm in}$~=~7\,-\,100~days \citep{Tokovinin2014}, which suppress very close binaries within $P$~$<$~7~days due to dynamical instability. This results in $p$(no~close\,$|$\,wide) = 0.98\,$\pm$\,0.01. Similarly, 5\%\,$\pm$\,1\% of solar-type primaries have inner binary companions across $P$~=~7\,-\,100~days (integral of red curve in Fig.~\ref{solarperiod}), providing  $p$(no~close)~=~0.95\,$\pm$\,0.01. For systems that do not have close binaries with $P$~=~7\,-\,100~days, we expect $p$(wide\,$|$\,no~close)~= 0.37\,$\times$\,0.98\,/\,0.95 = 38\%\,$\pm$\,5\% of very close binaries with $P$~$<$~7~days to have wide tertiary companions across $P_{\rm out}$~=~1\,-\,10$^4$~yr according to Eqn.~\ref{Bayes}. The observed wide companion fraction to very close binaries is a factor of $f_{\rm enhance}$ = (0.67\,$\pm$\,0.05)/(0.38\,$\pm$\,0.05) = 1.76\,$\pm$\,0.24 times larger than expectations at the 4.1$\sigma$ confidence level. As shown in Fig.~\ref{comp_enhance}, very close binaries with $M_2$~=~0.08\,-\,1.0\,\Msun\ secondaries exhibit a statistically significant excess of wide stellar companions.

\subsubsection{Close Brown Dwarf Companions}

We next examine the properties of close brown dwarf companions to determine if they are more similar to close binaries or hot Jupiters. \citet{Fontanive2019} investigated the hosts of close brown dwarf companions and very massive planets with $M_2$~=~7\,-\,60\,M$_{\rm J}$. Their survey includes mostly brown dwarf desert companions,  bridging the gap between hosts of genuine hot Jupiters with $M_{\rm 2}$~=~0.2\,-\,4\,M$_{\rm J}$ \citep{Ngo2016} and stellar-mass binaries with $M_2$~$>$~80\,M$_{\rm J}$ \citep{Tokovinin2006}. After accounting for the sensitivity of their AO survey, \citet{Fontanive2019} reported a companion fraction of 70\%\,$\pm$\,10\% across $a$~=~50\,-\,2,000~au (see their Table~8). This is measurably larger than the wide companion fraction of 47\%\,$\pm$\,7\% to hot Jupiter hosts across the same separation interval \citep{Ngo2016}. Based on Table~2 in \citet{Fontanive2019}, the mean metallicity of their 38 targets is $\langle$[Fe/H]$\rangle$~=~$-$0.12, which matches the mean metallicity of field binaries  ( $\langle$[Fe/H]$\rangle$~=~$-$0.15; \citealt{Raghavan2010,Moe2019}) and is substantially lower than the mean metallicity of hot Jupiter hosts  ($\langle$[Fe/H]$\rangle$~=~0.23; \citealt{Santos2004,Fischer2005,Buchhave2018}). The average host mass in the \citet{Fontanive2019} sample is $\langle M_1 \rangle$ = 1.2\,\Msun. For a magnitude-limited sample of comparable field stars, we estimate a companion fraction of $p_{\rm wide}$~=~0.29\,$\pm$\,0.04 across $a$~=~50\,-\,2,000~au. 

\citet{Fontanive2019} measured a slight deficit of close stellar companions within $a$~$<$~10~au to their hosts of close brown dwarf companions and very massive planets. Nonetheless, the deficit is not as significant as that observed for hot Jupiter hosts. For example, of their 38 targets, \citet{Fontanive2019} identified one likely and three confirmed MS companions with $a$~=~20\,-\,40~au (4/38 = 11\%\,$\pm$\,5\%). Given our fit to planet suppression, only 1\% of hot Jupiter hosts have MS companions across the same separation interval. Analytic models of disk fragmentation, accretion, and migration indicate that it is improbable to form a compact triple ($a_{\rm out}$~$<$~10~au) in which the inner companion is less than half the mass of the outer tertiary \citep{Tokovinin2019}. Hosts of close brown dwarf companions therefore likely have wide stellar companions that formed via turbulent core fragmentation, which tend to have $a$~$>$~10~au. Thus we estimate  $p$(no~close\,$|$\,wide)~=~0.90\,$\pm$\,0.04 and $p$(no~close)~=~0.75\,$\pm$\,0.05 for brown dwarf hosts, which is between our measurements for hot Jupiter hosts and stellar binaries. The expected companion frequency across $a$~=~50\,-\,2,000~au is $p$(wide\,$|$\,no~close) = 0.29\,$\times$\,0.90\,/\,0.75 = 0.35\,$\pm$\,0.05 according to Eqn.~\ref{Bayes}. Hosts of close companions with $M_2$~=~7\,-\,60\,M$_{\rm J}$ exhibit a factor of $f_{\rm enhance}$ = (0.70\,$\pm$\,0.10)/(0.35\,$\pm$\,0.05) = 2.00 $\pm$ 0.36 times the frequency of wide MS companions compared to expectations at the 3.2$\sigma$ significance level.  This is consistent with the wide companion enhancement factor of $f_{\rm enhance}$ = 1.76\,$\pm$\,0.24 measured for very close binaries, but is larger than the factor of $f_{\rm enhance}$ = 1.27\,$\pm$\,0.28 determined for hot Jupiter hosts (see Fig.~\ref{comp_enhance}). We conclude that close sub-stellar companions with $M_2$~$=$~7\,-\,60\,M$_{\rm J}$ formed similarly to close binaries (see more below). 

\subsubsection{Giant Planet KOIs}

In their combined RV and AO survey of giant planet KOIs, \citet{Wang2015c} discovered a deficit of close binaries within $a$~$<$~20~au relative to the field, and they found a wide binary fraction beyond $a$~$>$~200~au that matched their control sample. This is consistent with our interpretation that close binaries suppress planet formation while wide binaries have no effect. Conversely, after correcting for incompleteness, \citet{Wang2015c} reported a binary fraction to giant planet hosts across intermediate separations ($a$~=~20\,-\,200~au) that is 2.8\,$\pm$\,0.7 times larger than their field value\footnote{ The blue Wang~et~al. data points for $S_{\rm bin}$ in our Fig.~\ref{suppressfactor} were derived by combining their three surveys of small planets \citep{Wang2014b}, giant planets \citep{Wang2015c}, and transiting multi-planet systems \citep{Wang2015d}. Only their survey of giant planet hosts alone \citep{Wang2015c} reveal a spurious $<$\,3$\sigma$ excess of companions across $a$~=~20\,-\,80~au.}. However, neither their RV monitoring nor AO imaging were particularly sensitive to companions with $a$~=~20\,-\,80~au (see their Fig.~2), and so \citet{Wang2015c} applied large correction factors for incompleteness across this interval. We conclude that they over-compensated for incompleteness, leading to the spuriously large companion fraction across this narrow range of intermediate separations.

Based on Robo-AO observations of KOIs, \citet{Law2014} found that giant planet candidates ($R_{\rm p}$~$>$~3.9\Rearth) with $P_{\rm p}$~$<$~15~days have 2\,-\,3 times the frequency of wide stellar companions than either giant planets with longer periods or smaller planets in general (see their Fig.~11). However, a non-negligible fraction of KOIs are EB false positives, and {\it Kepler} hot Jupiter candidates in particular are most likely to be diluted EBs \citep{Fressin2013,Morton2016}. Very close binaries, including short-period EBs, exhibit a real excess of tertiary companions (\citealt{Tokovinin2006}; see above). A sample of giant planet candidates contaminated by EB false positives will therefore be artificially biased toward an enhanced wide binary fraction. In fact, their Robo-AO sample of giant planet candidates within $P$~$<$~15~days includes KOI-1152.01 with $R_2$~=~20\,\Rearth\ and KOI-1845.02 with $R_2$~=~21\,\Rearth, both of which are known EB false positives that have wide stellar companions \citep{Ofir2013,Law2014}. In addition,  Kepler-13b has $P$~=~1.76 days, $R_2$~=~25\,\Rearth,  $M_2$~=~10\,M$_{\rm J}$, and a wide stellar companion \citep{Szabo2011b,Shporer2011,Esteves2015}. Although technically below the deuterium-burning limit, Kepler-13b is not a typical hot Jupiter. Kepler-13b instead belongs to the class of very massive planets and brown dwarfs investigated in \citet{Fontanive2019}, which show a real excess of wide companions and formed similarly to very close binaries (see above). Only eight of the {\it Kepler} giant planet candidates with $P_{\rm p}$~$<$~15~days in the \citet{Law2014} Robo-AO survey have detected wide binary companions. After removing the three objects listed above, the wide binary fraction of short-period giant planet hosts is fully consistent with the measurements for the other types of KOIs. 

Finally, utilizing a larger sample of KOIs with Robo-AO observations, \citet{Ziegler2018} found that hosts of very hot giant planets with $P_{\rm p}$~$<$~3~days and $R_{\rm p}$~$>$~3.9\,\Rearth\ are four times more likely to have wide companions compared to hosts of smaller planets within the same period range at the 2.4$\sigma$ significance level. Beyond $P_{\rm p}$~$>$~3~days, however, hosts of both large and small planets in their sample have the same wide binary fraction. Hot Jupiters with $P_{\rm p}$~=~3\,-\,10~days therefore do not exhibit an excess of wide companions, consistent with our previous conclusions. The prompt factor of four boost in their wide companion fraction as the period of the giant planet falls below $P_{\rm p}$~$<$~3~days is substantially larger than our estimated factor of 1.6\,$\pm$\,0.4 increase based on the \citet{Ngo2016} sample of dynamically confirmed hot Jupiters. Instead, the variation reported by \citet{Ziegler2018} more closely resembles the trend observed for very close binaries (\citealt{Tokovinin2006}; see above), suggesting some giant planet KOIs with $P_{\rm p}$~$<$~3~days are actually brown dwarf or late-M companions. \citet{Ziegler2018} removed unconfirmed planets, but their sample of ``confirmed'' planets still includes planets validated only through indirect transit light curve analysis \citep{Morton2016}, i.e., no spectroscopic RV monitoring to directly measure their dynamical masses.  Moreover, the very massive Kepler-13b with $M_2$~=~10\,$M_{\rm J}$ is one of the only 21 validated giant planets with $P$~$<$~3~days in the expanded Robo-AO sample. Removing Kepler-13b would already make the inferred excess of wide companions statistically insignificant. In short, the various AO observations of giant planet KOIs support our conclusions that hosts of genuine hot Jupiters with $M_{\rm p}$~=~0.2\,-\,4\,$M_{\rm J}$ do not exhibit an excess of wide stellar companions. 

\subsubsection{Giant Planet TOIs}

\begin{figure}
\centerline{
\includegraphics[trim=0.7cm 2.7cm 8.8cm 1.1cm, clip=true, width=3.3in]{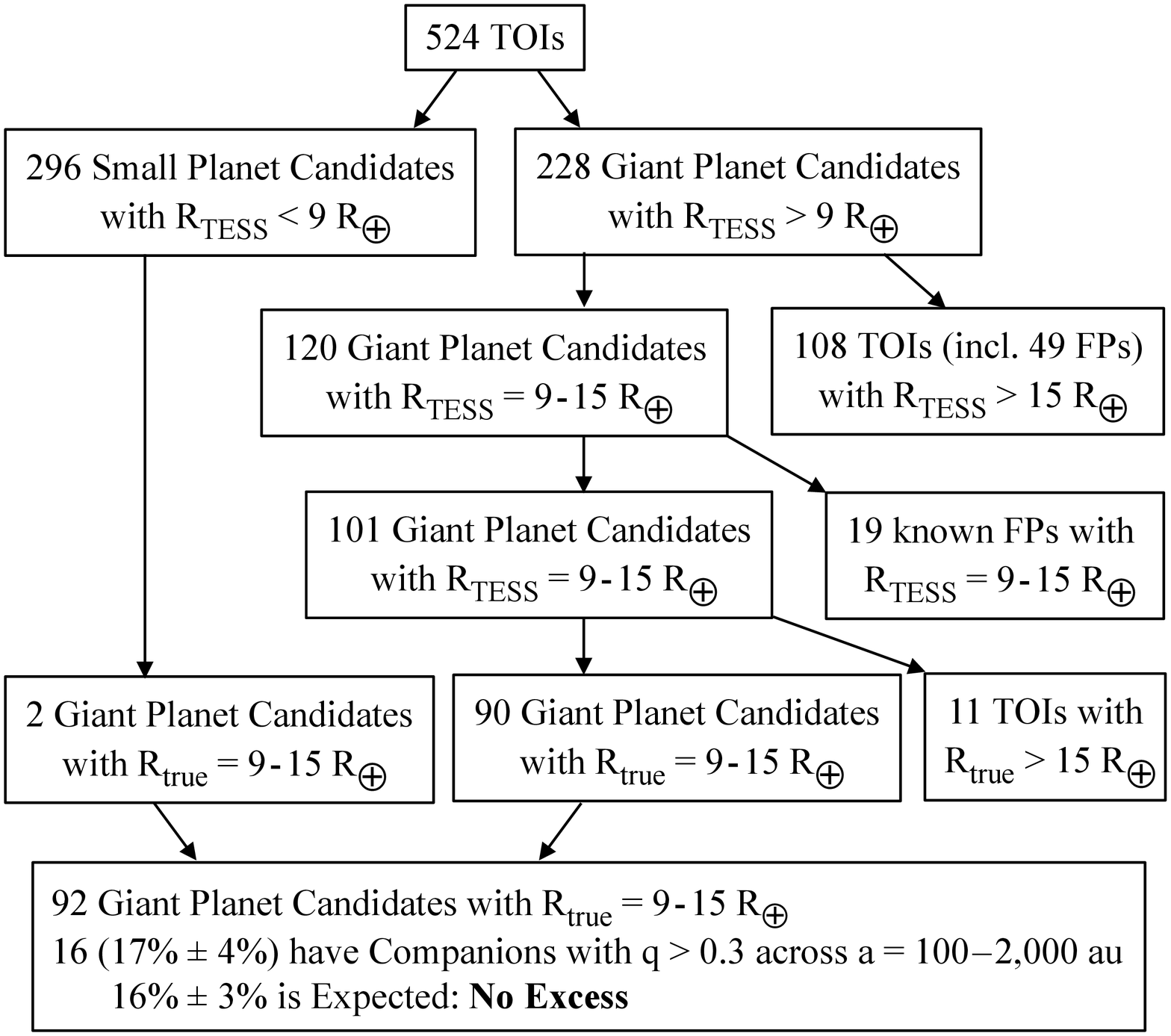}}
\caption{\citet{Ziegler2019} imaged 524 TOIs, 228 of which were originally classified as giant planet candidates with $R_{\rm p}$~$>$~9\,\Rearth. After excluding large transiting secondaries with $R_2$~$>$~15\,\Rearth\ (first step), removing known false positives (second step), and accounting for transit dilution by wide binaries (third step), there are 92 probable giant planet candidates with true radii $R_{\rm p}$~=~9\,-\,15\,\Rearth. Only 16 (17\%\,$\pm$\,4\%) have wide stellar companions with $q$~$>$~0.3 across $a$~=~100\,-\,2,000~au, fully consistent with the expected value of 16\%\,$\pm$\,3\%. The refined sample of probable {\it TESS} giant planet candidates confirms that wide binaries do not enhance the formation of hot Jupiters. }
\label{ZieglerFig}
\end{figure}

Most recently, \citet{Ziegler2019} utilized SOAR speckle imaging to resolve wide stellar companions to TOIs. They concluded that hosts of giant planet TOIs with $R_{\rm p}$~$>$~9\,\Rearth\ exhibit 3.5 times more wide companions compared to field stars.  TOIs, especially giant planet candidates, are heavily contaminated by EB false positives \citep{Sullivan2015}. In the following, we carefully distinguish probable {\it TESS} giant planets from likely EBs, and find that the former population does not exhibit an enhanced wide binary fraction while the latter population does. We illustrate our selection criteria as a flowchart in Fig.~\ref{ZieglerFig}. 

\citet{Ziegler2019} imaged 542 {\it TESS} planet candidates, including 524 official TOIs and 18 community objects of interest. They divided the 524 TOIs according to the planet radii reported by {\it TESS}, finding 244 small planet candidates with $R_{\rm p}$~$<$~9\,\Rearth, 199 large planet candidates with $R_{\rm p}$~$>$~9\,\Rearth, and the remaining 81 ambiguous planet candidates with radii not directly computed by the {\it TESS} pipeline. For the latter category, we rely on the supplementary planet radii measured by the EXOFAST code \citep{Eastman2013} and uploaded to the NASA Exoplanet Archive\footnote{https://exoplanetarchive.ipac.caltech.edu/}. We consider the additional 29 TOIs with EXOFAST radii $R_{\rm p}$~$>$~9\,\Rearth, bringing the total to 228 giant planet candidates. The fraction (228/524 = 44\%\,$\pm$\,2\%) of TOIs that are giant planet candidates is quite significant, further suggesting a large contamination by EB false positives. For comparison, only 4\% of the KOIs imaged by \citet{Kraus2016} have $R_{\rm p}$~$>$~9\,\Rearth\ (see bottom-left panel of Fig.~1 in \citealt{Ziegler2019}). 

Many of the giant planet TOIs have extremely large radii and uncertainties, e.g., $R_2$~=~30\,$\pm$\,10\,\Rearth. The goal of the {\it TESS} pipeline and TOI identification was to be complete toward all possible transiting exoplanets, not pure in minimizing contamination from EB false positives. Although a transiting $R_2$~=~30\,$\pm$\,10\,\Rearth\ object has a non-zero probability of being a giant planet, it is most likely an eclipsing M-dwarf companion, especially considering the Bayesian prior that there are substantially more EBs than transiting hot Jupiters. For reference, {\it Kepler} discovered 1,133 EBs with $P$~=~1\,-\,10 days \citep{Kirk2016} and 68 validated giant planets with $R_{\rm p}$~=~6\,-\,23\,\Rearth\ across the same period interval \citep{Morton2016}, only 8 of which are inflated hot Jupiters with $R_{\rm p}$~=~15\,-\,23\,\Rearth. We therefore remove the 108 giant planet candidates in the \citet{Ziegler2019} sample with $R_2$~$>$~15\,\Rearth. As of mid-September~2019, 49 of these 108 TOIs (45\%\,$\pm$\,5\%) with $R_2$~$>$~15\,\Rearth\ were already listed as false positives by the Exoplanet Follow-up Observing Program for TESS (ExoFOP-TESS\footnote{https://exofop.ipac.caltech.edu/tess/}), confirming our conclusion that a significant fraction of TOIs with $R_2$~$>$~15\Rearth\ are EBs. 

Meanwhile, of the remaining 228 - 108 = 120 TOIs with $R_{\rm p}$ = 9\,-\,15\,\Rearth, only 19 (16\%\,$\pm$\,3\%) have been listed as false positives by ExoFOP-TESS. This leaves 120-19 = 101 giant planet candidates with {\it TESS} and/or EXOFAST radii $R_{\rm p}$ = 9\,-\,15\,\Rearth. If their hosts have bright stellar companions that dilute the transits, however, then their true radii are systematically larger. For the TOIs with wide stellar companions detected by SOAR speckle imaging, \citet{Ziegler2019} listed the radius correction factors of the transiting objects assuming they orbit either the primaries or the secondaries. The Jovian planet occurrence rate increases with stellar mass \citep{Johnson2010,Bowler2010,Bonfils2013} and so we adopt the \citet{Ziegler2019}  radius correction factors assuming that giant planet TOIs in wide binaries orbit their respective primaries. Of our remaining 101 TOIs, 11 have sufficiently bright stellar companions such that their true radii are $R_2$~$>$~15\,\Rearth\ and therefore are most likely EBs. Conversely, there are only two TOIs with {\it TESS} radii $R_{\rm p}$~$<$~9\,\Rearth\ and bright companions such that their true radii are in the interval $R_{\rm p}$~$=$~9\,-\,15\,\Rearth. After excluding known false positives, accounting for transit dilution, and removing large transiting objects with $R_2$~$>$~15\,\Rearth\ that are most likely eclipsing M-dwarf companions, our culled final sample contains 92 probable giant planets with $R_{\rm p}$~=~9\,-\,15\,\Rearth, 74  (80\%\,$\pm$\,4\%) of which are hot Jupiters within $P_{\rm p}$~$<$~10~days.

Of the 92 {\it TESS} hosts of probable Jovian planets, only 16 have stellar companions detected across $a$~=~100\,-\,2,000~au with brightness contrasts $\Delta I$~$<$~5.1 mag \citep{Ziegler2019}. The SOAR speckle imaging survey was relatively complete in this parameter space, and most companions with $\Delta I$~$<$~5.1~mag within $a$~$<$~2,000~au are physically bound companions instead of background optical doubles (see Figs.~2\,-\,4 in \citealt{Ziegler2019}). As discussed in Section~\ref{HJq}, brightness contrasts of $\Delta I$~=~4.0 and 6.0 mag correspond to mass ratios of $q$ = 0.43 and 0.20, respectively. We estimate that $\Delta I$~=~5.1~mag maps to $q$~=~0.30. For our culled {\it TESS} sample, $F_{\rm Jovian,wide,q>0.3}$ = 16/92 = 17\%\,$\pm$\,4\% of the hosts to close Jovian planets have wide stellar companions with $a$~=~100\,-\,2,000~au and $q$~$>$~0.3. 

The {\it TESS} hosts of close Jovian planets are likely similar, on average, to the hot Jupiter hosts in the \citet{Ngo2016} sample. For a magnitude-limited sample of $M_1$~=~1.2\,\Msun\ primaries with [Fe/H]~=~0.2, $p$(wide)~=~21\%\,$\pm$\,4\% have MS (non-WD) companions across $a$~=~100\,-\,2,000~au. We adopt the same probabilities  $p$(no~close\,$|$\,wide)~=~0.85\,$\pm$\,0.05 and $p$(no~close)~=~0.64\,$\pm$\,0.06 due to suppression by close binaries as in Section~\ref{HJcomprate}. We therefore expect  $p$(wide\,$|$\,no~close)~= 0.21\,$\times$\,0.85\,/\,0.64 = 28\%\,$\pm$\,5\% of close Jovian hosts to have MS companions ($M_2$~$>$~0.08\,\Msun) across $a$~=~100\,-\,2,000~au according to Eqn.~\ref{Bayes}. We estimate that $F_{\rm q>0.3}$~=~57\%\,$\pm$\,5\% of wide stellar MS companions are above $q$~$>$~0.3 based on the measured mass-ratio distribution of wide companions to $M_1$~=~1.2\,\Msun\ primaries (Section~\ref{HJq}). We therefore expect $F_{\rm Jovian,wide,q>0.3}$ = (0.57\,$\pm$\,0.05)\,$\times$\,(0.28\,$\pm$\,0.05) = 16\%\,$\pm$\,3\% of close Jovian hosts to have MS companions above $q$~$>$~0.3 and across $a$~=~100\,-\,2,000~au. The wide companion enhancement factor of $f_{\rm enhance}$ = (0.17\,$\pm$\,0.04)/(0.16\,$\pm$\,0.03) = 1.07\,$\pm$\,0.28 is fully consistent with unity. As shown in Figs.~\ref{comp_enhance}\,-\,\ref{ZieglerFig}, our culled sample of probable close giant planets (mostly hot Jupiters) from the \citet{Ziegler2019} survey confirms our conclusion that wide binaries do not enhance the formation of close giant planets. 

Conversely, we examine the 108\,+\,19\,+\,11 = 138 false positive and/or TOIs with large transiting secondaries across $R_2$~=~15\,-\,40\,\Rearth. Most of these systems contain eclipsing late-M companions with  $M_2$~=~0.08\,-\,0.4\,\Msun. Of these 138 EBs, \citet{Ziegler2019} identified 29 tertiary companions across $a$~=~100\,-\,2,000~au with $\Delta I$~$<$~5.1~mag ($q$~=~$M_3$/$M_1$~$>$~0.3), providing $F_{\rm EB,wide,q>0.3}$ = 29/138 = 0.21\,$\pm$\,0.04. We adopt the same field wide binary fraction of $p$(wide)~=~0.21\,$\pm$\,0.04 as above, but probabilities $p$(no~close\,$|$\,wide)~=~0.93\,$\pm$\,0.03 and $p$(no~close)~=~0.85\,$\pm$\,0.04 that are halfway between our estimates for brown dwarf hosts in the \citet{Fontanive2019} sample and close GKM secondaries in the \citet{Tokovinin2006} survey. Eqn.~\ref{Bayes} provides $p$(wide\,$|$\,no~close) = 0.21\,$\times$\,0.93\,/\,0.85 =  0.23\,$\pm$\,0.04. Given $F_{\rm q>0.3}$~=~57\%\,$\pm$\,5\%, then we expect that $F_{\rm EB,wide,q>0.3}$ = (0.57\,$\pm$\,0.05)\,$\times$\,(0.23\,$\pm$\,0.04) = 13\%\,$\pm$\,2\% of EBs with late-M secondaries should have tertiary companions with $q$~=~$M_3$/$M_1$~$>$~0.3 across $a$~=~100\,-\,2,000~au. The observed wide companion fraction is $f_{\rm enhance}$ = (0.21\,$\pm$\,0.04)/(0.13\,$\pm$\,0.02) = 1.62\,$\pm$\,0.34 times larger than expectations at the 1.9$\sigma$ significance level. This is consistent with the wide companion enhancement factors of $f_{\rm enhance}$~=~2.00\,$\pm$\,0.36  and 1.76\,$\pm$\,0.24 we measured above for hosts of close brown dwarfs and GKM companions, respectively.
By separating the \citet{Ziegler2019} sample according to $R_2$, we find that hosts of probable hot Jupiters do not show an excess of wide companions, whereas probable EBs with late-M companions preferentially reside in triples (see Fig.~\ref{comp_enhance}). 

\subsection{Core Accretion versus Disk Fragmentation}
\label{Fragmentation}

Across wide separations of $a$~=~10\,-\,100~au, there is a continuous distribution of sub-stellar companion masses \citep{Wagner2019,Nielsen2019}. In particular, \citet{Wagner2019} found a break in the distribution near $M_2$~=~13\,M$_{\rm J}$, whereby wide brown dwarf companions above $M_2$~$>$~13\,M$_{\rm J}$ follow a uniform mass-ratio distribution while giant planets below $M_2$~$<$~13\,M$_{\rm J}$ are skewed toward smaller masses according to $N$~$\propto$~$M_2^{-1.3}$. They argued that wide brown dwarf companions above the break formed via turbulent core fragmentation and/or fragmentation of gravitationally unstable disks, whereas giant planets below the break formed via core accretion. The break at $M_2$~=~13~M$_{\rm J}$ happens to coincide with the deuterium-burning limit, the conventional division separating  brown dwarfs and planets.

At closer separations of $a$~$<$~0.5~au, however, the companion mass distribution is neither monotonic nor continuous.  First, there is the brown dwarf desert, but the desert is not completely dry \citep{Grether2006}. \citet{Csizmadia2015} showed that 0.2\% of solar-type primaries have brown dwarf companions within $P$~$<$~10~days, which is six times lower than the hot Jupiter occurrence rate and 15 times lower than the frequency of stellar companions within $P$~$<$~10~days. Second, there is a near complete absence of close companions with $M_2$~=~5\,-\,7\,M$_{\rm J}$ \citep{Hebrard2011,Schlaufman2018}. Hosts of genuine hot Jupiters with $M_{\rm p}$~=~0.2\,-\,5\,M$_{\rm J}$ below the mass gap are metal-rich with $\langle$[Fe/H]$\rangle$~=~0.23, consistent with expectations from core accretion theory \citep{Santos2004,Fischer2005,Buchhave2018}. Meanwhile, both close binaries and hosts of close brown dwarfs have mean metallicities of $\langle$[Fe/H]$\rangle$ = $-$0.15, consistent with fragmentation models of gravitationally unstable disks \citep{Raghavan2010,Ma2014,Moe2019}. As shown above, the mean metallicity of the 38 hosts of close brown dwarfs and very massive planets with $M_2$~=~7\,-\,60\,M$_{\rm J}$ in the \citet{Fontanive2019} sample is $\langle$[Fe/H]$\rangle$ = $-$0.12. Even their subset of 22 companions with $M_2$~=~7\,-\,13\,M$_{\rm J}$, which are technically below the deuterium-burning limit, has a mean metallicity of $\langle$[Fe/H]$\rangle$ = $-$0.06 that is discrepant with typical hot Jupiters and more consistent with very close binaries. Based on a two-dimensional clustering algorithm,  \citet{Schlaufman2018} identified a clear deficit across $M_2$~=~4\,-\,9\,M$_{\rm J}$ in which hot Jupiters below $M_2$~$<$~4\,$M_{\rm J}$ favored metal-rich hosts whereas sub-stellar companions above $M_2$~$>$~9\,M$_{\rm J}$ orbited metal-poor stars. Unlike directly imaged planets, where mapping between brightness and mass is model dependent, the masses of close sub-stellar companions are dynamically measured, and so the short-period mass gap is robust. We too advocate a bifurcation at $M_2$~=~6\,M$_{\rm J}$ for close sub-stellar companions within $a$~$<$~0.5~au, which cleanly separates genuine hot Jupiters that formed via core accretion from binaries, brown dwarfs, and very massive planets that formed via gravitational disk instability. We delineate the mass deficit across $M_2$~=~4\,-\,9\,M$_{\rm J}$ in Fig.~\ref{comp_enhance}. 

We combine our two measurements of the wide companion enhancement factors for the \citet{Ngo2016} sample of hot Jupiters and the \citet{Ziegler2019} subset of probable giant planet candidates with $R_{\rm p}$~=~ 9\,-\,15\,\Rearth. The weighted average is $f_{\rm enhance}$~=~1.17\,$\pm$\,0.19, fully consistent with unity (see grey region in Fig.~\ref{comp_enhance}). Hot Jupiters formed via core accretion relatively independently of wide stellar companions, and they favor metal-rich hosts. Meanwhile, our three measurements for companions above the mass gap yield a weighted average of $f_{\rm enhance}$ = 1.79\,$\pm$\,0.18, well above unity at the 5.2$\sigma$ confidence level. Close binaries and sub-stellar companions with $M_2$~$>$~7\,M$_{\rm J}$ formed via disk fragmentation, are relatively metal-poor, and exhibit a statistically significant excess of wide tertiary companions. 

The average formation and migration timescales of hot Jupiters and very close binaries may also be different. The very close binary fraction ($P$~$<$~10~days) of T~Tauri stars matches the field value, demonstrating the majority migrated during the embedded Class 0/I phase \citep{Melo2003,Prato2007,Moe2018,Kounkel2019}. Only 20\%\,$\pm$\,10\% of very close binaries migrated during the MS phase via Kozai-Lidov cycles in misaligned triples coupled to tidal friction \citep{Moe2018}. The dominant migration mechanism of hot Jupiters is continuously debated (see \citealt{Dawson2018} for a review). There is a least one hot Jupiter, V830~Tau~b, with mass $M_{\rm p}$~=~0.8\,M$_{\rm J}$ and period $P_{\rm p}$~=~4.9~day that orbits a T~Tauri star, suggesting it migrated via planet-disk interactions \citep{Donati2016}. Clearly, a much larger sample is needed to compare the occurrence rates of hot Jupiters orbiting T~Tauri stars versus field stars, thereby revealing their dominant migration mechanism. Moreover, it is imperative to compare similar types of hot Jupiters. For example, CI~Tau~b is another sub-stellar companion with $P$~=~9.0~days orbiting a T~Tauri star \citep{JohnsKrull2016,Flagg2019}. At $M_2$~=~11.6\,M$_{\rm J}$, the companion is just below the deuterium-burning limit, but well above the mass gap across $M_2$~=~5\,-\,7\,M$_{\rm J}$. We argue that CI~Tau~b is a sub-stellar companion that formed via gravitational disk instability and migrated during the Class 0/I phase, just like the majority of very close binaries. It should not be conflated with typical hot Jupiters with $M_{\rm p}$~=~0.2\,-\,4\,M$_{\rm J}$,  which formed via core accretion and may potentially migrate on different timescales. 

\section{Summary}
\label{Summary}

We list our main results and conclusions as follows:

{\it Planet suppression by close binaries.} We synthesized several RV and imaging surveys of planet hosts \citep{Knutson2014,Wang2014b, Wang2015c, Wang2015d, Ngo2016, Kraus2016, Matson2018,Ziegler2019}. We determined that planet suppression is a gradual function of binary separation (not a step function as previously modeled), such that binaries within $a$~$<$~1~au fully suppress S-type planets, binaries with $a$~=~10~au host close planets at 15$_{-12}^{+17}$\% the occurrence rate of single stars, and wide binaries beyond $a$~$>$~200$_{-120}^{+200}$~au have no effect on close planet formation (Section~\ref{Suppression} and Fig.~\ref{suppressfactor}). Both solar-type and M-dwarf binaries and both hosts of small and large planets all exhibit similar planet suppression factors with respect to binary orbital separation.  Unlike their stellar-mass counterparts, brown dwarf companions within $a$~$<$~5~au do not strongly suppress S-type planets \citep{Triaud2017,Rey2018}. 
There are currently only two known S-type planets in which the measured orbital periods of the stellar-mass companions are within $P$~$<$~10~yr, both of which happen to be extremely eccentric warm Jupiters (Kepler-420b, \citealt{Santerne2014}; Kepler-693b, \citealt{Masuda2017}). 

{\it Fraction of stars without close planets}. In magnitude-limited surveys, 43\%\,$\pm$\,7\% of solar-type primaries do not host close planets due to suppression by close binaries (Section~\ref{Implications} and Figs.~\ref{closebin}\,-\,\ref{solarperiod}). This is more than double the value of 19\% reported by \citet{Kraus2016} for three reasons: (1) solar-type stars have more late-M and WD companions than those identified in the \citet{Raghavan2010} survey \citep{Moe2017}, (2) RV monitoring of planet hosts reveals that the suppression factor tapers to $S_{\rm bin}$~$<$~15\% within $a$~$<$~10~au \citep{Knutson2014,Wang2014b,Ngo2016}, and (3) Malmquist bias in a magnitude-limited survey increases the close binary fraction by a factor of $f_{\rm Malmquist}$~=~1.3\,$\pm$\,0.1 compared to a volume-limited sample. 

{\it RV versus transit methods.} By removing spectroscopic binaries from their samples, RV searches for giant planets boost their detection rates by a factor of 1.8\,$\pm$\,0.2 compared to transiting surveys (Sections~\ref{HJRate}-\ref{GiantPlanet} and Figs.~\ref{ratio}-\ref{HJ}). This selection bias fully accounts for the  observed factor of two discrepancy in hot Jupiter occurrence rates measured from RV surveys (1.1\%; \citealt{Marcy2005,Mayor2011,Wright2012}) versus {\it Kepler}  and {\it TESS} (0.5\%; \citealt{Howard2012,Fressin2013,Santerne2016,Zhou2019}). The occurrence rates of long-period giant planets inferred from RV surveys ($M_{\rm p}$~$>$~0.1\,M$_{\rm J}$) and transit methods ($R_{\rm p}$~$>$~5\,\Rearth) appear to be consistent with each other \citep{Fernandes2019}. However, the {\it Kepler} sample of long-period giant planets is contaminated by EB false positives \citep{Santerne2016} and ``super-puffs'' that are actually low-mass non-Jovian planets \citep{Masuda2014,Lopez2014,Lee2016}. The occurrence rate of Neptunes ($R_{\rm p}$~=~2\,-\,6\,\Rearth) within $P_{\rm p}$~$<$~50~days is 16\%\,$\pm$\,2\% based on {\it Kepler} observations \citep{Howard2012,Fressin2013,Mulders2015a}, but RV surveys show that 28\%\,$\pm$\,5\% of solar-type stars have Neptunes ($M_{\rm p}$\,sin\,$i$ = 3\,-\,30\,\Mearth) within the same period range \citep{Mayor2011}. This discrepancy provides further confirmation that the large binary fraction in the {\it Kepler} survey diminishes planet occurrence rates by a factor of two. 

{\it Frequency $\eta_{\oplus}$ of habitable earth-sized planets orbiting solar-type stars}. After accounting for both planet suppression by close binaries and transit dilution by wide binaries, the occurrence rate of small planets orbiting single G-dwarfs is a factor of 2.1\,$\pm$\,0.3 times larger than the rate inferred from all {\it Kepler} G-dwarfs (Section~\ref{SmallPlanet} and Fig.~\ref{smallplanetFig}). We therefore also expect $\eta_{\oplus}$ for single G-dwarfs to be a factor of 2.1\,$\pm$\,0.3 times larger than the frequency estimated for all G-dwarfs by previous studies. The strong dependence of $\eta_{\oplus}$ on binary status has significant implications for expected yields and target prioritization of direct planet imaging surveys. The binary star $\alpha$~Centauri AB likely does not host an earth-sized planet in the habitable zone, while the single star $\tau$\,Ceti probably does (see also Paper II). 

{\it Trends with host mass.} According to RV surveys, the giant planet occurrence rate increases monotonically with host mass
\citep{Johnson2010}, but transit surveys show that both the hot Jupiter and overall giant planet occurrence rates decrease across $M_1$~=~0.8\,-\,2.3\,\Msun \citep{Fressin2013,Zhou2019}. We resolve this discrepancy  by accounting for the larger close binary fraction of AF dwarfs within the {\it Kepler} sample (Section~\ref{GiantPlanet} and Fig.~\ref{HJ}). The occurrence rate of small planets within $P_{\rm p}$~$<$~50~days decreases by a factor of 3.0\,-\,3.5 between {\it Kepler} M-dwarf and F-dwarf hosts \citep{Mulders2015a,Mulders2015b}. Binaries account for half (but not all) of this observed trend, i.e., single M-dwarfs host small, close planets at 1.9\,$\pm$\,0.4 times the occurrence rate of single F-dwarfs (Section~\ref{SmallPlanet} and Fig.~\ref{smallplanetFig}).

{\it Mass-ratio distribution of wide companions to hot Jupiter hosts.} Although the overall mass-ratio distribution of solar-type type binaries is uniform with a small excess of twins, wide solar-type binaries are weighted toward small mass ratios \citep{Moe2017}, consistent with the observed distribution of wide companions to hot Jupiter hosts (Section~\ref{HJq} and Fig.~\ref{q_HJ}). Unlike small planets, hot Jupiters with deep, frequent transits are relatively immune to photometric dilution by wide binaries. 

{\it No excess of wide stellar companions to hot Jupiter hosts.} The close binary fraction of hot Jupiter hosts is smaller compared to field stars, and so both the wide and single star fractions of hot Jupiter hosts must increase in parallel above the field values in order to compensate (Section~\ref{HJcomprate} and Fig.~\ref{HJ_Fig}). We expect the wide binary fraction of hot Jupiter hosts to be 37\%\,$\pm$\,7\% across $a$~=~50\,-\,2,000~au, which is fully consistent (1.0$\sigma$) with the completeness-corrected value of 47\%\,$\pm$\,7\% reported by \citet{Ngo2016}. Meanwhile, very close binaries and hosts of brown dwarf companions exhibit a real excess of wide tertiary companions \citep{Tokovinin2006,Fontanive2019}. Samples of giant planet KOIs and TOIs \citep{Law2014,Wang2015c,Ziegler2018,Ziegler2019} are contaminated by EB false positives, which leads to the spurious enhancement of wide stellar companions (Section~\ref{HJvBin} and Figs.~\ref{comp_enhance}\,-\,\ref{ZieglerFig}).

{\it Formation of hot Jupiters versus close binaries and brown dwarf companions.} Within $a$~$<$~0.5~au, there is a gap in companion masses across $M_2$~=~5\,-\,7\,M$_{\rm J}$ \citep{Hebrard2011,Schlaufman2018}, illustrating two different formation mechanisms (Section~\ref{Fragmentation} and Fig.~\ref{comp_enhance}). Typical hot Jupiters, which have $M_{\rm p}$~=~0.2\,-\,4\,M$_{\rm J}$ below the mass gap, probably formed via core accretion, have metal-rich hosts ($\langle$[Fe/H]$\rangle$~=~0.23), and do not exhibit a statistically significant excess of wide stellar companions. Meanwhile, very close binaries, brown dwarf companions, and massive planets with $M_2$~$>$~7\,\Msun\ above the mass gap likely formed via fragmentation of gravitationally unstable disks, are relatively metal-poor ($\langle$[Fe/H]$\rangle$~=~$-$0.15), and exhibit a 5.2$\sigma$ excess of wide stellar companions.

MM and KMK acknowledge financial support from NASA grant ATP-170070\,(80NSSC18K0726) and Heising-Simons Foundation grant 2018-1034. We are thankful for illuminating discussions with Trent~Dupuy, Jim~Fuller, Heather~Knutson, Adam~Kraus, Eve~Lee, Michael~Meyer, Gijs~Mulders, Erik~Petigura, Andrei~Tokovinin, Ji~Wang, and Andrew~Youdin.  We also thank the anonymous referee who provided suggestions that improved the quality and credibility of the manuscript.

{\it Data Availability:}  All data analysis presented in this study utilized previous observations that were published in the cited papers. There is no new data to report.

\bibliographystyle{mnras}       
\bibliography{moe_biblio}

\begin{thebibliography}{}
\makeatletter
\relax
\def\mn@urlcharsother{\let\do\@makeother \do\$\do\&\do\#\do\^\do\_\do\%\do\~}
\def\mn@doi{\begingroup\mn@urlcharsother \@ifnextchar [ {\mn@doi@}
  {\mn@doi@[]}}
\def\mn@doi@[#1]#2{\def\@tempa{#1}\ifx\@tempa\@empty \href
  {http://dx.doi.org/#2} {doi:#2}\else \href {http://dx.doi.org/#2} {#1}\fi
  \endgroup}
\def\mn@eprint#1#2{\mn@eprint@#1:#2::\@nil}
\def\mn@eprint@arXiv#1{\href {http://arxiv.org/abs/#1} {{\tt arXiv:#1}}}
\def\mn@eprint@dblp#1{\href {http://dblp.uni-trier.de/rec/bibtex/#1.xml}
  {dblp:#1}}
\def\mn@eprint@#1:#2:#3:#4\@nil{\def\@tempa {#1}\def\@tempb {#2}\def\@tempc
  {#3}\ifx \@tempc \@empty \let \@tempc \@tempb \let \@tempb \@tempa \fi \ifx
  \@tempb \@empty \def\@tempb {arXiv}\fi \@ifundefined
  {mn@eprint@\@tempb}{\@tempb:\@tempc}{\expandafter \expandafter \csname
  mn@eprint@\@tempb\endcsname \expandafter{\@tempc}}}

\bibitem[\protect\citeauthoryear{{Abt}, {Gomez}  \& {Levy}}{{Abt}
  et~al.}{1990}]{Abt1990}
{Abt} H.~A.,  {Gomez} A.~E.,   {Levy} S.~G.,  1990, \mn@doi [\apjs]
  {10.1086/191508}, \href {http://adsabs.harvard.edu/abs/1990ApJS...74..551A}
  {74, 551}

\bibitem[\protect\citeauthoryear{{Alonso} et~al.,}{{Alonso}
  et~al.}{2004}]{Alonso2004}
{Alonso} R.,  et~al., 2004, \mn@doi [\apjl] {10.1086/425256}, \href
  {https://ui.adsabs.harvard.edu/abs/2004ApJ...613L.153A} {613, L153}

\bibitem[\protect\citeauthoryear{{Artymowicz} \& {Lubow}}{{Artymowicz} \&
  {Lubow}}{1994}]{Artymowicz1994}
{Artymowicz} P.,  {Lubow} S.~H.,  1994, \mn@doi [\apj] {10.1086/173679}, \href
  {https://ui.adsabs.harvard.edu/abs/1994ApJ...421..651A} {421, 651}

\bibitem[\protect\citeauthoryear{{Badenes} et~al.,}{{Badenes}
  et~al.}{2018}]{Badenes2018}
{Badenes} C.,  et~al., 2018, \mn@doi [\apj] {10.3847/1538-4357/aaa765}, \href
  {https://ui.adsabs.harvard.edu/abs/2018ApJ...854..147B} {854, 147}

\bibitem[\protect\citeauthoryear{{Bakos}, {Noyes}, {Kov{\'a}cs}, {Stanek},
  {Sasselov}  \& {Domsa}}{{Bakos} et~al.}{2004}]{Bakos2004}
{Bakos} G.,  {Noyes} R.~W.,  {Kov{\'a}cs} G.,  {Stanek} K.~Z.,  {Sasselov}
  D.~D.,   {Domsa} I.,  2004, \mn@doi [\pasp] {10.1086/382735}, \href
  {https://ui.adsabs.harvard.edu/abs/2004PASP..116..266B} {116, 266}

\bibitem[\protect\citeauthoryear{{Basri} \& {Reiners}}{{Basri} \&
  {Reiners}}{2006}]{Basri2006}
{Basri} G.,  {Reiners} A.,  2006, \mn@doi [\aj] {10.1086/505198}, \href
  {https://ui.adsabs.harvard.edu/abs/2006AJ....132..663B} {132, 663}

\bibitem[\protect\citeauthoryear{{Batalha} et~al.,}{{Batalha}
  et~al.}{2010}]{Batalha2010}
{Batalha} N.~M.,  et~al., 2010, \mn@doi [\apjl] {10.1088/2041-8205/713/2/L109},
  \href {https://ui.adsabs.harvard.edu/abs/2010ApJ...713L.109B} {713, L109}

\bibitem[\protect\citeauthoryear{{Bate} \& {Bonnell}}{{Bate} \&
  {Bonnell}}{1997}]{Bate1997}
{Bate} M.~R.,  {Bonnell} I.~A.,  1997, \mn@doi [\mnras]
  {10.1093/mnras/285.1.33}, \href
  {https://ui.adsabs.harvard.edu/abs/1997MNRAS.285...33B} {285, 33}

\bibitem[\protect\citeauthoryear{{Bergfors} et~al.,}{{Bergfors}
  et~al.}{2010}]{Bergfors2010}
{Bergfors} C.,  et~al., 2010, \mn@doi [\aap] {10.1051/0004-6361/201014114},
  \href {https://ui.adsabs.harvard.edu/abs/2010A&A...520A..54B} {520, A54}

\bibitem[\protect\citeauthoryear{{Bonfils} et~al.,}{{Bonfils}
  et~al.}{2013}]{Bonfils2013}
{Bonfils} X.,  et~al., 2013, \mn@doi [\aap] {10.1051/0004-6361/201014704},
  \href {https://ui.adsabs.harvard.edu/abs/2013A&A...549A.109B} {549, A109}

\bibitem[\protect\citeauthoryear{{Borkovits}, {Hajdu}, {Sztakovics},
  {Rappaport}, {Levine}, {B{\'\i}r{\'o}}  \& {Klagyivik}}{{Borkovits}
  et~al.}{2016}]{Borkovits2016}
{Borkovits} T.,  {Hajdu} T.,  {Sztakovics} J.,  {Rappaport} S.,  {Levine} A.,
  {B{\'\i}r{\'o}} I.~B.,   {Klagyivik} P.,  2016, \mn@doi [\mnras]
  {10.1093/mnras/stv2530}, \href
  {https://ui.adsabs.harvard.edu/abs/2016MNRAS.455.4136B} {455, 4136}

\bibitem[\protect\citeauthoryear{{Bouma}, {Masuda}  \& {Winn}}{{Bouma}
  et~al.}{2018}]{Bouma2018}
{Bouma} L.~G.,  {Masuda} K.,   {Winn} J.~N.,  2018, \mn@doi [\aj]
  {10.3847/1538-3881/aabfb8}, \href
  {https://ui.adsabs.harvard.edu/abs/2018AJ....155..244B} {155, 244}

\bibitem[\protect\citeauthoryear{{Bowler} et~al.,}{{Bowler}
  et~al.}{2010}]{Bowler2010}
{Bowler} B.~P.,  et~al., 2010, \mn@doi [\apj] {10.1088/0004-637X/709/1/396},
  \href {https://ui.adsabs.harvard.edu/abs/2010ApJ...709..396B} {709, 396}

\bibitem[\protect\citeauthoryear{{Branch}}{{Branch}}{1976}]{Branch1976}
{Branch} D.,  1976, \mn@doi [\apj] {10.1086/154841}, \href
  {https://ui.adsabs.harvard.edu/abs/1976ApJ...210..392B} {210, 392}

\bibitem[\protect\citeauthoryear{{Bromley} \& {Kenyon}}{{Bromley} \&
  {Kenyon}}{2015}]{Bromley2015}
{Bromley} B.~C.,  {Kenyon} S.~J.,  2015, \mn@doi [\apj]
  {10.1088/0004-637X/806/1/98}, \href
  {https://ui.adsabs.harvard.edu/abs/2015ApJ...806...98B} {806, 98}

\bibitem[\protect\citeauthoryear{{Bryan} et~al.,}{{Bryan}
  et~al.}{2016}]{Bryan2016}
{Bryan} M.~L.,  et~al., 2016, \mn@doi [\apj] {10.3847/0004-637X/821/2/89},
  \href {https://ui.adsabs.harvard.edu/abs/2016ApJ...821...89B} {821, 89}

\bibitem[\protect\citeauthoryear{{Buchhave}, {Bitsch}, {Johansen}, {Latham},
  {Bizzarro}, {Bieryla}  \& {Kipping}}{{Buchhave} et~al.}{2018}]{Buchhave2018}
{Buchhave} L.~A.,  {Bitsch} B.,  {Johansen} A.,  {Latham} D.~W.,  {Bizzarro}
  M.,  {Bieryla} A.,   {Kipping} D.~M.,  2018, \mn@doi [\apj]
  {10.3847/1538-4357/aaafca}, \href
  {https://ui.adsabs.harvard.edu/abs/2018ApJ...856...37B} {856, 37}

\bibitem[\protect\citeauthoryear{{Burgasser}, {Kirkpatrick}, {Reid}, {Brown},
  {Miskey}  \& {Gizis}}{{Burgasser} et~al.}{2003}]{Burgasser2003}
{Burgasser} A.~J.,  {Kirkpatrick} J.~D.,  {Reid} I.~N.,  {Brown} M.~E.,
  {Miskey} C.~L.,   {Gizis} J.~E.,  2003, \mn@doi [\apj] {10.1086/346263},
  \href {https://ui.adsabs.harvard.edu/abs/2003ApJ...586..512B} {586, 512}

\bibitem[\protect\citeauthoryear{{Campante} et~al.,}{{Campante}
  et~al.}{2015}]{Campante2015}
{Campante} T.~L.,  et~al., 2015, \mn@doi [\apj] {10.1088/0004-637X/799/2/170},
  \href {https://ui.adsabs.harvard.edu/abs/2015ApJ...799..170C} {799, 170}

\bibitem[\protect\citeauthoryear{{Carrera}, {Johansen}  \& {Davies}}{{Carrera}
  et~al.}{2015}]{Carrera2015}
{Carrera} D.,  {Johansen} A.,   {Davies} M.~B.,  2015, \mn@doi [\aap]
  {10.1051/0004-6361/201425120}, \href
  {https://ui.adsabs.harvard.edu/abs/2015A&A...579A..43C} {579, A43}

\bibitem[\protect\citeauthoryear{{Cheetham}, {Kraus}, {Ireland}, {Cieza},
  {Rizzuto}  \& {Tuthill}}{{Cheetham} et~al.}{2015}]{Cheetham2015}
{Cheetham} A.~C.,  {Kraus} A.~L.,  {Ireland} M.~J.,  {Cieza} L.,  {Rizzuto}
  A.~C.,   {Tuthill} P.~G.,  2015, \mn@doi [\apj] {10.1088/0004-637X/813/2/83},
  \href {https://ui.adsabs.harvard.edu/abs/2015ApJ...813...83C} {813, 83}

\bibitem[\protect\citeauthoryear{{Chini}, {Fuhrmann}, {Barr}, {Pozo},
  {Westhues}  \& {Hodapp}}{{Chini} et~al.}{2014}]{Chini2014}
{Chini} R.,  {Fuhrmann} K.,  {Barr} A.,  {Pozo} F.,  {Westhues} C.,   {Hodapp}
  K.,  2014, \mn@doi [\mnras] {10.1093/mnras/stt1953}, \href
  {https://ui.adsabs.harvard.edu/abs/2014MNRAS.437..879C} {437, 879}

\bibitem[\protect\citeauthoryear{{Clark}, {Blake}  \& {Knapp}}{{Clark}
  et~al.}{2012}]{Clark2012}
{Clark} B.~M.,  {Blake} C.~H.,   {Knapp} G.~R.,  2012, \mn@doi [\apj]
  {10.1088/0004-637X/744/2/119}, \href
  {https://ui.adsabs.harvard.edu/abs/2012ApJ...744..119C} {744, 119}

\bibitem[\protect\citeauthoryear{{Csizmadia} et~al.,}{{Csizmadia}
  et~al.}{2015}]{Csizmadia2015}
{Csizmadia} S.,  et~al., 2015, \mn@doi [\aap] {10.1051/0004-6361/201526763},
  \href {https://ui.adsabs.harvard.edu/abs/2015A&A...584A..13C} {584, A13}

\bibitem[\protect\citeauthoryear{{Czekala}, {Chiang}, {Andrews}, {Jensen},
  {Torres}, {Wilner}, {Stassun}  \& {Macintosh}}{{Czekala}
  et~al.}{2019}]{Czekala2019}
{Czekala} I.,  {Chiang} E.,  {Andrews} S.~M.,  {Jensen} E. L.~N.,  {Torres} G.,
   {Wilner} D.~J.,  {Stassun} K.~G.,   {Macintosh} B.,  2019, \mn@doi [\apj]
  {10.3847/1538-4357/ab287b}, \href
  {https://ui.adsabs.harvard.edu/abs/2019ApJ...883...22C} {883, 22}

\bibitem[\protect\citeauthoryear{{Dawson} \& {Johnson}}{{Dawson} \&
  {Johnson}}{2018}]{Dawson2018}
{Dawson} R.~I.,  {Johnson} J.~A.,  2018, \mn@doi [\araa]
  {10.1146/annurev-astro-081817-051853}, \href
  {https://ui.adsabs.harvard.edu/abs/2018ARA&A..56..175D} {56, 175}

\bibitem[\protect\citeauthoryear{{De Rosa} et~al.,}{{De Rosa}
  et~al.}{2014}]{DeRosa2014}
{De Rosa} R.~J.,  et~al., 2014, \mn@doi [\mnras] {10.1093/mnras/stt1932}, \href
  {https://ui.adsabs.harvard.edu/abs/2014MNRAS.437.1216D} {437, 1216}

\bibitem[\protect\citeauthoryear{{Dieterich}, {Henry}, {Golimowski}, {Krist}
  \& {Tanner}}{{Dieterich} et~al.}{2012}]{Dieterich2012}
{Dieterich} S.~B.,  {Henry} T.~J.,  {Golimowski} D.~A.,  {Krist} J.~E.,
  {Tanner} A.~M.,  2012, \mn@doi [\aj] {10.1088/0004-6256/144/2/64}, \href
  {https://ui.adsabs.harvard.edu/abs/2012AJ....144...64D} {144, 64}

\bibitem[\protect\citeauthoryear{{Donati} et~al.,}{{Donati}
  et~al.}{2016}]{Donati2016}
{Donati} J.~F.,  et~al., 2016, \mn@doi [\nat] {10.1038/nature18305}, \href
  {https://ui.adsabs.harvard.edu/abs/2016Natur.534..662D} {534, 662}

\bibitem[\protect\citeauthoryear{{Dong} et~al.,}{{Dong}
  et~al.}{2014}]{Dong2014}
{Dong} S.,  et~al., 2014, \mn@doi [\apjl] {10.1088/2041-8205/789/1/L3}, \href
  {https://ui.adsabs.harvard.edu/abs/2014ApJ...789L...3D} {789, L3}

\bibitem[\protect\citeauthoryear{{Dong}, {Zhu}, {Rafikov}  \& {Stone}}{{Dong}
  et~al.}{2015}]{Dong2015}
{Dong} R.,  {Zhu} Z.,  {Rafikov} R.~R.,   {Stone} J.~M.,  2015, \mn@doi [\apjl]
  {10.1088/2041-8205/809/1/L5}, \href
  {https://ui.adsabs.harvard.edu/abs/2015ApJ...809L...5D} {809, L5}

\bibitem[\protect\citeauthoryear{{Dressing} \& {Charbonneau}}{{Dressing} \&
  {Charbonneau}}{2015}]{Dressing2015}
{Dressing} C.~D.,  {Charbonneau} D.,  2015, \mn@doi [\apj]
  {10.1088/0004-637X/807/1/45}, \href
  {https://ui.adsabs.harvard.edu/abs/2015ApJ...807...45D} {807, 45}

\bibitem[\protect\citeauthoryear{{Duch{\^e}ne} \& {Kraus}}{{Duch{\^e}ne} \&
  {Kraus}}{2013}]{Duchene2013}
{Duch{\^e}ne} G.,  {Kraus} A.,  2013, \mn@doi [\araa]
  {10.1146/annurev-astro-081710-102602}, 51, 269

\bibitem[\protect\citeauthoryear{{Dupuy}, {Kratter}, {Kraus}, {Isaacson},
  {Mann}, {Ireland}, {Howard}  \& {Huber}}{{Dupuy} et~al.}{2016}]{Dupuy2016}
{Dupuy} T.~J.,  {Kratter} K.~M.,  {Kraus} A.~L.,  {Isaacson} H.,  {Mann} A.~W.,
   {Ireland} M.~J.,  {Howard} A.~W.,   {Huber} D.,  2016, \mn@doi [\apj]
  {10.3847/0004-637X/817/1/80}, \href
  {https://ui.adsabs.harvard.edu/abs/2016ApJ...817...80D} {817, 80}

\bibitem[\protect\citeauthoryear{{Duquennoy} \& {Mayor}}{{Duquennoy} \&
  {Mayor}}{1991}]{Duquennoy1991}
{Duquennoy} A.,  {Mayor} M.,  1991, \aap, \href
  {https://ui.adsabs.harvard.edu/abs/1991A&A...248..485D} {500, 337}

\bibitem[\protect\citeauthoryear{{Eastman}, {Gaudi}  \& {Agol}}{{Eastman}
  et~al.}{2013}]{Eastman2013}
{Eastman} J.,  {Gaudi} B.~S.,   {Agol} E.,  2013, \mn@doi [\pasp]
  {10.1086/669497}, \href
  {https://ui.adsabs.harvard.edu/abs/2013PASP..125...83E} {125, 83}

\bibitem[\protect\citeauthoryear{{El-Badry} \& {Rix}}{{El-Badry} \&
  {Rix}}{2018}]{ElBadry2018}
{El-Badry} K.,  {Rix} H.-W.,  2018, \mn@doi [\mnras] {10.1093/mnras/sty2186},
  \href {https://ui.adsabs.harvard.edu/abs/2018MNRAS.480.4884E} {480, 4884}

\bibitem[\protect\citeauthoryear{{El-Badry} \& {Rix}}{{El-Badry} \&
  {Rix}}{2019}]{ElBadry2019a}
{El-Badry} K.,  {Rix} H.-W.,  2019, \mn@doi [\mnras] {10.1093/mnrasl/sly206},
  \href {https://ui.adsabs.harvard.edu/abs/2019MNRAS.482L.139E} {482, L139}

\bibitem[\protect\citeauthoryear{{El-Badry}, {Rix}, {Tian}, {Duch{\^e}ne}  \&
  {Moe}}{{El-Badry} et~al.}{2019}]{ElBadry2019b}
{El-Badry} K.,  {Rix} H.-W.,  {Tian} H.,  {Duch{\^e}ne} G.,   {Moe} M.,  2019,
  \mn@doi [\mnras] {10.1093/mnras/stz2480}, \href
  {https://ui.adsabs.harvard.edu/abs/2019MNRAS.489.5822E} {489, 5822}

\bibitem[\protect\citeauthoryear{{Esteves}, {De Mooij}  \&
  {Jayawardhana}}{{Esteves} et~al.}{2015}]{Esteves2015}
{Esteves} L.~J.,  {De Mooij} E. J.~W.,   {Jayawardhana} R.,  2015, \mn@doi
  [\apj] {10.1088/0004-637X/804/2/150}, \href
  {https://ui.adsabs.harvard.edu/abs/2015ApJ...804..150E} {804, 150}

\bibitem[\protect\citeauthoryear{{Evans} et~al.,}{{Evans}
  et~al.}{2016}]{Evans2016}
{Evans} D.~F.,  et~al., 2016, \mn@doi [\aap] {10.1051/0004-6361/201527970},
  \href {https://ui.adsabs.harvard.edu/abs/2016A&A...589A..58E} {589, A58}

\bibitem[\protect\citeauthoryear{{Evans} et~al.,}{{Evans}
  et~al.}{2018}]{Evans2018}
{Evans} D.~F.,  et~al., 2018, \mn@doi [\aap] {10.1051/0004-6361/201731855},
  \href {https://ui.adsabs.harvard.edu/abs/2018A&A...610A..20E} {610, A20}

\bibitem[\protect\citeauthoryear{{Fabrycky} \& {Tremaine}}{{Fabrycky} \&
  {Tremaine}}{2007}]{Fabrycky2007}
{Fabrycky} D.,  {Tremaine} S.,  2007, \mn@doi [\apj] {10.1086/521702}, \href
  {https://ui.adsabs.harvard.edu/abs/2007ApJ...669.1298F} {669, 1298}

\bibitem[\protect\citeauthoryear{{Feng}, {Tuomi}, {Jones}, {Barnes},
  {Anglada-Escud{\'e}}, {Vogt}  \& {Butler}}{{Feng} et~al.}{2017}]{Feng2017}
{Feng} F.,  {Tuomi} M.,  {Jones} H.~R.~A.,  {Barnes} J.,  {Anglada-Escud{\'e}}
  G.,  {Vogt} S.~S.,   {Butler} R.~P.,  2017, \mn@doi [\aj]
  {10.3847/1538-3881/aa83b4}, \href
  {https://ui.adsabs.harvard.edu/abs/2017AJ....154..135F} {154, 135}

\bibitem[\protect\citeauthoryear{{Fernandes}, {Mulders}, {Pascucci},
  {Mordasini}  \& {Emsenhuber}}{{Fernandes} et~al.}{2019}]{Fernandes2019}
{Fernandes} R.~B.,  {Mulders} G.~D.,  {Pascucci} I.,  {Mordasini} C.,
  {Emsenhuber} A.,  2019, \mn@doi [\apj] {10.3847/1538-4357/ab0300}, \href
  {https://ui.adsabs.harvard.edu/abs/2019ApJ...874...81F} {874, 81}

\bibitem[\protect\citeauthoryear{{Fischer} \& {Marcy}}{{Fischer} \&
  {Marcy}}{1992}]{Fischer1992}
{Fischer} D.~A.,  {Marcy} G.~W.,  1992, \mn@doi [\apj] {10.1086/171708}, \href
  {https://ui.adsabs.harvard.edu/abs/1992ApJ...396..178F} {396, 178}

\bibitem[\protect\citeauthoryear{{Fischer} \& {Valenti}}{{Fischer} \&
  {Valenti}}{2005}]{Fischer2005}
{Fischer} D.~A.,  {Valenti} J.,  2005, \mn@doi [\apj] {10.1086/428383}, \href
  {https://ui.adsabs.harvard.edu/abs/2005ApJ...622.1102F} {622, 1102}

\bibitem[\protect\citeauthoryear{{Flagg}, {Johns-Krull}, {Nofi}, {Llama},
  {Prato}, {Sullivan}, {Jaffe}  \& {Mace}}{{Flagg} et~al.}{2019}]{Flagg2019}
{Flagg} L.,  {Johns-Krull} C.~M.,  {Nofi} L.,  {Llama} J.,  {Prato} L.,
  {Sullivan} K.,  {Jaffe} D.~T.,   {Mace} G.,  2019, \mn@doi [\apjl]
  {10.3847/2041-8213/ab276d}, \href
  {https://ui.adsabs.harvard.edu/abs/2019ApJ...878L..37F} {878, L37}

\bibitem[\protect\citeauthoryear{{Fontanive}, {Rice}, {Bonavita}, {Lopez},
  {Mu{\v{z}}i{\'c}}, {}  \& {Biller}}{{Fontanive} et~al.}{2019}]{Fontanive2019}
{Fontanive} C.,  {Rice} K.,  {Bonavita} M.,  {Lopez} E.,  {Mu{\v{z}}i{\'c}} {}
  K.,   {Biller} B.,  2019, \mn@doi [\mnras] {10.1093/mnras/stz671}, \href
  {https://ui.adsabs.harvard.edu/abs/2019MNRAS.485.4967F} {485, 4967}

\bibitem[\protect\citeauthoryear{{Fragione}}{{Fragione}}{2019}]{Fragione2019}
{Fragione} G.,  2019, \mn@doi [\mnras] {10.1093/mnras/sty3367}, \href
  {https://ui.adsabs.harvard.edu/abs/2019MNRAS.483.3465F} {483, 3465}

\bibitem[\protect\citeauthoryear{{Fressin} et~al.,}{{Fressin}
  et~al.}{2013}]{Fressin2013}
{Fressin} F.,  et~al., 2013, \mn@doi [\apj] {10.1088/0004-637X/766/2/81}, \href
  {https://ui.adsabs.harvard.edu/abs/2013ApJ...766...81F} {766, 81}

\bibitem[\protect\citeauthoryear{{Gaidos}, {Mann}, {Kraus}  \&
  {Ireland}}{{Gaidos} et~al.}{2016}]{Gaidos2016}
{Gaidos} E.,  {Mann} A.~W.,  {Kraus} A.~L.,   {Ireland} M.,  2016, \mn@doi
  [\mnras] {10.1093/mnras/stw097}, \href
  {https://ui.adsabs.harvard.edu/abs/2016MNRAS.457.2877G} {457, 2877}

\bibitem[\protect\citeauthoryear{{Gong} \& {Ji}}{{Gong} \&
  {Ji}}{2018}]{Gong2018}
{Gong} Y.-X.,  {Ji} J.,  2018, \mn@doi [\mnras] {10.1093/mnras/sty1300}, \href
  {https://ui.adsabs.harvard.edu/abs/2018MNRAS.478.4565G} {478, 4565}

\bibitem[\protect\citeauthoryear{{Grether} \& {Lineweaver}}{{Grether} \&
  {Lineweaver}}{2006}]{Grether2006}
{Grether} D.,  {Lineweaver} C.~H.,  2006, \mn@doi [\apj] {10.1086/500161},
  \href {https://ui.adsabs.harvard.edu/abs/2006ApJ...640.1051G} {640, 1051}

\bibitem[\protect\citeauthoryear{{Gullikson}, {Kraus}  \&
  {Dodson-Robinson}}{{Gullikson} et~al.}{2016}]{Gullikson2016}
{Gullikson} K.,  {Kraus} A.,   {Dodson-Robinson} S.,  2016, \mn@doi [\aj]
  {10.3847/0004-6256/152/2/40}, \href
  {https://ui.adsabs.harvard.edu/abs/2016AJ....152...40G} {152, 40}

\bibitem[\protect\citeauthoryear{{Guo}, {Johnson}, {Mann}, {Kraus}, {Curtis}
  \& {Latham}}{{Guo} et~al.}{2017}]{Guo2017}
{Guo} X.,  {Johnson} J.~A.,  {Mann} A.~W.,  {Kraus} A.~L.,  {Curtis} J.~L.,
  {Latham} D.~W.,  2017, \mn@doi [\apj] {10.3847/1538-4357/aa6004}, \href
  {https://ui.adsabs.harvard.edu/abs/2017ApJ...838...25G} {838, 25}

\bibitem[\protect\citeauthoryear{{Haghighipour} \& {Raymond}}{{Haghighipour} \&
  {Raymond}}{2007}]{Haghighipour2007}
{Haghighipour} N.,  {Raymond} S.~N.,  2007, \mn@doi [\apj] {10.1086/520501},
  \href {https://ui.adsabs.harvard.edu/abs/2007ApJ...666..436H} {666, 436}

\bibitem[\protect\citeauthoryear{{Harris}, {Andrews}, {Wilner}  \&
  {Kraus}}{{Harris} et~al.}{2012}]{Harris2012}
{Harris} R.~J.,  {Andrews} S.~M.,  {Wilner} D.~J.,   {Kraus} A.~L.,  2012,
  \mn@doi [\apj] {10.1088/0004-637X/751/2/115}, \href
  {https://ui.adsabs.harvard.edu/abs/2012ApJ...751..115H} {751, 115}

\bibitem[\protect\citeauthoryear{{Hatzes} \& {Rauer}}{{Hatzes} \&
  {Rauer}}{2015}]{Hatzes2015}
{Hatzes} A.~P.,  {Rauer} H.,  2015, \mn@doi [\apjl]
  {10.1088/2041-8205/810/2/L25}, \href
  {https://ui.adsabs.harvard.edu/abs/2015ApJ...810L..25H} {810, L25}

\bibitem[\protect\citeauthoryear{{H{\'e}brard} et~al.,}{{H{\'e}brard}
  et~al.}{2011}]{Hebrard2011}
{H{\'e}brard} G.,  et~al., 2011, \mn@doi [\aap] {10.1051/0004-6361/201117192},
  \href {https://ui.adsabs.harvard.edu/abs/2011A&A...533A.130H} {533, A130}

\bibitem[\protect\citeauthoryear{{Holberg}, {Oswalt}, {Sion}  \&
  {McCook}}{{Holberg} et~al.}{2016}]{Holberg2016}
{Holberg} J.~B.,  {Oswalt} T.~D.,  {Sion} E.~M.,   {McCook} G.~P.,  2016,
  \mn@doi [\mnras] {10.1093/mnras/stw1357}, \href
  {https://ui.adsabs.harvard.edu/abs/2016MNRAS.462.2295H} {462, 2295}

\bibitem[\protect\citeauthoryear{{Holman} \& {Wiegert}}{{Holman} \&
  {Wiegert}}{1999}]{Holman1999}
{Holman} M.~J.,  {Wiegert} P.~A.,  1999, \mn@doi [\aj] {10.1086/300695}, \href
  {https://ui.adsabs.harvard.edu/abs/1999AJ....117..621H} {117, 621}

\bibitem[\protect\citeauthoryear{{Howard} et~al.,}{{Howard}
  et~al.}{2012}]{Howard2012}
{Howard} A.~W.,  et~al., 2012, \mn@doi [\apjs] {10.1088/0067-0049/201/2/15},
  \href {https://ui.adsabs.harvard.edu/abs/2012ApJS..201...15H} {201, 15}

\bibitem[\protect\citeauthoryear{{Janson} et~al.,}{{Janson}
  et~al.}{2012}]{Janson2012}
{Janson} M.,  et~al., 2012, \mn@doi [\apj] {10.1088/0004-637X/754/1/44}, \href
  {https://ui.adsabs.harvard.edu/abs/2012ApJ...754...44J} {754, 44}

\bibitem[\protect\citeauthoryear{{Johns-Krull} et~al.,}{{Johns-Krull}
  et~al.}{2016}]{JohnsKrull2016}
{Johns-Krull} C.~M.,  et~al., 2016, \mn@doi [\apj]
  {10.3847/0004-637X/826/2/206}, \href
  {https://ui.adsabs.harvard.edu/abs/2016ApJ...826..206J} {826, 206}

\bibitem[\protect\citeauthoryear{{Johnson}, {Butler}, {Marcy}, {Fischer},
  {Vogt}, {Wright}  \& {Peek}}{{Johnson} et~al.}{2007}]{Johnson2007}
{Johnson} J.~A.,  {Butler} R.~P.,  {Marcy} G.~W.,  {Fischer} D.~A.,  {Vogt}
  S.~S.,  {Wright} J.~T.,   {Peek} K. M.~G.,  2007, \mn@doi [\apj]
  {10.1086/521720}, \href
  {https://ui.adsabs.harvard.edu/abs/2007ApJ...670..833J} {670, 833}

\bibitem[\protect\citeauthoryear{{Johnson}, {Aller}, {Howard}  \&
  {Crepp}}{{Johnson} et~al.}{2010}]{Johnson2010}
{Johnson} J.~A.,  {Aller} K.~M.,  {Howard} A.~W.,   {Crepp} J.~R.,  2010,
  \mn@doi [\pasp] {10.1086/655775}, \href
  {https://ui.adsabs.harvard.edu/abs/2010PASP..122..905J} {122, 905}

\bibitem[\protect\citeauthoryear{{Jontof-Hutter}, {Lissauer}, {Rowe}  \&
  {Fabrycky}}{{Jontof-Hutter} et~al.}{2014}]{JontofHutter2014}
{Jontof-Hutter} D.,  {Lissauer} J.~J.,  {Rowe} J.~F.,   {Fabrycky} D.~C.,
  2014, \mn@doi [\apj] {10.1088/0004-637X/785/1/15}, \href
  {https://ui.adsabs.harvard.edu/abs/2014ApJ...785...15J} {785, 15}

\bibitem[\protect\citeauthoryear{{Kirk} et~al.,}{{Kirk}
  et~al.}{2016}]{Kirk2016}
{Kirk} B.,  et~al., 2016, \mn@doi [\aj] {10.3847/0004-6256/151/3/68}, \href
  {https://ui.adsabs.harvard.edu/abs/2016AJ....151...68K} {151, 68}

\bibitem[\protect\citeauthoryear{{Kiseleva}, {Eggleton}  \&
  {Mikkola}}{{Kiseleva} et~al.}{1998}]{Kiseleva1998}
{Kiseleva} L.~G.,  {Eggleton} P.~P.,   {Mikkola} S.,  1998, \mn@doi [\mnras]
  {10.1046/j.1365-8711.1998.01903.x}, \href
  {https://ui.adsabs.harvard.edu/abs/1998MNRAS.300..292K} {300, 292}

\bibitem[\protect\citeauthoryear{{Knutson} et~al.,}{{Knutson}
  et~al.}{2014}]{Knutson2014}
{Knutson} H.~A.,  et~al., 2014, \mn@doi [\apj] {10.1088/0004-637X/785/2/126},
  \href {https://ui.adsabs.harvard.edu/abs/2014ApJ...785..126K} {785, 126}

\bibitem[\protect\citeauthoryear{{Kobulnicky} \& {Fryer}}{{Kobulnicky} \&
  {Fryer}}{2007}]{Kobulnicky2007}
{Kobulnicky} H.~A.,  {Fryer} C.~L.,  2007, \mn@doi [\apj] {10.1086/522073},
  \href {https://ui.adsabs.harvard.edu/abs/2007ApJ...670..747K} {670, 747}

\bibitem[\protect\citeauthoryear{{Kounkel} et~al.,}{{Kounkel}
  et~al.}{2019}]{Kounkel2019}
{Kounkel} M.,  et~al., 2019, \mn@doi [\aj] {10.3847/1538-3881/ab13b1}, \href
  {https://ui.adsabs.harvard.edu/abs/2019AJ....157..196K} {157, 196}

\bibitem[\protect\citeauthoryear{{Kouwenhoven}, {Brown}, {Portegies Zwart}  \&
  {Kaper}}{{Kouwenhoven} et~al.}{2007}]{Kouwenhoven2007}
{Kouwenhoven} M.~B.~N.,  {Brown} A.~G.~A.,  {Portegies Zwart} S.~F.,   {Kaper}
  L.,  2007, \mn@doi [\aap] {10.1051/0004-6361:20077719}, \href
  {https://ui.adsabs.harvard.edu/abs/2007A&A...474...77K} {474, 77}

\bibitem[\protect\citeauthoryear{{Kratter}}{{Kratter}}{2017}]{Kratter2017}
{Kratter} K.~M.,  2017, in {Pessah} M.,  {Gressel} O.,  eds,  Astrophysics and
  Space Science Library Vol. 445, Astrophysics and Space Science Library.
  p.~315, \mn@doi{10.1007/978-3-319-60609-5_11}

\bibitem[\protect\citeauthoryear{{Kraus}, {Ireland}, {Martinache}  \&
  {Lloyd}}{{Kraus} et~al.}{2008}]{Kraus2008}
{Kraus} A.~L.,  {Ireland} M.~J.,  {Martinache} F.,   {Lloyd} J.~P.,  2008,
  \mn@doi [\apj] {10.1086/587435}, \href
  {https://ui.adsabs.harvard.edu/abs/2008ApJ...679..762K} {679, 762}

\bibitem[\protect\citeauthoryear{{Kraus}, {Ireland}, {Martinache}  \&
  {Hillenbrand}}{{Kraus} et~al.}{2011}]{Kraus2011}
{Kraus} A.~L.,  {Ireland} M.~J.,  {Martinache} F.,   {Hillenbrand} L.~A.,
  2011, \mn@doi [\apj] {10.1088/0004-637X/731/1/8}, \href
  {https://ui.adsabs.harvard.edu/abs/2011ApJ...731....8K} {731, 8}

\bibitem[\protect\citeauthoryear{{Kraus}, {Ireland}, {Hillenbrand}  \&
  {Martinache}}{{Kraus} et~al.}{2012}]{Kraus2012}
{Kraus} A.~L.,  {Ireland} M.~J.,  {Hillenbrand} L.~A.,   {Martinache} F.,
  2012, \mn@doi [\apj] {10.1088/0004-637X/745/1/19}, \href
  {https://ui.adsabs.harvard.edu/abs/2012ApJ...745...19K} {745, 19}

\bibitem[\protect\citeauthoryear{{Kraus}, {Ireland}, {Huber}, {Mann}  \&
  {Dupuy}}{{Kraus} et~al.}{2016}]{Kraus2016}
{Kraus} A.~L.,  {Ireland} M.~J.,  {Huber} D.,  {Mann} A.~W.,   {Dupuy} T.~J.,
  2016, \mn@doi [\aj] {10.3847/0004-6256/152/1/8}, \href
  {http://adsabs.harvard.edu/abs/2016AJ....152....8K} {152, 8}

\bibitem[\protect\citeauthoryear{{Law}, {Hodgkin}  \& {Mackay}}{{Law}
  et~al.}{2008}]{Law2008}
{Law} N.~M.,  {Hodgkin} S.~T.,   {Mackay} C.~D.,  2008, \mn@doi [\mnras]
  {10.1111/j.1365-2966.2007.12675.x}, \href
  {https://ui.adsabs.harvard.edu/abs/2008MNRAS.384..150L} {384, 150}

\bibitem[\protect\citeauthoryear{{Law} et~al.,}{{Law} et~al.}{2014}]{Law2014}
{Law} N.~M.,  et~al., 2014, \mn@doi [\apj] {10.1088/0004-637X/791/1/35}, \href
  {https://ui.adsabs.harvard.edu/abs/2014ApJ...791...35L} {791, 35}

\bibitem[\protect\citeauthoryear{{Lawler} et~al.,}{{Lawler}
  et~al.}{2014}]{Lawler2014}
{Lawler} S.~M.,  et~al., 2014, \mn@doi [\mnras] {10.1093/mnras/stu1641}, \href
  {https://ui.adsabs.harvard.edu/abs/2014MNRAS.444.2665L} {444, 2665}

\bibitem[\protect\citeauthoryear{{Lee} \& {Chiang}}{{Lee} \&
  {Chiang}}{2016}]{Lee2016}
{Lee} E.~J.,  {Chiang} E.,  2016, \mn@doi [\apj] {10.3847/0004-637X/817/2/90},
  \href {https://ui.adsabs.harvard.edu/abs/2016ApJ...817...90L} {817, 90}

\bibitem[\protect\citeauthoryear{{Lee}, {Offner}, {Kratter}, {Smullen}  \&
  {Li}}{{Lee} et~al.}{2019}]{Lee2019}
{Lee} A.~T.,  {Offner} S. S.~R.,  {Kratter} K.~M.,  {Smullen} R.~A.,   {Li}
  P.~S.,  2019, arXiv e-prints, \href
  {https://ui.adsabs.harvard.edu/abs/2019arXiv191107863L} {p. arXiv:1911.07863}

\bibitem[\protect\citeauthoryear{{Lloyd}}{{Lloyd}}{2011}]{Lloyd2011}
{Lloyd} J.~P.,  2011, \mn@doi [\apjl] {10.1088/2041-8205/739/2/L49}, \href
  {https://ui.adsabs.harvard.edu/abs/2011ApJ...739L..49L} {739, L49}

\bibitem[\protect\citeauthoryear{{Lopez} \& {Fortney}}{{Lopez} \&
  {Fortney}}{2014}]{Lopez2014}
{Lopez} E.~D.,  {Fortney} J.~J.,  2014, \mn@doi [\apj]
  {10.1088/0004-637X/792/1/1}, \href
  {https://ui.adsabs.harvard.edu/abs/2014ApJ...792....1L} {792, 1}

\bibitem[\protect\citeauthoryear{{Ma} \& {Ge}}{{Ma} \& {Ge}}{2014}]{Ma2014}
{Ma} B.,  {Ge} J.,  2014, \mn@doi [\mnras] {10.1093/mnras/stu134}, \href
  {https://ui.adsabs.harvard.edu/abs/2014MNRAS.439.2781M} {439, 2781}

\bibitem[\protect\citeauthoryear{{MacGregor}, {Lawler}, {Wilner}, {Matthews},
  {Kennedy}, {Booth}  \& {Di Francesco}}{{MacGregor}
  et~al.}{2016}]{MacGregor2016}
{MacGregor} M.~A.,  {Lawler} S.~M.,  {Wilner} D.~J.,  {Matthews} B.~C.,
  {Kennedy} G.~M.,  {Booth} M.,   {Di Francesco} J.,  2016, \mn@doi [\apj]
  {10.3847/0004-637X/828/2/113}, \href
  {https://ui.adsabs.harvard.edu/abs/2016ApJ...828..113M} {828, 113}

\bibitem[\protect\citeauthoryear{{Marcy}, {Butler}, {Fischer}, {Vogt},
  {Wright}, {Tinney}  \& {Jones}}{{Marcy} et~al.}{2005}]{Marcy2005}
{Marcy} G.,  {Butler} R.~P.,  {Fischer} D.,  {Vogt} S.,  {Wright} J.~T.,
  {Tinney} C.~G.,   {Jones} H.~R.~A.,  2005, \mn@doi [Progress of Theoretical
  Physics Supplement] {10.1143/PTPS.158.24}, \href
  {https://ui.adsabs.harvard.edu/abs/2005PThPS.158...24M} {158, 24}

\bibitem[\protect\citeauthoryear{{Martin}}{{Martin}}{2018}]{Martin2018}
{Martin} D.~V.,  2018, preprint, \href
  {http://adsabs.harvard.edu/abs/2018arXiv180208693M} {} (\mn@eprint {arXiv}
  {1802.08693})

\bibitem[\protect\citeauthoryear{{Masuda}}{{Masuda}}{2014}]{Masuda2014}
{Masuda} K.,  2014, \mn@doi [\apj] {10.1088/0004-637X/783/1/53}, \href
  {https://ui.adsabs.harvard.edu/abs/2014ApJ...783...53M} {783, 53}

\bibitem[\protect\citeauthoryear{{Masuda}}{{Masuda}}{2017}]{Masuda2017}
{Masuda} K.,  2017, \mn@doi [\aj] {10.3847/1538-3881/aa7aeb}, \href
  {https://ui.adsabs.harvard.edu/abs/2017AJ....154...64M} {154, 64}

\bibitem[\protect\citeauthoryear{{Matson}, {Howell}, {Horch}  \&
  {Everett}}{{Matson} et~al.}{2018}]{Matson2018}
{Matson} R.~A.,  {Howell} S.~B.,  {Horch} E.~P.,   {Everett} M.~E.,  2018,
  \mn@doi [\aj] {10.3847/1538-3881/aac778}, \href
  {https://ui.adsabs.harvard.edu/abs/2018AJ....156...31M} {156, 31}

\bibitem[\protect\citeauthoryear{{Matsuo}, {Shibai}, {Ootsubo}  \&
  {Tamura}}{{Matsuo} et~al.}{2007}]{Matsuo2007}
{Matsuo} T.,  {Shibai} H.,  {Ootsubo} T.,   {Tamura} M.,  2007, \mn@doi [\apj]
  {10.1086/517964}, \href
  {https://ui.adsabs.harvard.edu/abs/2007ApJ...662.1282M} {662, 1282}

\bibitem[\protect\citeauthoryear{{Mayor} et~al.,}{{Mayor}
  et~al.}{2011}]{Mayor2011}
{Mayor} M.,  et~al., 2011, arXiv e-prints, \href
  {https://ui.adsabs.harvard.edu/abs/2011arXiv1109.2497M} {p. arXiv:1109.2497}

\bibitem[\protect\citeauthoryear{{Mazeh}, {Holczer}  \& {Faigler}}{{Mazeh}
  et~al.}{2016}]{Mazeh2016}
{Mazeh} T.,  {Holczer} T.,   {Faigler} S.,  2016, \mn@doi [\aap]
  {10.1051/0004-6361/201528065}, \href
  {https://ui.adsabs.harvard.edu/abs/2016A&A...589A..75M} {589, A75}

\bibitem[\protect\citeauthoryear{{Melo}}{{Melo}}{2003}]{Melo2003}
{Melo} C.~H.~F.,  2003, \mn@doi [\aap] {10.1051/0004-6361:20031242}, \href
  {https://ui.adsabs.harvard.edu/abs/2003A&A...410..269M} {410, 269}

\bibitem[\protect\citeauthoryear{{Moe} \& {Di Stefano}}{{Moe} \& {Di
  Stefano}}{2013}]{Moe2013}
{Moe} M.,  {Di Stefano} R.,  2013, \mn@doi [\apj] {10.1088/0004-637X/778/2/95},
  \href {https://ui.adsabs.harvard.edu/abs/2013ApJ...778...95M} {778, 95}

\bibitem[\protect\citeauthoryear{{Moe} \& {Di Stefano}}{{Moe} \& {Di
  Stefano}}{2015a}]{Moe2015a}
{Moe} M.,  {Di Stefano} R.,  2015a, \mn@doi [\apj]
  {10.1088/0004-637X/801/2/113}, \href
  {https://ui.adsabs.harvard.edu/abs/2015ApJ...801..113M} {801, 113}

\bibitem[\protect\citeauthoryear{{Moe} \& {Di Stefano}}{{Moe} \& {Di
  Stefano}}{2015b}]{Moe2015b}
{Moe} M.,  {Di Stefano} R.,  2015b, \mn@doi [\apj]
  {10.1088/0004-637X/810/1/61}, \href
  {https://ui.adsabs.harvard.edu/abs/2015ApJ...810...61M} {810, 61}

\bibitem[\protect\citeauthoryear{{Moe} \& {Di Stefano}}{{Moe} \& {Di
  Stefano}}{2017}]{Moe2017}
{Moe} M.,  {Di Stefano} R.,  2017, \mn@doi [\apjs] {10.3847/1538-4365/aa6fb6},
  \href {http://adsabs.harvard.edu/abs/2017ApJS..230...15M} {230, 15}

\bibitem[\protect\citeauthoryear{{Moe} \& {Kratter}}{{Moe} \&
  {Kratter}}{2018}]{Moe2018}
{Moe} M.,  {Kratter} K.~M.,  2018, \mn@doi [\apj] {10.3847/1538-4357/aaa6d2},
  \href {https://ui.adsabs.harvard.edu/abs/2018ApJ...854...44M} {854, 44}

\bibitem[\protect\citeauthoryear{{Moe}, {Kratter}  \& {Badenes}}{{Moe}
  et~al.}{2019}]{Moe2019}
{Moe} M.,  {Kratter} K.~M.,   {Badenes} C.,  2019, \mn@doi [\apj]
  {10.3847/1538-4357/ab0d88}, \href
  {https://ui.adsabs.harvard.edu/abs/2019ApJ...875...61M} {875, 61}

\bibitem[\protect\citeauthoryear{{Mordasini}, {Alibert}, {Benz}  \&
  {Naef}}{{Mordasini} et~al.}{2008}]{Mordasini2008}
{Mordasini} C.,  {Alibert} Y.,  {Benz} W.,   {Naef} D.,  2008, in {Fischer} D.,
   {Rasio} F.~A.,  {Thorsett} S.~E.,   {Wolszczan} A.,  eds,  Astronomical
  Society of the Pacific Conference Series Vol. 398, Extreme Solar Systems.
  p.~235 (\mn@eprint {arXiv} {0710.5667})

\bibitem[\protect\citeauthoryear{{Morton} \& {Swift}}{{Morton} \&
  {Swift}}{2014}]{Morton2014}
{Morton} T.~D.,  {Swift} J.,  2014, \mn@doi [\apj]
  {10.1088/0004-637X/791/1/10}, \href
  {https://ui.adsabs.harvard.edu/abs/2014ApJ...791...10M} {791, 10}

\bibitem[\protect\citeauthoryear{{Morton}, {Bryson}, {Coughlin}, {Rowe},
  {Ravichandran}, {Petigura}, {Haas}  \& {Batalha}}{{Morton}
  et~al.}{2016}]{Morton2016}
{Morton} T.~D.,  {Bryson} S.~T.,  {Coughlin} J.~L.,  {Rowe} J.~F.,
  {Ravichandran} G.,  {Petigura} E.~A.,  {Haas} M.~R.,   {Batalha} N.~M.,
  2016, \mn@doi [\apj] {10.3847/0004-637X/822/2/86}, \href
  {https://ui.adsabs.harvard.edu/abs/2016ApJ...822...86M} {822, 86}

\bibitem[\protect\citeauthoryear{{Mulders}, {Pascucci}  \& {Apai}}{{Mulders}
  et~al.}{2015a}]{Mulders2015a}
{Mulders} G.~D.,  {Pascucci} I.,   {Apai} D.,  2015a, \mn@doi [\apj]
  {10.1088/0004-637X/798/2/112}, \href
  {https://ui.adsabs.harvard.edu/abs/2015ApJ...798..112M} {798, 112}

\bibitem[\protect\citeauthoryear{{Mulders}, {Pascucci}  \& {Apai}}{{Mulders}
  et~al.}{2015b}]{Mulders2015b}
{Mulders} G.~D.,  {Pascucci} I.,   {Apai} D.,  2015b, \mn@doi [\apj]
  {10.1088/0004-637X/814/2/130}, \href
  {https://ui.adsabs.harvard.edu/abs/2015ApJ...814..130M} {814, 130}

\bibitem[\protect\citeauthoryear{{Murphy}, {Moe}, {Kurtz}, {Bedding},
  {Shibahashi}  \& {Boffin}}{{Murphy} et~al.}{2018}]{Murphy2018}
{Murphy} S.~J.,  {Moe} M.,  {Kurtz} D.~W.,  {Bedding} T.~R.,  {Shibahashi} H.,
   {Boffin} H. M.~J.,  2018, \mn@doi [\mnras] {10.1093/mnras/stx3049}, \href
  {https://ui.adsabs.harvard.edu/abs/2018MNRAS.474.4322M} {474, 4322}

\bibitem[\protect\citeauthoryear{{Naoz} \& {Fabrycky}}{{Naoz} \&
  {Fabrycky}}{2014}]{Naoz2014}
{Naoz} S.,  {Fabrycky} D.~C.,  2014, \mn@doi [\apj]
  {10.1088/0004-637X/793/2/137}, \href
  {https://ui.adsabs.harvard.edu/abs/2014ApJ...793..137N} {793, 137}

\bibitem[\protect\citeauthoryear{{Ngo} et~al.,}{{Ngo} et~al.}{2015}]{Ngo2015}
{Ngo} H.,  et~al., 2015, \mn@doi [\apj] {10.1088/0004-637X/800/2/138}, \href
  {https://ui.adsabs.harvard.edu/abs/2015ApJ...800..138N} {800, 138}

\bibitem[\protect\citeauthoryear{{Ngo} et~al.,}{{Ngo} et~al.}{2016}]{Ngo2016}
{Ngo} H.,  et~al., 2016, \mn@doi [\apj] {10.3847/0004-637X/827/1/8}, \href
  {https://ui.adsabs.harvard.edu/abs/2016ApJ...827....8N} {827, 8}

\bibitem[\protect\citeauthoryear{{Nielsen} et~al.,}{{Nielsen}
  et~al.}{2019}]{Nielsen2019}
{Nielsen} E.~L.,  et~al., 2019, \mn@doi [\aj] {10.3847/1538-3881/ab16e9}, \href
  {https://ui.adsabs.harvard.edu/abs/2019AJ....158...13N} {158, 13}

\bibitem[\protect\citeauthoryear{{Nordstr{\"o}m} et~al.,}{{Nordstr{\"o}m}
  et~al.}{2004}]{Nordstrom2004}
{Nordstr{\"o}m} B.,  et~al., 2004, \mn@doi [\aap] {10.1051/0004-6361:20035959},
  \href {https://ui.adsabs.harvard.edu/abs/2004A&A...418..989N} {418, 989}

\bibitem[\protect\citeauthoryear{{North} et~al.,}{{North}
  et~al.}{2017}]{North2017}
{North} T. S.~H.,  et~al., 2017, \mn@doi [\mnras] {10.1093/mnras/stx2009},
  \href {https://ui.adsabs.harvard.edu/abs/2017MNRAS.472.1866N} {472, 1866}

\bibitem[\protect\citeauthoryear{{Ofir} \& {Dreizler}}{{Ofir} \&
  {Dreizler}}{2013}]{Ofir2013}
{Ofir} A.,  {Dreizler} S.,  2013, \mn@doi [\aap] {10.1051/0004-6361/201219877},
  \href {https://ui.adsabs.harvard.edu/abs/2013A&A...555A..58O} {555, A58}

\bibitem[\protect\citeauthoryear{{Ofir}, {Dreizler}, {Zechmeister}  \&
  {Husser}}{{Ofir} et~al.}{2014}]{Ofir2014}
{Ofir} A.,  {Dreizler} S.,  {Zechmeister} M.,   {Husser} T.-O.,  2014, \mn@doi
  [\aap] {10.1051/0004-6361/201220935}, \href
  {https://ui.adsabs.harvard.edu/abs/2014A&A...561A.103O} {561, A103}

\bibitem[\protect\citeauthoryear{{{\"O}pik}}{{{\"O}pik}}{1924}]{Opik1924}
{{\"O}pik} E.,  1924, Publications of the Tartu Astrofizica Observatory, \href
  {https://ui.adsabs.harvard.edu/abs/1924PTarO..25f...1O} {25, 1}

\bibitem[\protect\citeauthoryear{{Owen} \& {Lai}}{{Owen} \&
  {Lai}}{2018}]{Owen2018}
{Owen} J.~E.,  {Lai} D.,  2018, \mn@doi [\mnras] {10.1093/mnras/sty1760}, \href
  {https://ui.adsabs.harvard.edu/abs/2018MNRAS.479.5012O} {479, 5012}

\bibitem[\protect\citeauthoryear{{Petigura}, {Marcy}  \& {Howard}}{{Petigura}
  et~al.}{2013}]{Petigura2013}
{Petigura} E.~A.,  {Marcy} G.~W.,   {Howard} A.~W.,  2013, \mn@doi [\apj]
  {10.1088/0004-637X/770/1/69}, \href
  {https://ui.adsabs.harvard.edu/abs/2013ApJ...770...69P} {770, 69}

\bibitem[\protect\citeauthoryear{{Petigura} et~al.,}{{Petigura}
  et~al.}{2018}]{Petigura2018}
{Petigura} E.~A.,  et~al., 2018, \mn@doi [\aj] {10.3847/1538-3881/aaa54c},
  \href {https://ui.adsabs.harvard.edu/abs/2018AJ....155...89P} {155, 89}

\bibitem[\protect\citeauthoryear{{Pollacco} et~al.,}{{Pollacco}
  et~al.}{2006}]{Pollacco2006}
{Pollacco} D.~L.,  et~al., 2006, \mn@doi [\pasp] {10.1086/508556}, \href
  {https://ui.adsabs.harvard.edu/abs/2006PASP..118.1407P} {118, 1407}

\bibitem[\protect\citeauthoryear{{Pourbaix} et~al.,}{{Pourbaix}
  et~al.}{2002}]{Pourbaix2002}
{Pourbaix} D.,  et~al., 2002, \mn@doi [\aap] {10.1051/0004-6361:20020287},
  \href {https://ui.adsabs.harvard.edu/abs/2002A&A...386..280P} {386, 280}

\bibitem[\protect\citeauthoryear{{Prato}}{{Prato}}{2007}]{Prato2007}
{Prato} L.,  2007, \mn@doi [\apj] {10.1086/510882}, \href
  {https://ui.adsabs.harvard.edu/abs/2007ApJ...657..338P} {657, 338}

\bibitem[\protect\citeauthoryear{{Rafikov} \& {Silsbee}}{{Rafikov} \&
  {Silsbee}}{2015a}]{Rafikov2015a}
{Rafikov} R.~R.,  {Silsbee} K.,  2015a, \mn@doi [\apj]
  {10.1088/0004-637X/798/2/69}, \href
  {https://ui.adsabs.harvard.edu/abs/2015ApJ...798...69R} {798, 69}

\bibitem[\protect\citeauthoryear{{Rafikov} \& {Silsbee}}{{Rafikov} \&
  {Silsbee}}{2015b}]{Rafikov2015b}
{Rafikov} R.~R.,  {Silsbee} K.,  2015b, \mn@doi [\apj]
  {10.1088/0004-637X/798/2/70}, \href
  {https://ui.adsabs.harvard.edu/abs/2015ApJ...798...70R} {798, 70}

\bibitem[\protect\citeauthoryear{{Raghavan} et~al.,}{{Raghavan}
  et~al.}{2010}]{Raghavan2010}
{Raghavan} D.,  et~al., 2010, \mn@doi [\apjs] {10.1088/0067-0049/190/1/1},
  \href {http://adsabs.harvard.edu/abs/2010ApJS..190....1R} {190, 1}

\bibitem[\protect\citeauthoryear{{Reffert}, {Bergmann}, {Quirrenbach},
  {Trifonov}  \& {K{\"u}nstler}}{{Reffert} et~al.}{2015}]{Reffert2015}
{Reffert} S.,  {Bergmann} C.,  {Quirrenbach} A.,  {Trifonov} T.,
  {K{\"u}nstler} A.,  2015, \mn@doi [\aap] {10.1051/0004-6361/201322360}, \href
  {https://ui.adsabs.harvard.edu/abs/2015A&A...574A.116R} {574, A116}

\bibitem[\protect\citeauthoryear{{Rey} et~al.,}{{Rey} et~al.}{2018}]{Rey2018}
{Rey} J.,  et~al., 2018, \mn@doi [\aap] {10.1051/0004-6361/201833180}, \href
  {https://ui.adsabs.harvard.edu/abs/2018A&A...619A.115R} {619, A115}

\bibitem[\protect\citeauthoryear{{Rice}, {Lodato}, {Pringle}, {Armitage}  \&
  {Bonnell}}{{Rice} et~al.}{2006}]{Rice2006}
{Rice} W.~K.~M.,  {Lodato} G.,  {Pringle} J.~E.,  {Armitage} P.~J.,   {Bonnell}
  I.~A.,  2006, \mn@doi [\mnras] {10.1111/j.1745-3933.2006.00215.x}, \href
  {https://ui.adsabs.harvard.edu/abs/2006MNRAS.372L...9R} {372, L9}

\bibitem[\protect\citeauthoryear{{Rizzuto} et~al.,}{{Rizzuto}
  et~al.}{2013}]{Rizzuto2013}
{Rizzuto} A.~C.,  et~al., 2013, \mn@doi [\mnras] {10.1093/mnras/stt1690}, \href
  {https://ui.adsabs.harvard.edu/abs/2013MNRAS.436.1694R} {436, 1694}

\bibitem[\protect\citeauthoryear{{Sana} et~al.,}{{Sana}
  et~al.}{2012}]{Sana2012}
{Sana} H.,  et~al., 2012, \mn@doi [Science] {10.1126/science.1223344}, 337, 444

\bibitem[\protect\citeauthoryear{{Sana} et~al.,}{{Sana}
  et~al.}{2014}]{Sana2014}
{Sana} H.,  et~al., 2014, \mn@doi [\apjs] {10.1088/0067-0049/215/1/15}, \href
  {https://ui.adsabs.harvard.edu/abs/2014ApJS..215...15S} {215, 15}

\bibitem[\protect\citeauthoryear{{Santerne} et~al.,}{{Santerne}
  et~al.}{2014}]{Santerne2014}
{Santerne} A.,  et~al., 2014, \mn@doi [\aap] {10.1051/0004-6361/201424158},
  \href {https://ui.adsabs.harvard.edu/abs/2014A&A...571A..37S} {571, A37}

\bibitem[\protect\citeauthoryear{{Santerne} et~al.,}{{Santerne}
  et~al.}{2016}]{Santerne2016}
{Santerne} A.,  et~al., 2016, \mn@doi [\aap] {10.1051/0004-6361/201527329},
  \href {https://ui.adsabs.harvard.edu/abs/2016A&A...587A..64S} {587, A64}

\bibitem[\protect\citeauthoryear{{Santos}, {Israelian}  \& {Mayor}}{{Santos}
  et~al.}{2004}]{Santos2004}
{Santos} N.~C.,  {Israelian} G.,   {Mayor} M.,  2004, \mn@doi [\aap]
  {10.1051/0004-6361:20034469}, \href
  {https://ui.adsabs.harvard.edu/abs/2004A&A...415.1153S} {415, 1153}

\bibitem[\protect\citeauthoryear{{Schlaufman}}{{Schlaufman}}{2018}]{Schlaufman2018}
{Schlaufman} K.~C.,  2018, \mn@doi [\apj] {10.3847/1538-4357/aa961c}, \href
  {https://ui.adsabs.harvard.edu/abs/2018ApJ...853...37S} {853, 37}

\bibitem[\protect\citeauthoryear{{Shahaf} \& {Mazeh}}{{Shahaf} \&
  {Mazeh}}{2019}]{Shahaf2019}
{Shahaf} S.,  {Mazeh} T.,  2019, \mn@doi [\mnras] {10.1093/mnras/stz1517},
  \href {https://ui.adsabs.harvard.edu/abs/2019MNRAS.487.3356S} {487, 3356}

\bibitem[\protect\citeauthoryear{{Shan}, {Johnson}  \& {Morton}}{{Shan}
  et~al.}{2015}]{Shan2015}
{Shan} Y.,  {Johnson} J.~A.,   {Morton} T.~D.,  2015, \mn@doi [\apj]
  {10.1088/0004-637X/813/1/75}, \href
  {https://ui.adsabs.harvard.edu/abs/2015ApJ...813...75S} {813, 75}

\bibitem[\protect\citeauthoryear{{Shatsky} \& {Tokovinin}}{{Shatsky} \&
  {Tokovinin}}{2002}]{Shatsky2002}
{Shatsky} N.,  {Tokovinin} A.,  2002, \mn@doi [\aap]
  {10.1051/0004-6361:20011542}, \href
  {https://ui.adsabs.harvard.edu/abs/2002A&A...382...92S} {382, 92}

\bibitem[\protect\citeauthoryear{{Shporer} et~al.,}{{Shporer}
  et~al.}{2011}]{Shporer2011}
{Shporer} A.,  et~al., 2011, \mn@doi [\aj] {10.1088/0004-6256/142/6/195}, \href
  {https://ui.adsabs.harvard.edu/abs/2011AJ....142..195S} {142, 195}

\bibitem[\protect\citeauthoryear{{Silsbee} \& {Rafikov}}{{Silsbee} \&
  {Rafikov}}{2015}]{Silsbee2015}
{Silsbee} K.,  {Rafikov} R.~R.,  2015, \mn@doi [\apj]
  {10.1088/0004-637X/798/2/71}, \href
  {https://ui.adsabs.harvard.edu/abs/2015ApJ...798...71S} {798, 71}

\bibitem[\protect\citeauthoryear{{Sullivan} et~al.,}{{Sullivan}
  et~al.}{2015}]{Sullivan2015}
{Sullivan} P.~W.,  et~al., 2015, \mn@doi [\apj] {10.1088/0004-637X/809/1/77},
  \href {https://ui.adsabs.harvard.edu/abs/2015ApJ...809...77S} {809, 77}

\bibitem[\protect\citeauthoryear{{Szab{\'o}} \& {K{\'a}lm{\'a}n}}{{Szab{\'o}}
  \& {K{\'a}lm{\'a}n}}{2019}]{Szabo2019}
{Szab{\'o}} G.~M.,  {K{\'a}lm{\'a}n} S.,  2019, \mn@doi [\mnras]
  {10.1093/mnrasl/slz036}, \href
  {https://ui.adsabs.harvard.edu/abs/2019MNRAS.485L.116S} {485, L116}

\bibitem[\protect\citeauthoryear{{Szab{\'o}} \& {Kiss}}{{Szab{\'o}} \&
  {Kiss}}{2011}]{Szabo2011}
{Szab{\'o}} G.~M.,  {Kiss} L.~L.,  2011, \mn@doi [\apjl]
  {10.1088/2041-8205/727/2/L44}, \href
  {https://ui.adsabs.harvard.edu/abs/2011ApJ...727L..44S} {727, L44}

\bibitem[\protect\citeauthoryear{{Szab{\'o}} et~al.,}{{Szab{\'o}}
  et~al.}{2011}]{Szabo2011b}
{Szab{\'o}} G.~M.,  et~al., 2011, \mn@doi [\apjl] {10.1088/2041-8205/736/1/L4},
  \href {https://ui.adsabs.harvard.edu/abs/2011ApJ...736L...4S} {736, L4}

\bibitem[\protect\citeauthoryear{{Thebault} \& {Haghighipour}}{{Thebault} \&
  {Haghighipour}}{2015}]{Thebault2015}
{Thebault} P.,  {Haghighipour} N.,  2015, {Planet Formation in Binaries}.
Springer, pp 309--340, \mn@doi{10.1007/978-3-662-45052-9_13}

\bibitem[\protect\citeauthoryear{{Th{\'e}bault}, {Marzari}  \&
  {Scholl}}{{Th{\'e}bault} et~al.}{2008}]{Thebault2008}
{Th{\'e}bault} P.,  {Marzari} F.,   {Scholl} H.,  2008, \mn@doi [\mnras]
  {10.1111/j.1365-2966.2008.13536.x}, \href
  {https://ui.adsabs.harvard.edu/abs/2008MNRAS.388.1528T} {388, 1528}

\bibitem[\protect\citeauthoryear{{Tobin} et~al.,}{{Tobin}
  et~al.}{2016}]{Tobin2016}
{Tobin} J.~J.,  et~al., 2016, \mn@doi [\apj] {10.3847/0004-637X/818/1/73},
  \href {https://ui.adsabs.harvard.edu/abs/2016ApJ...818...73T} {818, 73}

\bibitem[\protect\citeauthoryear{{Tokovinin}}{{Tokovinin}}{2000}]{Tokovinin2000}
{Tokovinin} A.~A.,  2000, \aap, \href
  {https://ui.adsabs.harvard.edu/abs/2000A&A...360..997T} {360, 997}

\bibitem[\protect\citeauthoryear{{Tokovinin}}{{Tokovinin}}{2014}]{Tokovinin2014}
{Tokovinin} A.,  2014, \mn@doi [\aj] {10.1088/0004-6256/147/4/86}, \href
  {https://ui.adsabs.harvard.edu/abs/2014AJ....147...86T} {147, 86}

\bibitem[\protect\citeauthoryear{{Tokovinin} \& {Moe}}{{Tokovinin} \&
  {Moe}}{2019}]{Tokovinin2019}
{Tokovinin} A.,  {Moe} M.,  2019, arXiv e-prints, \href
  {https://ui.adsabs.harvard.edu/abs/2019arXiv191001522T} {p. arXiv:1910.01522}

\bibitem[\protect\citeauthoryear{{Tokovinin}, {Thomas}, {Sterzik}  \&
  {Udry}}{{Tokovinin} et~al.}{2006}]{Tokovinin2006}
{Tokovinin} A.,  {Thomas} S.,  {Sterzik} M.,   {Udry} S.,  2006, \mn@doi [\aap]
  {10.1051/0004-6361:20054427}, \href
  {https://ui.adsabs.harvard.edu/abs/2006A&A...450..681T} {450, 681}

\bibitem[\protect\citeauthoryear{{Toonen}, {Hollands}, {G{\"a}nsicke}  \&
  {Boekholt}}{{Toonen} et~al.}{2017}]{Toonen2017}
{Toonen} S.,  {Hollands} M.,  {G{\"a}nsicke} B.~T.,   {Boekholt} T.,  2017,
  \mn@doi [\aap] {10.1051/0004-6361/201629978}, \href
  {https://ui.adsabs.harvard.edu/abs/2017A&A...602A..16T} {602, A16}

\bibitem[\protect\citeauthoryear{{Triaud} et~al.,}{{Triaud}
  et~al.}{2017}]{Triaud2017}
{Triaud} A. H.~M.~J.,  et~al., 2017, \mn@doi [\mnras] {10.1093/mnras/stx154},
  \href {https://ui.adsabs.harvard.edu/abs/2017MNRAS.467.1714T} {467, 1714}

\bibitem[\protect\citeauthoryear{{Wagner}, {Apai}  \& {Kratter}}{{Wagner}
  et~al.}{2019}]{Wagner2019}
{Wagner} K.,  {Apai} D.,   {Kratter} K.~M.,  2019, \mn@doi [\apj]
  {10.3847/1538-4357/ab1904}, \href
  {https://ui.adsabs.harvard.edu/abs/2019ApJ...877...46W} {877, 46}

\bibitem[\protect\citeauthoryear{{Wang}, {Xie}, {Barclay}  \& {Fischer}}{{Wang}
  et~al.}{2014a}]{Wang2014a}
{Wang} J.,  {Xie} J.-W.,  {Barclay} T.,   {Fischer} D.~A.,  2014a, \mn@doi
  [\apj] {10.1088/0004-637X/783/1/4}, \href
  {https://ui.adsabs.harvard.edu/abs/2014ApJ...783....4W} {783, 4}

\bibitem[\protect\citeauthoryear{{Wang}, {Fischer}, {Xie}  \& {Ciardi}}{{Wang}
  et~al.}{2014b}]{Wang2014b}
{Wang} J.,  {Fischer} D.~A.,  {Xie} J.-W.,   {Ciardi} D.~R.,  2014b, \mn@doi
  [\apj] {10.1088/0004-637X/791/2/111}, \href
  {https://ui.adsabs.harvard.edu/abs/2014ApJ...791..111W} {791, 111}

\bibitem[\protect\citeauthoryear{{Wang}, {Fischer}, {Horch}  \& {Huang}}{{Wang}
  et~al.}{2015a}]{Wang2015b}
{Wang} J.,  {Fischer} D.~A.,  {Horch} E.~P.,   {Huang} X.,  2015a, \mn@doi
  [\apj] {10.1088/0004-637X/799/2/229}, \href
  {https://ui.adsabs.harvard.edu/abs/2015ApJ...799..229W} {799, 229}

\bibitem[\protect\citeauthoryear{{Wang}, {Fischer}, {Horch}  \& {Xie}}{{Wang}
  et~al.}{2015b}]{Wang2015c}
{Wang} J.,  {Fischer} D.~A.,  {Horch} E.~P.,   {Xie} J.-W.,  2015b, \mn@doi
  [\apj] {10.1088/0004-637X/806/2/248}, \href
  {https://ui.adsabs.harvard.edu/abs/2015ApJ...806..248W} {806, 248}

\bibitem[\protect\citeauthoryear{{Wang}, {Fischer}, {Xie}  \& {Ciardi}}{{Wang}
  et~al.}{2015c}]{Wang2015d}
{Wang} J.,  {Fischer} D.~A.,  {Xie} J.-W.,   {Ciardi} D.~R.,  2015c, \mn@doi
  [\apj] {10.1088/0004-637X/813/2/130}, \href
  {https://ui.adsabs.harvard.edu/abs/2015ApJ...813..130W} {813, 130}

\bibitem[\protect\citeauthoryear{{Ward-Duong} et~al.,}{{Ward-Duong}
  et~al.}{2015}]{WardDuong2015}
{Ward-Duong} K.,  et~al., 2015, \mn@doi [\mnras] {10.1093/mnras/stv384}, \href
  {https://ui.adsabs.harvard.edu/abs/2015MNRAS.449.2618W} {449, 2618}

\bibitem[\protect\citeauthoryear{{White} \& {Ghez}}{{White} \&
  {Ghez}}{2001}]{White2001}
{White} R.~J.,  {Ghez} A.~M.,  2001, \mn@doi [\apj] {10.1086/321542}, \href
  {https://ui.adsabs.harvard.edu/abs/2001ApJ...556..265W} {556, 265}

\bibitem[\protect\citeauthoryear{{Winn} \& {Fabrycky}}{{Winn} \&
  {Fabrycky}}{2015}]{Winn2015}
{Winn} J.~N.,  {Fabrycky} D.~C.,  2015, \mn@doi [\araa]
  {10.1146/annurev-astro-082214-122246}, \href
  {https://ui.adsabs.harvard.edu/abs/2015ARA&A..53..409W} {53, 409}

\bibitem[\protect\citeauthoryear{{Winters} et~al.,}{{Winters}
  et~al.}{2019}]{Winters2019}
{Winters} J.~G.,  et~al., 2019, \mn@doi [\aj] {10.3847/1538-3881/ab05dc}, \href
  {https://ui.adsabs.harvard.edu/abs/2019AJ....157..216W} {157, 216}

\bibitem[\protect\citeauthoryear{{Wright}, {Marcy}, {Howard}, {Johnson},
  {Morton}  \& {Fischer}}{{Wright} et~al.}{2012}]{Wright2012}
{Wright} J.~T.,  {Marcy} G.~W.,  {Howard} A.~W.,  {Johnson} J.~A.,  {Morton}
  T.~D.,   {Fischer} D.~A.,  2012, \mn@doi [\apj]
  {10.1088/0004-637X/753/2/160}, \href
  {https://ui.adsabs.harvard.edu/abs/2012ApJ...753..160W} {753, 160}

\bibitem[\protect\citeauthoryear{{Xie}, {Zhou}  \& {Ge}}{{Xie}
  et~al.}{2010}]{Xie2010}
{Xie} J.-W.,  {Zhou} J.-L.,   {Ge} J.,  2010, \mn@doi [\apj]
  {10.1088/0004-637X/708/2/1566}, \href
  {https://ui.adsabs.harvard.edu/abs/2010ApJ...708.1566X} {708, 1566}

\bibitem[\protect\citeauthoryear{{Zhou} et~al.,}{{Zhou}
  et~al.}{2019}]{Zhou2019}
{Zhou} G.,  et~al., 2019, \mn@doi [\aj] {10.3847/1538-3881/ab36b5}, \href
  {https://ui.adsabs.harvard.edu/abs/2019AJ....158..141Z} {158, 141}

\bibitem[\protect\citeauthoryear{{Ziegler} et~al.,}{{Ziegler}
  et~al.}{2018}]{Ziegler2018}
{Ziegler} C.,  et~al., 2018, \mn@doi [\aj] {10.3847/1538-3881/aace59}, \href
  {https://ui.adsabs.harvard.edu/abs/2018AJ....156...83Z} {156, 83}

\bibitem[\protect\citeauthoryear{{Ziegler}, {Tokovinin}, {Brice{\~n}o}, {Mang},
  {Law}  \& {Mann}}{{Ziegler} et~al.}{2020}]{Ziegler2019}
{Ziegler} C.,  {Tokovinin} A.,  {Brice{\~n}o} C.,  {Mang} J.,  {Law} N.,
  {Mann} A.~W.,  2020, \mn@doi [\aj] {10.3847/1538-3881/ab55e9}, \href
  {https://ui.adsabs.harvard.edu/abs/2020AJ....159...19Z} {159, 19}

\bibitem[\protect\citeauthoryear{{Zong} et~al.,}{{Zong}
  et~al.}{2018}]{Zong2018}
{Zong} W.,  et~al., 2018, \mn@doi [\apjs] {10.3847/1538-4365/aadf81}, \href
  {https://ui.adsabs.harvard.edu/abs/2018ApJS..238...30Z} {238, 30}

\makeatother
\end{thebibliography}

\appendix

\section{Close Binary Fraction for Field Metallicity}
\label{Appendix}

\subsection{Solar-type Binaries}
\label{Solar}

 We measure the binary fractions $F_{\rm a<10au}$ and $F_{\rm a<100au}$ of field solar-type stars, making sure to include all stellar-mass companions corrected down to $M_2$~$=$~0.08\,\Msun, including WD companions. It is difficult to detect WD companions to solar-type and especially early-type primaries without multi-epoch RV observations or high-resolution imaging \citep{Holberg2016,Toonen2017}. Transiting surveys like {\it Kepler} and {\it TESS} therefore targeted AFGK-dwarfs with no measurable bias against such Sirius-like binaries. Similarly, all the multiplicity surveys of solar-type stars investigated below did not exclude systems with known WD companions. However, \citet{Duquennoy1991} explicitly accounted for incompleteness of faint WD companions, leading to their larger bias-corrected binary fraction, whereas \citet{Raghavan2010} did not.
 
 According to the \citet{Duquennoy1991} bias-corrected period distribution of solar-type binaries (their Fig.~7; 164 primaries), which includes both MS and WD companions, there are 35 and 62 binaries with periods below log\,$P$\,(days)~$<$~4.0 ($a$~$<$~10~au) and log\,$P$\,(days)~$<$~5.5 ($a$~$<$~100~au), respectively. This yields $F_{\rm a<10au}$ = 35/164 = 0.21\,$\pm$\,0.04 and $F_{\rm a<100au}$ = 62/164 = 0.38\,$\pm$\,0.05. The \citet{Duquennoy1991} survey spans spectral types F7-G9 and luminosity classes IV-VI ($M_1$~=~0.90\,-\,1.21\,\Msun), providing an average mass of $\langle M_1 \rangle$~=~1.02\,\Msun\ (see Fig.~\ref{closebin}).  

 Within the \citet{Raghavan2010} 25-pc sample of 454 solar-type primaries, 65 and 110 have known inner binary companions with estimated separations below $a$~$<$~10~au and $a$~$<$~100~au, respectively. These numbers do not include outer tertiaries in hierarchical triples nor triples in A-(Ba,Bb) configurations in which a solar-type primary orbits a close pair of M-dwarfs (see Section~8 in \citealt{Moe2017} for a detailed discussion).  The {\it uncorrected} binary fractions are therefore $F_{\rm a<10au}$~=~65/454~=~0.14\,$\pm$\,0.02 and $F_{\rm a<100au}$~=~110/454 = 0.24\,$\pm$\,0.02, respectively. However, the \citet{Raghavan2010} survey is measurably incomplete \citep{Chini2014,Moe2017}.  In particular, \citet{Moe2017} emphasized that the majority of late-M and WD companions with intermediate separations of $a$~$\approx$~5\,-\,30~au were missed by \citet{Raghavan2010}.  After accounting for incompleteness, \citet{Moe2017} reported a corrected MS binary fraction of 15\%\,$\pm$\,3\% across 0.2~$<$~log\,$P$\,(days)~$<$~3.7 ( $a$~=~0.03\,-\,6\,au) and $q$~=~$M_2$/$M_1$~$>$~0.1 ($M_2$~$>$~0.10\,\Msun; see their Fig.~42 and Table~13).  Considering the few additional binaries with $a$~$<$~0.03~au, $a$~=~6\,-\,10~au, and $M_2$~=~0.08\,-\,0.10\,\Msun, then we estimate that 17\%\,$\pm$\,3\% of solar-type stars have MS companions below $a$~$<$~10~au.  \citet{Moe2017} also demonstrated that 30\%\,$\pm$\,10\% of SB1s, i.e., 20\%\,$\pm$\,6\% of all close solar-type binaries, contain WD companions (see also \citealt{Murphy2018}).  The total solar-type binary fraction below $a$~$<$~10~au, including MS and WD companions, is therefore $F_{\rm a<10au}$~=~0.22\,$\pm$\,0.04.  Making similar corrections to the \citet{Raghavan2010} sample of slightly wider binaries, we estimate $F_{\rm a<100au}$ = 0.37\,$\pm$\,0.05.  The \citet{Raghavan2010} 25-pc survey covers spectral types F6-K3 and luminosity classes IV-V ($M_1$~=~0.75\,-\,1.25\,\Msun), yielding a mean metallicity of $\langle$[Fe/H]$\rangle$~=~$-$0.15 and an average primary mass of $\langle M_1 \rangle$~=~0.95\,\Msun\ (see Fig.~\ref{closebin}).

\citet{Moe2019} recently compiled a variety of solar-type binary surveys and reported a bias-corrected close binary fraction of $F_{\rm a<10au}$~=~0.24\,$\pm$\,0.04 for the mean metallicity of the field.  Their tightest constraints derive from the APOGEE RV variability survey \citep{Badenes2018}, which includes both MS and WD companions. Anchoring the observed log-normal period distribution of solar-type stars to $F_{\rm a<10au}$~=~0.24\,$\pm$\,0.04, then 18\%\,$\pm$\,3\% of solar-type stars have companions across $a$~=~10\,-\,100~au. About 20\%\,$\pm$\,6\% of such companions with intermediate separations are outer tertiaries in hierarchical triples \citep{Raghavan2010,Tokovinin2014}, and so only 15\%\,$\pm$\,3\% of solar-type stars have inner binary companions across $a$~=~10\,-\,100~au.  We display both $F_{\rm a<10au}$ = 0.24\,$\pm$\,0.04 and $F_{\rm a<100au}$ = $F_{\rm a<10au}$ + 0.15\,$\pm$\,0.03 = 0.39\,$\pm$\,0.05 at $M_1$~=~1.00\,\Msun\ in Fig.~\ref{closebin}. 

The \citet{Tokovinin2014} 67-pc sample of F/G-type multiples provide important consistency checks for our adopted model of close binaries. We ignore the F0-F4 stars above the Kraft break, as it is more difficult to detect RV variations from rotationally broadened profiles. We analyze the 4,494 F5-G9\,IV/V primaries in the \citet{Tokovinin2014} sample, providing a median spectral type of G1\,V ($M_1$~=~1.03\,\Msun). We focus on SBs with known orbital periods across $P$~=~1\,-\,100~days, which are relatively complete within $d$~$<$~30~pc given the average 0.3~km~s$^{-1}$ sensitivity and cadence of the RV observations (see below). In Fig.~\ref{TokFig}, we plot the primary SB fraction $F_{\rm SB;1-100}$ across $P$~=~1\,-\,100~days as a cumulative function of distance (blue histogram). With increasing distance, the SB sample becomes less complete, falling from  $F_{\rm SB;1-100}$~=~5.9\%\,$\pm$\,1.1\% within $d$~$<$~30~pc to 3.6\,\% $\pm$\,0.3\% within $d$~$<$~65~pc. 

\citet{Tokovinin2014} also reported the number $N_{\rm RV}$ of RV epochs for each primary. Even $N_{\rm RV}$~=~3 epochs are sufficient in detecting RV variability from the majority of binaries within $P$~$<$~100~days, but it typically requires $N_{\rm RV}$~>~7 epochs to fit unique orbital periods and to be complete toward all stellar-mass companions. We display the fraction of primaries with $N_{\rm RV}$~$>$~5 epochs as the red histogram in Fig.~\ref{TokFig}, where the upper and lower error bars correspond to $N_{\rm RV}$~$>$~3 and $>$~7, respectively. About 90\% of the F5-G9\,IV/V primaries within $d$~$<$~30~pc have $N_{\rm RV}$~$>$~5 epochs, whereas only 54\% of all primaries in the 67-pc sample are this complete. Dividing the observed SB fraction by the completeness fraction yields the corrected SB fraction (black histogram), which is consistent with $F_{\rm SB;1-100}$~=~6.0\% across all distances. 

Some studies of planet suppression by close binaries, e.g., \citet{Ngo2016}, scaled the log-normal period distribution of all companions to the observed binary fraction of $F_{\rm bin}$~=~0.46. This inconsistency results in a close SB fraction of $F_{\rm SB;1-100}$~=~3.9\% that underestimates the true value by a factor of 1.6. In our volume-limited model of solar-type systems (Fig.~\ref{solarperiod}), we instead scale the log-normal period distribution to a multiplicity frequency of $f_{\rm mult}$~=~0.67 companions per primary, which includes both inner binaries and outer tertiaries in triples. Integrating our distribution across $P$ =~1\,-\,100~days yields a close SB fraction of $F_{\rm SB;1-100}$~=~6.1\% (green dashed line in Fig.~\ref{TokFig}), matching both the observed value within $d$~$<$~30~pc and the completeness-corrected value for larger volumes.

\begin{figure}
\centerline{
\includegraphics[trim=0.3cm 0.3cm 0.2cm 0.4cm, clip=true, width=3.3in]{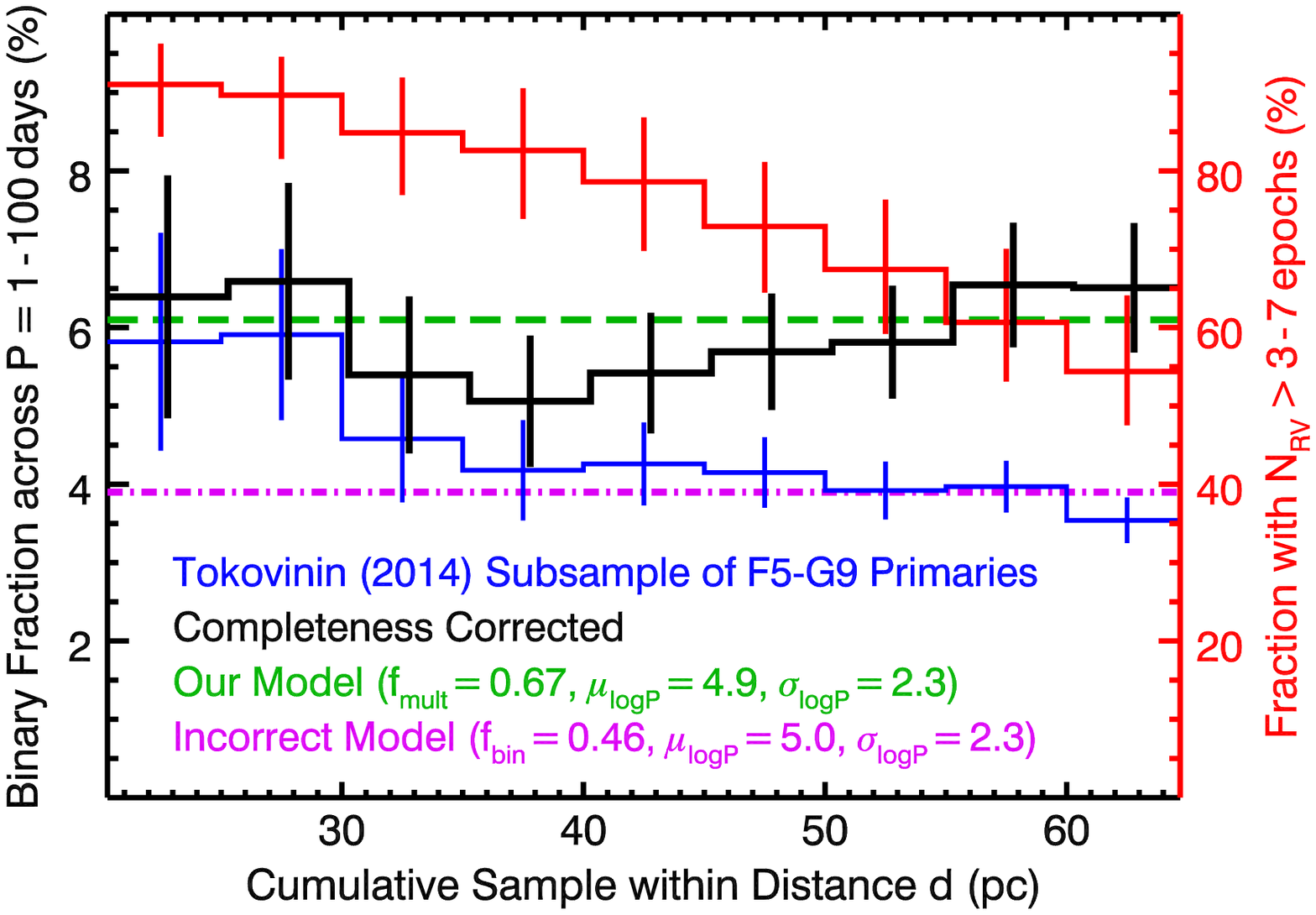}}
\caption{The close SB fraction across $P$~=~1\,-\,100~days of solar-type primaries in the \citet{Tokovinin2014} sample as a cumulative function of distance (blue). We also display the fraction of primaries with $N_{\rm RV}$~$>$~5\,$\pm$\,2 RV epochs (red; right axis). Dividing the observed SB fraction by the completeness fraction yields the bias-corrected SB fraction (black). Some previous studies of planet suppression by close binaries incorrectly scaled the period distribution of all MS companions, including outer tertiaries, to the observed binary fraction, which significantly underestimates the true close binary fraction (dash-dotted magenta). In our model of solar-type systems, we instead normalize the canonical log-normal period distribution to a bias-corrected multiplicity frequency of $f_{\rm mult}$~=~0.67 (including both MS and WD companions), which reproduces the observed binary fraction across $P$~=~1\,-\,100~days (dashed green).  }
\label{TokFig}
\end{figure}

\subsection{Early-type Binaries}

The binary fraction within $a$~$<$~100~au substantially increases with primary mass, nearly reaching 100\% for O/early-B primaries \citep{Abt1990,Sana2012,Duchene2013,Sana2014,Moe2017,Moe2019}.  \citet{Moe2013} estimated that the frequency of very close companions ($P$~$<$~20~days) with mass ratios $q$~$>$~0.1 scales as $F$~$\propto$~$M_1^{0.7}$.  The slope is slightly steeper if we include extreme mass-ratio companions with $q$~=~0.05\,-\,0.10 that closely orbit early-type primaries \citep{Moe2015a,Murphy2018}.  We estimate that the ratio of the A-type ($M_1$~=~2.0\,\Msun) to G-type ($M_1$~=~1.0\,\Msun) very close binary fraction is $R_{\rm A/G}$($a$\,$<$\,0.2\,au)~=~1.7\,$\pm$\,0.3.  Across intermediate periods $P$~=~100\,-\,1,500 days ($a$~=~0.5\,-\,3~au), \citet{Murphy2018} reported that the binary fraction of late-A stars is 2.1\,$\pm$\,0.3 times larger than that observed for solar-type stars across the same period interval. The ratio $R_{\rm A/G}$($a$\,$=$\,0.5\,-\,3\,au)~=~2.1\,$\pm$\,0.3 already incorporates both WD and MS companions corrected down to $M_2$~=~0.08\,\Msun.   At wider separations ($a$~$>$~50~au), stellar companions with $q$~$>$~0.1 orbiting A-type MS stars are only 30\% more frequent compared to solar-type primaries \citep{DeRosa2014,Moe2017}. Accounting for late-M and WD companions, we adopt $R_{\rm A/G}$($a$\,$>$\,50\,au)~=~1.4\,$\pm$\,0.2.  We interpolate $R_{\rm A/G}$ with respect to log\,$a$.  

The separation distribution of companions to A-type primaries is $f_{\rm A}$ = $R_{\rm A/G}$\,$f_{\rm G}$, where $f_{\rm G}$ follows the canonical G-type log-normal separation distribution scaled to $F_{\rm a<10au}$~=~0.23 (Section \ref{Solar}). The total frequency of companions to A-type stars is 0.45\,$\pm$\,0.06 below $a$~$<$~10~au  and 0.26\,$\pm$\,0.04 across $a$~$=$~10\,-\,100~au.  Based on the discussion of intermediate-mass multiples in \citet{Moe2017}, we estimate that 10\%\,$\pm$\,5\% of the companions below $a$~$<$~10~au and 35\%\,$\pm$\,10\% of the companions across $a$~=~10\,-\,100~au are outer tertiaries in hierarchical triples. For $M_1$~=~2.0\,\Msun, we adopt $F_{\rm a<10au}$ = (0.90\,$\pm$\,0.05)(0.45\,$\pm$\,0.06) = 0.41\,$\pm$\,0.06 and $F_{\rm a<100au}$ = $F_{\rm a<10au}$ + (0.65\,$\pm$\,0.10)(0.26\,$\pm$\,0.04) = 0.58\,$\pm$\,0.08.

The binary fraction of B-type MS stars ($M_1$~=~2.5\,-\,17\,\Msun) is 70\%\,-\,95\%, depending on the mass, and most wide companions ($a$~$>$~100~au) to B-type MS primaries are  outer tertiaries in hierarchical triples \citep{Abt1990, Shatsky2002, Kouwenhoven2007, Kobulnicky2007, Rizzuto2013, Moe2017}.  Hence, the binary fraction of B-type MS stars below $a$~$<$~100~au is 70\%\,-\,90\%, increasing slightly with $M_1$. Based on the compilation of surveys investigated in \citet{Moe2017}, \citet{Moe2019} reported $F_{\rm a<10au}$~=~0.70\,$\pm$\,0.11 and $F_{\rm a<100au}$~$=$~0.90\,$\pm$\,0.10 for $M_1$~=~10\,\Msun.  Following a similar method, we calculate $F_{\rm a<10au}$~=~0.57\,$\pm$\,0.10 and $F_{\rm a<100au}$~$=$~0.77\,$\pm$\,0.12 for $M_1$~=~5\,\Msun\ (see Fig.~\ref{closebin}). The relatively large error bars derive from the uncertainties in the frequency of close M-dwarf companions to B-type primaries ( $q$~$<$~0.1) and the occurrence rate of single runaway B-type stars.

\subsection{M-dwarf Binaries}

We next utilize several surveys of M-dwarf binaries to measure the close binary fraction of low-mass stars. Unlike our measurements for solar-type and early-type primaries, we exclude WD companions to M-dwarfs. WD companions typically dominate at blue wavelengths compared to M-dwarfs. For example, the nearest M-dwarf + WD binary is Stein~2051, which contains a M4V star with B~=~12.8~mag and a DC5 WD with B~=~12.7~mag (SIMBAD\footnote{https://simbad.u-strasbg.fr/simbad/}). Unresolved WD companions to M-dwarfs are thus easily discernible based on their composite spectra or observed colors. Planet surveys therefore rarely target M-dwarfs with close WD companions. Similarly, multiplicity surveys of M-dwarfs, such as the \citet{Winters2019} volume-limited sample, also purposefully exclude systems with known WD companions.

\citet{Fischer1992} examined 60 M-dwarfs, mostly M2V-M5V ($\langle M_1 \rangle$ = 0.34\,\Msun), within 20~pc for spectroscopic and visual companions.  After correcting for incompleteness down to $M_2$~=~0.08\,\Msun, they reported an overall binary fraction of 0.42\,$\pm$\,0.09.  In their visual binary sample of 58 M-dwarfs, they found 15 (26\%\,$\pm$\,6\%) and 9 (16\%\,$\pm$\,5\%) binaries beyond $a$~$>$~10~au and $a$~$>$~100~au, respectively.  This provides $F_{\rm a<10au}$ = 0.42\,$-$\,0.26 = 0.16\,$\pm$\,0.05 and $F_{\rm a<100au}$= 0.42\,$-$\,0.16 = 0.26\,$\pm$\,0.06 (see Fig.~\ref{closebin}). 

We examine different imaging surveys below that spatially resolve companions to nearby M-dwarfs.  Although such visual companions comprise the majority of nearby M-dwarf binaries, we must account for the few M-dwarf companions within $a$~$<$~3 au that can be identified only with spectroscopic or eclipsing techniques.  After correcting for incompleteness in their radial velocity observations, \citet{Fischer1992} estimated that 2\%\,$\pm$\,1\% and 6\%\,$\pm$\,3\% of early-M dwarfs have stellar companions below $a$~$<$~0.4~au and $a$~$<$~3~au, respectively.  Similarly, \citet{Clark2012} estimated that 1\% of late-M primaries and 3\% of early-M primaries have stellar companions with very close separations inside of $a$~$<$~0.4~au. \citet{Shan2015} measured the EB fraction of M-dwarfs in the {\it Kepler} field. After accounting for the geometrical probability of eclipses, they found that 11\%\,$\pm$\,3\% of early-M stars have companions within $a$~$<$~0.4~au.  Their result is a factor of three times larger than that measured by the previously cited spectroscopic surveys of early-M dwarfs.  \citet{Shan2015} noted their value may be overestimated due to various selection effects, e.g., Malmquist bias. In the following, we consider that 4\%\,$\pm$\,2\% of late-M stars and 8\%\,$\pm$\,3\% of early-M stars have stellar companions below $a$~$<$~3~au.  For reference, 12\%\,$\pm$\,3\% of solar-type stars have MS (non-WD) companions below this separation limit (see Section~\ref{Solar}).  

 \citet{Bergfors2010} performed a lucky imaging survey of 108 M0-M6 dwarfs within 52 pc (median spectral type of M3V corresponding to $M_1$~=~0.36\,\Msun). They measured a corrected binary fraction of 32\%\,$\pm$\,6\% across $a$ = 3\,-\,180~au. We remove the 12 companions (11\%\,$\pm$\,3\%) that have projected separations  beyond $\rho$~$>$~80~au ($a$~$>$~100~au) and/or have spectral types later than M8 ($M_2$~$<$~0.08\,\Msun; their Table~3).  We add the expected 8\%\,$\pm$\,3\% of stellar companions below $a$~$<$~3~au, resulting in $F_{\rm a<100au}$~=~0.29\,$\pm$\,0.06. \citet{Bergfors2010} resolved five stellar companions (5\%\,$\pm$\,2\%) across projected separations of $\rho$ = 3\,-\,10 au.  Accounting for the 8\%\,$\pm$\,3\% of early-M stars with companions below $a$~$<$~3~au, we estimate $F_{\rm a<10au}$~=~0.13\,$\pm$\,0.04.

\citet{Janson2012} also utilized lucky imaging to identify visual companions for a much larger sample of 701 M-type and 60 late-K stars.  Based on their constrained volume-limited sample of 337 M0-M5 primaries (median spectral type of M2.5V corresponding to $M_1$ = 0.42\,\Msun), they found 85 multiples and reported a corrected stellar binary fraction of 27\%\,$\pm$\,3\% across $a$~=~3\,-\,230 au.  Based on the listed parameters in their Table~3, we find that 6\%\,$\pm$\,1\%, 15\%\,$\pm$\,2\%, and 6\%\,$\pm$\,1\% (totaling their reported 27\%\,$\pm$\,3\%) of the early-M primaries in their constrained sample have inner binary companions across $a$~=~3\,-\,10\,au, 10\,-\,100\,au, and 100\,-\,230\,au, respectively.  After adding the estimated 8\%\,$\pm$\,3\% of systems with companions below $a$~$<$~3~au, we adopt $F_{\rm a<10au}$~=~0.14\,$\pm$\,0.03 and $F_{\rm a<100au}$~=~0.29\,$\pm$\,0.04.  

Utilizing high-resolution {\it HST} imaging, \citet{Dieterich2012} searched for companions to 255 stars within 10~pc.  Based on their subsample of 126 M-dwarfs (median spectral type of M3.5V corresponding to $M_1$~=~0.26\,\Msun), they reported a corrected stellar companion fraction of 10\%\,$\pm$\,3\% and a brown dwarf companion fraction of $<$\,2\% across $a$~=~5\,-\,70~au.  Of their 11 detections across $a$~=~5\,-\,70~au, five have short separations within $a$~=~5\,-\,10~au (Table~5 in \citealt{Dieterich2012}).  We estimate that an additional 8\%\,$\pm$\,3\% of mid-M stars have stellar companions below $a$~$<$~5~au and an additional 1\% across $a$ = 70\,-\,100~au, resulting in $F_{\rm a<10au}$~=~0.12\,$\pm$\,0.04 and $F_{\rm a<100au}$~=~0.19\,$\pm$\,0.05. 

\citet{WardDuong2015} incorporated both adaptive optics and common proper motion to search for wide companions to 245 K7-M6 dwarfs (average primary mass of $M_1$ = 0.44\,\Msun\ according to their Fig.~4). They reported a corrected binary fraction of 24\%\,$\pm$\,3\% across $a$~=~3\,-\,10,000~au. In their adaptive optics subsample of 196 objects, we count 12 companions and 20 companions across projected separations of $\rho$~=~3\,-\,10~au and $\rho$ = 10\,-\,100~au, respectively (their Fig.~4).  This provides a binary fraction of (12+20)/196 = 16\%\,$\pm$\,3\% across $a$ = 3\,-\,100~au, which is consistent with their bias-corrected separation distribution presented in their Fig.~16. \citet{WardDuong2015} also resolved 10 companions across small projected separations of $\rho$~=~0.5\,-\,3~au, but such very close companions are incomplete in their adaptive optics survey (see their Fig.~12).  The observed number of very close companions provides a firm lower limit of $>$\,5\% for the binary fraction below $a$~$<$~3~au.  We therefore adopt 9\%\,$\pm$\,3\% for the frequency of companions within $a$~$<$~3~au to early-M/late-K stars, resulting in $F_{\rm a<10au}$~=~0.15\,$\pm$\,0.03 and $F_{\rm a<100au}$~=~0.25\,$\pm$\,0.04. 

\citet{Winters2019} recently compiled a list of all known stellar companions to M-dwarfs within 25~pc by combining an exhaustive literature search with their own high-contrast imaging surveys. We focus on their subset of 188 M-dwarf systems within 10~pc ($\langle M_1 \rangle$~=~0.30\,\Msun\ based on their Table 4), which is relatively complete toward stellar companions across all orbital periods. In this subsample, \citet{Winters2019} reported 56 M-dwarf pairs, which provides a total stellar binary fraction of 56/188 = 30\%\,$\pm$\,4\% that is consistent with their bias-corrected 25-pc value of 27\%. Adopting their conversion factor $a$~=~1.26$\rho$ between orbital and projected separations, then there are 32 M-dwarf binaries with $a$~$<$~10~au and only 12 additional companions with $a$~=~10\,-\,100~au in their 10-pc subsample, resulting in $F_{\rm a<10au}$ = 32/188 = 0.17\,$\pm$\,0.03 and $F_{\rm a<100au}$ = 44/188 = 0.23\,$\pm$\,0.03, respectively. All six surveys of early-M and mid-M dwarfs result in close binary fractions that are consistent with each other (see Fig.~\ref{closebin}). 

Finally, we quantify the frequency of close late-M + late-M binaries.  \citet{Basri2006} performed a spectroscopic RV variability survey of 53 very low-mass objects with spectral types later than M5.  Of the 34 stars with spectral types M5-M8 in their sample ($M_1$~=~0.08\,-\,0.15\,\Msun; median of $M_1$ = 0.11\,\Msun), only two (6\%\,$\pm$\,4\%) are spectroscopic binaries with $a$~$<$~6~au.  Based on a literature review, \citet{Basri2006} estimated that the fraction of low-mass stars with companions across $a$~=~6\,-\,20 au is similar to that below $a$~$<$~6~au, and they concluded that the frequency at wider separations beyond $a$~$>$~20~au is negligible.  We therefore estimate that $F_{\rm a<10au}$~=~0.10\,$\pm$\,0.05 and $F_{\rm a<100au}$~=~0.12\,$\pm$\,0.06 of $M_1$~=~0.11\,\Msun\ primaries have $M_2$~=~0.08\,-\,0.11\,\Msun\ stellar MS companions within $a$~$<$~10~au and $a$~$<$~100~au, respectively.  

\citet{Law2008} performed a lucky imaging survey of 77 low-mass stars across a narrow interval of primary spectral types M4.5-M5.5 ($M_1$ = 0.12\,-\,0.18\,\Msun; median of 0.15\,\Msun).  They detected 12 and 9 stellar late-M companions across projected separations of $\rho$ = 2\,-\,10~au and $\rho$~=~10\,-\,80~au, respectively.  Because late-M binaries are significantly weighted toward twin mass ratios, the observed magnitude-limited close binary fraction is $f_{\rm Malmquist}$~=~2.0 times the true volume-limited close binary fraction  \citep{Burgasser2003,Law2008}. After accounting for Malmquist bias in their magnitude-limited survey, \citet{Law2008} reported a 14\%\,$\pm$\,5\% binary star fraction across $a$~=~2\,-\,80~au. After adding the 3\%\,$\pm$\,2\% of stellar companions with very close separations below $a$~$<$~2~au, we estimate $F_{\rm a<10au}$~=~0.11\,$\pm$\,0.04 and $F_{\rm a<100au}$~=~0.17\,$\pm$\,0.05 for $M_1$ = 0.15\,\Msun\ (see Fig.~\ref{closebin}).

\bsp
\label{lastpage}
\end{document}